\documentclass[transmag]{IEEEtran}
\pdfoutput=1
\IEEEoverridecommandlockouts
\usepackage[noadjust]{cite}
\usepackage[mathscr]{euscript}
\usepackage{algorithm}
\usepackage[noend]{algpseudocode}
\usepackage{array}
\usepackage{mdwmath}
\usepackage{mathtools}
\usepackage{eqparbox}
\usepackage{mdframed}
\usepackage{steinmetz}
\usepackage{stfloats}
\usepackage{graphicx}
\usepackage{url}
\usepackage{tikz}
\usepackage{pgfplots}
\usepackage{subcaption}
\graphicspath{{../pdf/}{../jpeg/}}
\DeclareGraphicsExtensions{.pdf,.jpeg,.png}
\usepackage{epstopdf}
\usepackage{amsfonts}

\newtheorem{lemma}{Lemma}

\usepackage{latexsym}
\usepackage{amsfonts,amssymb,amsmath}
\usepackage{hyperref}
\def\BibTeX{{\rm B\kern-.05em{\sc i\kern-.025em b}\kern-.08em T\kern-.1667em\lower.7ex\hbox{E}\kern-.125emX}}
\begin{document}
\title{Intelligent Reflecting Surface Assisted Multi-User MISO Communication: Channel Estimation and Beamforming Design}
\author{Qurrat-Ul-Ain Nadeem, \IEEEmembership{Member, IEEE},  Hibatallah Alwazani, Abla Kammoun, \IEEEmembership{Member, IEEE}, Anas \\ Chaaban, \IEEEmembership{Senior Member, IEEE},
        M{\'e}rouane Debbah, \IEEEmembership{Fellow, IEEE}, and Mohamed-Slim Alouini, \IEEEmembership{Fellow, IEEE}
				\thanks{Q.-U.-A. Nadeem, H. Alwazani, and A. Chaaban are with School of Engineering, The University of British Columbia, Kelowna, Canada (email: \{qurrat.nadeem, hibat97, anas.chaaban\}@ubc.ca)}
\thanks{A. Kammoun and M.-S. Alouini are with the Computer, Electrical and Mathematical Sciences and Engineering (CEMSE) Division, King Abdullah University of Science and Technology (KAUST), Thuwal, Saudi Arabia 23955-6900 (e-mail: \{abla.kammoun,slim.alouini\}@kaust.edu.sa)}
\thanks{M. Debbah is jointly with Universit{\'e} Paris-Saclay, CNRS, CentraleSup{\'e}lec, 91190 Gif-sur-Yvette, France and The Huawei Mathematical and Algorithmic Sciences Lab, 92100 Boulogne, Billancourt, France (e-mail: merouane.debbah@centralesupelec.fr).}
}

\IEEEtitleabstractindextext{\begin{abstract}
The concept of reconfiguring wireless propagation environments using intelligent reflecting surfaces (IRS)s has recently emerged, where an IRS comprises of a large number of passive reflecting elements that can smartly reflect the impinging electromagnetic waves for performance enhancement. Previous works have shown promising gains assuming the availability of perfect channel state information (CSI) at the base station (BS) and the IRS, which is impractical due to the passive nature of the reflecting elements.  This paper makes one of the preliminary contributions of studying an IRS-assisted multi-user multiple-input single-output (MISO) communication system under imperfect CSI. Different from the few recent works that develop  least-squares (LS) estimates of the IRS-assisted channel vectors, we exploit the  prior knowledge of the large-scale fading statistics at the BS to  derive the Bayesian minimum mean squared error (MMSE) channel estimates under a protocol in which the IRS applies a set of optimal phase shifts vectors over multiple channel estimation sub-phases. The resulting mean squared error (MSE) is both analytically and numerically shown to be lower than that achieved by the LS estimates. Joint designs for the precoding and power allocation at the BS and reflect beamforming at the IRS are proposed to maximize the minimum user signal-to-interference-plus-noise ratio (SINR) subject to a transmit power constraint. Performance evaluation results illustrate the efficiency of the proposed system and study its susceptibility to channel estimation errors.

\end{abstract}

}
\maketitle 

\section{Introduction}

\IEEEPARstart{M}{assive}  multiple-input multiple-output (MIMO) communication, millimeter wave (mmWave) communication, and network densification are some of the main technological advancements that are leading the emergence of Fifth Generation (5G) mobile communication systems. However, these technologies face two main practical limitations. First, they consume a lot of power, which is a critical issue for practical implementation  and second, they struggle to provide the users with uninterrupted connectivity and a guaranteed quality of service (QoS) in harsh propagation environments, due to the lack of control over the wireless propagation channel. For example: the network's total energy consumption scales linearly as more base stations (BS)s are added to densify the network, while each active antenna element in a massive MIMO array is connected to a radio frequency (RF) chain comprising of several active components, rendering the total cost and energy consumption to be very high. Moreover, massive MIMO performance is known to suffer when the propagation environment exhibits poor scattering conditions \cite{ourworkTWC}, whereas, communication at mmWave frequencies suffers from high path and penetration losses. These two limitations have resulted in the need for green and sustainable future cellular networks, where the network operator has some control over the propagation environment.

An emerging concept that addresses this need is that of a smart radio environment, where the wireless propagation environment is turned into an intelligent reconfigurable space that plays an active role in transferring radio signals from the transmitter to the receiver \cite{SRE,SRE1,LISA, basar}. This concept is enabled by the use of intelligent reflecting surfaces (IRSs) in the environment, that shape the impinging electromagnetic (EM) waves in desired ways in a passive manner, without generating new radio signals and thereby without incurring any additional power consumption. Several current research activities  focus on developing different converging solutions to implement these IRSs, including fabricating new meta-surfaces and reflect arrays, making them re-configurable, implementing testbeds and generating experimental results \cite{SRE, RA, RAAA, LIS_mag1, LIS4,MS_nature, Viso,LIS_mag2}.

 Very recently, works approaching this subject from the wireless communication design and analysis perspective have appeared, which view the IRS as a planar array of a large number of passive reflecting elements,  each of which can independently induce a phase shift onto the incident EM waves and reflect them passively. By adaptively and intelligently adjusting the phase shifts of all the IRS elements, referred to as passive beamforming or reflect beamforming \cite{LIS, LIS_new}, desired communication objectives can be realized.  In the last year, several joint designs for precoding at the BS and phase shifts matrix at the IRS have been proposed to achieve different communication goals, for example: maximize the system's energy efficiency subject to the individual signal-to-interference-plus-noise ratio (SINR) constraints at the users in \cite{8741198}, maximize the minimum user rate subject to a transmit power constraint in the asymptotic regime in \cite{LIS_jour}, minimize the transmit power at the BS subject to users' individual SINR constraints in \cite{LIS_new} and maximize the sum-rate subject to a transmit power constraint in \cite{RISn, WSR}. Moreover the use of IRS to maximize the minimum secrecy rate for physical layer security has been studied in \cite{phy} and to assist in simultaneous wireless information and power transfer has been studied in \cite{SWIPT}. IRSs have also found applications in wide-band orthogonal frequency division multiplexing (OFDM) systems in \cite{IRSOFDM} and  non-orthogonal multiple-access systems in \cite{NOMA}. 

A vast majority of the existing  works assume the availability of perfect perfect channel state information (CSI) to design the precoding vectors at the BS and phase shifts matrix at the IRS. However, this assumption is highly unlikely to hold in practice for an IRS-assisted system. This is because as opposed to conventional multi-antanna and relay-assisted communication systems, where channels can be estimated by actively sending, receiving and processing pilot symbols, the IRS has no radio resources of its own to send and receive pilot symbols and no signal processing capability to estimate the channels. Therefore, it is critical to re-evaluate the promising gains shown by IRS-assisted communication systems under an imperfect CSI model. 

Recently \cite{LS} and \cite{LS1} have proposed channel estimation protocols for an IRS-assisted single-user MISO system based on the least squares (LS) estimation criteria, where the former paper  estimates the IRS-assisted channels one-by-one by keeping one IRS element active and the other elements off in each sub-phase of the channel estimation period, while the latter improves this protocol by keeping all the IRS elements active and reflecting throughout the channel estimation phase, under an optimal solution for the IRS phase shifts matrix. The method in \cite{LS, LS1} is extended in \cite{esto, est_new}, where the authors derive LS estimates for a single-user system assuming that the surface can be divided into multiple sub-surfaces of  adjacent strongly correlated reflecting elements that apply the same reflection coefficient. The work is also extended in  \cite{est_new1}, that aims to reduce the channel training time by developing a three stage channel estimation protocol which exploits the strong correlation in the IRS-assisted channels due to the common BS-to-IRS channel. However, the protocol assumes an ideal environment where there is no received noise at the BS in the channel estimation phase, which is definitely not going to hold under any practical setting.  Channel estimation using compressive sensing and deep learning techniques have been proposed in \cite{CS}  for a single-user system by requiring a few elements of the IRS to be active. The authors in \cite{CE_MU} focus on an IRS-assisted multi-user MISO system and leverage the sparsity of the cascaded channel, which consists of the BS-IRS channel and the IRS-user channel, to formulate the channel estimation problem as a sparse channel matrix recovery problem using compressive sensing techniques. The problem is solved using a two-step procedure based multi-user joint channel estimator.  The authors in \cite{cas}  exploit the rank-deficient structure of the massive MIMO channel to formulate and solve the cascaded channel estimation problem. To induce sparsity,  some randomly selected IRS elements are switched off at each time.  

With the exception of \cite{LS, LS1,esto, est_new} that derive LS channel estimates for a single-user IRS-assisted system, the proposed protocols are based on approximate algorithms that do not yield analytical expressions for the channel estimates which could facilitate future theoretical analysis of IRS-assisted systems.  Moreover, the contributions of most of these works are limited to developing  channel estimation protocols and numerically evaluating them in terms of the mean squared error (MSE). They do not utilize the estimates to develop joint precoding and reflect beamforming designs for different downlink communication scenarios of interest, where the downlink rate loss caused by channel training is an important issue especially in IRS-assisted systems. The most notable work that proposes beamforming design under imperfect CSI is  \cite{WSR}, that deals with the sum-rate maximization problem under a transmit power constraint by modeling the true channel coefficients as realizations from the sample space that is dominated by the knowledge of the imperfect CSI. However, the authors do not exploit any practical channel estimation protocol but rather assume a distribution for the channel estimation noise in the development of their algorithms.  

Motivated by these gaps in research,  we study the channel estimation and beamforming design problem for an IRS-assisted multi-user MISO communication system. We first outline the IRS-assisted system model, considering correlated Rayleigh fading channels between the IRS and the users, which are practically more relevant than the independent Rayleigh fading channels considered in most existing works. We then propose an optimal minimum mean squared error (MMSE) based channel estimation protocol to estimate the direct BS-to-users channel vectors as well as the cascaded channel vectors consisting of the BS-to-IRS link and the IRS-to-users links.  The proposed protocol divides the channel estimation phase into multiple sub-phases, where in each sub-phase an optimal reflect beamforming vector is applied across the IRS elements. It turns out that the optimal IRS configuration in the training phase is to choose the reflect beamforming vectors as columns of the discrete Fourier transform (DFT) matrix. The proposed DFT-MMSE estimation protocol utilizes prior information on the large-scale fading statistics, that change very  slowly as compared to the fast-fading process and can be easily tracked at the BS \cite{usergroups, ourworkTWC, MMS}, to derive closed-form expressions of the MMSE estimates of the direct channel and the IRS-assisted channels. A detailed analytical comparison in terms of the normalized MSE  confirms the superiority of the MMSE-DFT protocol over the LS-DFT protocol in \cite{LS1} and the LS-ON/OFF protocol in \cite{LS}. 

To study the performance of the IRS-assisted communication system, we focus on solving the maximization of the minimum SINR (max-min SINR) problem  by jointly designing the precoding vectors and power allocation at the BS and the phase shifts vector at the IRS, subject to a transmit power constraint at the BS and non-convex unit-modulus constraints on the IRS elements. The objective function is also non-convex in which the precoding vectors, allocated powers and phase shifts are coupled and no optimal design is known. We tackle the problem using alternating optimization (AO) where the precoding vectors and allocated powers at the BS are optimized iteratively with the phase shifts at the IRS, until convergence is achieved. For fixed IRS phase shifts vector, the optimal solution to the max-min SINR sub-problem with respect to precoding vectors and allocated powers is given by the optimal linear precoder (OLP) \cite{MMS}, while for fixed precoding and power allocation, the solution to max-min SINR sub-problem with respect to IRS phase shifts is obtained by applying semi-definite relaxation and solving the resulting fractional optimization problem optimally using the generalized Dinkelbach's algorithm. The proposed AO algorithm is proved to converge. We then extend the AO algorithm to the imperfect CSI scenario, where the MMSE estimates are utilized to design the precoder and the IRS phase shifts vector. The max-min SINR problem has only been dealt with in the context of IRS-assisted systems in \cite{LIS_jour}, where the authors approximate and solve this problem in the asymptotic regime under perfect CSI using project gradient ascent. Our work accounts for CSI errors and focuses on the exact problem. Simulation results are provided towards the end of the work that show the IRS-assisted system to be highly efficient but also sensitive to CSI errors as compared to the conventional MISO communication system. 

The paper is organized as follows. The communication model for an IRS-assisted MISO system is introduced in Section II.  We propose and analyze the MMSE-DFT channel estimation protocol  in Section III. Joint design for precoding vectors and power allocation at the BS and phase shifts vector at the IRS are developed to solve the max-min SINR problem in Section IV. Simulation results are provided in Section V and conclusions are presented in Section VI. 

\textit{Notation:} The following notation is used throughout this work. The notation $x \in [a,b]$ implies that the scalar $x$ lies in the closed interval between $a$ and $b$ as $a\leq x \leq b$. Boldface lower-case and upper-case characters denote vectors and matrices respectively. The notations $\mathbf{x}\in \mathbb{C}^{N\times 1}$ and $\mathbf{X}\in \mathbb{C}^{N\times N}$ represent a vector of dimension $N$ and a matrix of dimension $N\times N$ respectively with complex entries. The superscripts $(\cdot)^{T}$ and $(\cdot)^{H}$ represent the transpose and conjugate transpose respectively, $\mathbb{E}[\cdot]$ represents the expectation and $\log(\cdot)$ represents the logarithm.  The operators $\text{tr}(\textbf{X})$ and $||\textbf{X}||$ denote the trace and the spectral norm respectively of the matrix $\mathbf{X}$. Also $\mathbf{X}^{-1}$ denotes the inverse of a non-singular matrix $\mathbf{X}$.   The $N\times N$ identity matrix is denoted by $\textbf{I}_{N}$ and the $N\times N$ diagonal matrix of entries $\{x_{n}\}$ is denoted by $\textbf{X}=\text{diag}(x_{1}, x_{2},\dots, x_{N})$. A random vector $\textbf{x} \sim \mathcal{CN} (\textbf{m},\boldsymbol{\Phi})$ is complex Gaussian distributed with mean vector $\textbf{m}$ and covariance matrix $\boldsymbol{\Phi}$. The Kronecker product of two matrices $\mathbf{X}$ and $\mathbf{Y}$ is denoted as $\mathbf{X} \otimes \mathbf{Y}$.

\section{Communication Model}

 In this section, we outline the transmission model and channel model utilized to study the IRS-assisted system. To improve the clarity of mathematical exposition, the important symbols used in this section are listed in Table \ref{table1}. 

\begin{table}[!t]
\centering
\normalsize
\caption{Important symbols defining the communication model.}
\begin{tabular}{|l|l|}
\hline
  \textbf{Symbol} & \textbf{Definition} \\ 
\hline
$M$& Number of antennas at the BS.  \\ 
$K$ & Number of single-antenna users. \\
 $N$ & Number of IRS reflecting elements.  \\
$T$ & Symbols in each coherence interval. \\
 $\mathbf{x}\in \mathbb{C}^{M\times 1}$ & Precoded transmit signal vector.  \\ 
$p_k$& Allocated transmit power for user $k$.\\
$\mathbf{P}\in \mathbb{C}^{K\times K}$ & $\text{diag}(p_1,\dots, p_K)$. \\
$\mathbf{p}\in \mathbb{C}^{K\times 1} $& $[p_1, \dots, p_K]^T$.\\
 $\mathbf{g}_k\in \mathbb{C}^{M\times 1}$ & Precoding vector of user $k$.  \\ 
$\mathbf{G}\in \mathbb{C}^{M\times K}$ & $[\mathbf{g}_1, \dots, \mathbf{g}_K]$. \\
$s_k$ & Data symbol for user $k$. \\
$\mathbf{s}\in \mathbb{C}^{K\times 1}$ & $[s_1,\dots, s_K]^T$.\\
$P_{max}$ & Tx power budget at the BS. \\
$\sigma^2_n$ & Variance of received noise at user. \\
$\mathbf{h}_{d,k}\in \mathbb{C}^{M\times 1}$ & Direct BS-to-user-$k$ channel. \\
$\mathbf{H}_{1}\in \mathbb{C}^{M\times N}$ & BS-to-IRS channel.\\
$\mathbf{h}_{1,n}\in \mathbb{C}^{M\times 1}$ & Column vector $n$ of $\mathbf{H}_{1}$.\\
$\mathbf{h}_{2,k}\in \mathbb{C}^{M\times 1}$ & IRS-to-user-$k$ channel.\\
$\mathbf{H}_{0,k}\in \mathbb{C}^{M\times N}$ & Cascaded IRS-assisted channel given\\
& as $\mathbf{H}_{0,k}=\mathbf{H}_1\text{diag}(\mathbf{h}_{2,k}^{T})$.\\
$\mathbf{h}_{0,n,k}\in \mathbb{C}^{M\times 1}$ & Column vector $n$ of $\mathbf{H}_{0,k}$.\\
$\boldsymbol{\Theta} \in \mathbb{C}^{N\times N}$ & IRS reflection matrix given as $\boldsymbol{\Theta}=$\\
 & $\text{diag}(\alpha_1 \exp(j\theta_{1}), \dots, \alpha_N \exp(j\theta_{N}))$. \\
$\mathbf{v}\in \mathbb{C}^{N\times 1}$ & IRS reflect beamforming vector given as\\
& $\mathbf{v}=[\alpha_{1} \exp(j\theta_{1}), \dots, \alpha_{N} \exp(j\theta_{N})]^T$. \\
$v_n$ & Element $n$ of $\mathbf{v}$.\\
$\theta_{n} \in [0,2\pi ]$ & Phase-shift applied by IRS element $n$. \\
$\alpha_{n} \in [0,1]$ & Reflection coefficient of element $n$. \\
$\beta_1$ & Path loss factor for $\mathbf{H}_1$.\\
$\beta_{d,k}$ & Path loss factor for $\mathbf{h}_{d,k}$.\\
$\beta_{2,k}$ & Path loss factor for $\mathbf{h}_{2,k}$.\\
$\beta_k$& Product of $\beta_1$ and $\beta_{2,k}$.\\
$\mathbf{R}_{BS_k} \in \mathbb{C}^{M\times M}$ & Correlation matrix at BS w.r.t. user $k$.\\
$\mathbf{R}_{IRS_k} \in \mathbb{C}^{N\times N}$ & Correlation matrix at IRS w.r.t. user $k$.\\
$d_{BS}$ & Inter-element separation at BS. \\
$d_{IRS}$ & Inter-element separation at IRS. \\
$\phi_{LoS}$ & LoS azimuth angle for BS-to-IRS link. \\
$\theta_{LoS}$& LoS elevation angle for BS-to-IRS link.\\
$\gamma_k$ & SINR of user $k$. \\
$R_k$ & Rate of user $k$.\\
\hline
\end{tabular}
\label{table1}
\end{table}

\begin{figure}[!t]
\centering
\includegraphics[scale=.26]{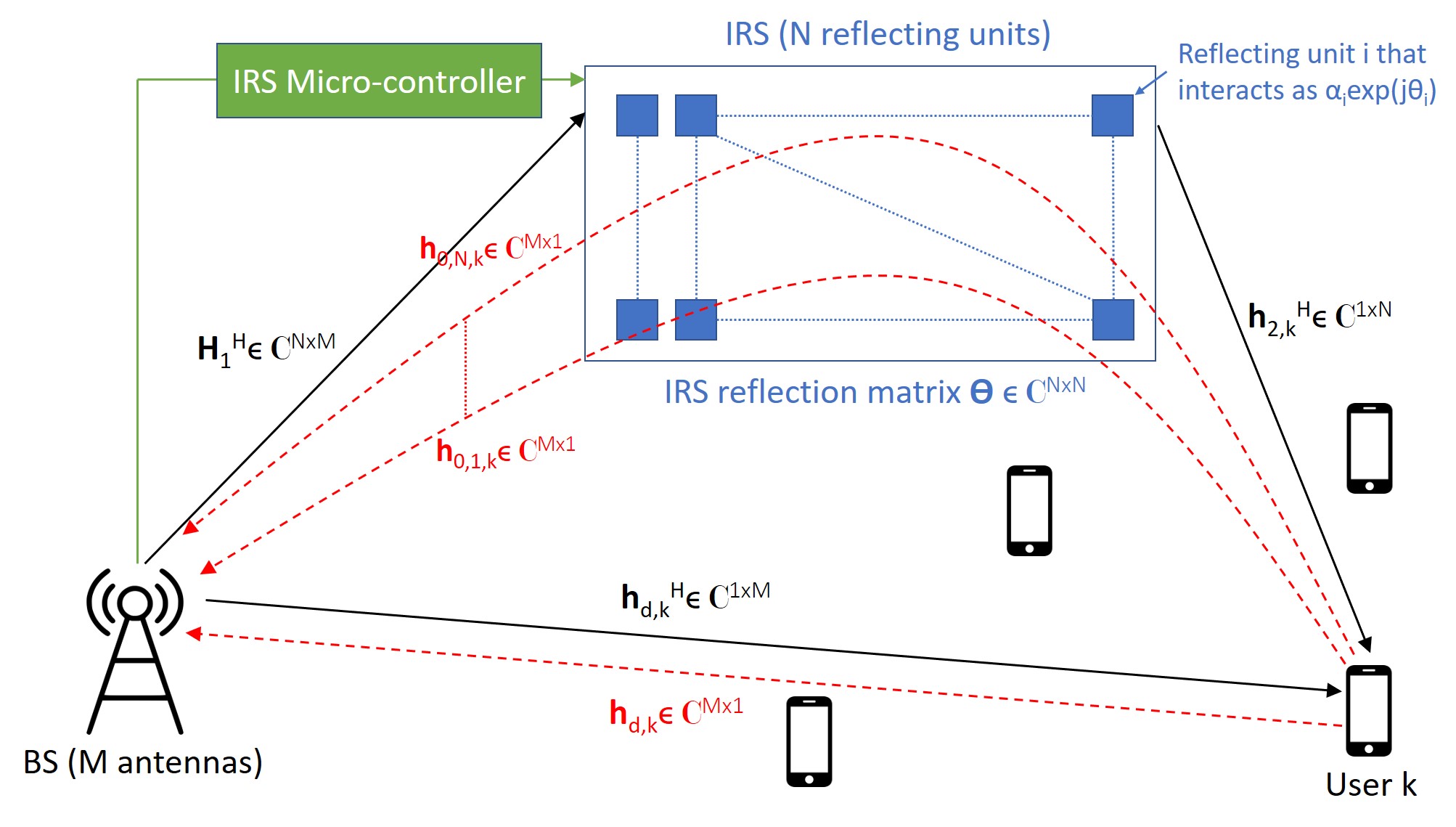}
\caption{IRS-assisted multi-user MISO system. Red dashed lines represent the uplink channel vectors estimated in the proposed protocol.}
\label{SU1}
\end{figure}

\subsection{Transmission Model}
The proposed IRS-assisted multi-user MISO system is illustrated in Fig. \ref{SU1}, which consists of a BS equipped with $M$ antennas serving $K$ single-antenna  users. This communication is assisted by an IRS, comprising of $N$ nearly passive reflecting elements which introduce phase shifts onto the incoming signal waves. The IRS is attached to the facade of a building located in the line-of-sight (LoS) of the BS. The reflection configuration of the IRS, that governs the phase shifts applied by individual IRS elements, is controlled by a micro-controller, which gets this information from the BS over a  backhaul link.

The BS employs Gaussian codebooks and linear precoding, where $p_k$, $\mathbf{g}_k\in \mathbb{C}^{M \times 1}$ and $s_k\in \mathcal{CN}(0,1)$ are the allocated power, digital precoding vector and data symbol of user $k$ respectively. Based on these definitions, the transmit signal vector $\mathbf{x}\in \mathbb{C}^{M\times 1}$ is given as
\begin{align}
&\mathbf{x}=\sum_{k=1}^{K} \sqrt{\frac{p_{k}}{K}} \mathbf{g}_{k} s_{k}.
\end{align}
 Given $s_k$'s are independent and identically distributed (i.i.d.) $\mathcal{CN}(0,1)$ variables, $\mathbf{x}$ has to satisfy the average transmit (Tx) power per user constraint as
\begin{align}
\label{p_cons}
& \mathbb{E}[||\mathbf{x}||^{2}]=\frac{1}{K}\text{tr}(\mathbf{P} \mathbf{G}^{H} \mathbf{G}) \leq  P_{max},
\end{align}
where $P_{max}>0$ is the Tx power constraint at the BS,  $\mathbf{P}=\text{diag}(p_1, \dots, p_K)\in \mathbb{C}^{K\times K}$ is the power allocation matrix, $\mathbf{G}=[\mathbf{g}_1, \dots, \mathbf{g}_K]\in \mathbb{C}^{M\times K}$ is the precoding matrix, and $\mathbf{s}=[s_1,\dots, s_K]^T$ is the vector of users' data symbols. 

We consider the block-fading model for the channels, which stay constant over the coherence interval of length $T$ symbols. The received complex baseband signal $y_k(t) \in \mathbb{C}$ at user $k$ in time-slot $t$ is given as
\begin{align}
y_{k}(t)&=(\mathbf{h}_{d,k}^H+\mathbf{h}_{2,k}^H \boldsymbol{\Theta}^H \mathbf{H}_{1}^H)\mathbf{x}(t)+n_{k}(t), \nonumber \\
\label{ch1}
&=(\mathbf{h}_{d,k}^H+\mathbf{v}^H \mathbf{H}_{0,k}^H)\mathbf{x}(t)+n_{k}(t), \hspace{.02in} t=1,\dots, T,
\end{align}
where $\mathbf{H}_{1}=[\mathbf{h}_{1,1},\dots, \mathbf{h}_{1,N}] \in \mathbb{C}^{M\times N}$ is  the LoS channel between the BS and the IRS, $\mathbf{h}_{2,k} \in \mathbb{C}^{N\times 1}$ is the channel between the IRS and user $k$, $\mathbf{h}_{d,k} \in \mathbb{C}^{M\times 1}$ is the direct channel between the BS and user $k$ and $n_{k}(t)\sim \mathcal{CN}(0,\sigma_n^2)$ is the noise at the user. The IRS is represented by the diagonal matrix $\boldsymbol{\Theta}=\text{diag}(\alpha_1 \exp(j\theta_{1}), \dots, \alpha_N \exp(j\theta_{N}))$, where $\theta_{n} \in [0,2\pi ]$ and $\alpha_{n} \in [0,1]$ represent the phase-shift and the amplitude coefficient for element $n$ respectively. Note that $\boldsymbol{\Theta}$ is not updated on a symbol-duration level, but rather on a coherence-time level, i.e. after every $T$ symbols. 

The uplink channel through the IRS given by $\mathbf{H}_{1} \boldsymbol{\Theta} \mathbf{h}_{2,k}$ can be equivalently expressed as $\textbf{H}_{0,k} \textbf{v}$, where $\mathbf{v}=[\alpha_{1} \exp(j\theta_{1}), \dots, \alpha_{N} \exp(j\theta_{N})]^T \in \mathbb{C}^{N\times 1}$ is the reflect beamforming vector of the IRS and $\mathbf{H}_{0,k}=\mathbf{H}_1\text{diag}(\mathbf{h}_{2,k}^{T})\in \mathbb{C}^{M\times N}$ is the cascaded channel matrix. The cascaded matrix $\mathbf{H}_{0,k}$ has $N$ column vectors of dimension $M$, where each column vector $\mathbf{h}_{0,n,k}$, $n=1,\dots , N$, can be written as $\mathbf{h}_{0,n,k}=\mathbf{h}_{1,n}\mathbf{h}_{2,k}(n)$, where $\mathbf{h}_{2,k}(n)$ denotes element $n$ of $\mathbf{h}_{2,k}$. This formulation in \eqref{ch1} enables the separation of the  response of the IRS in $\mathbf{v}$ from the cascaded channel outside the IRS control in  $\mathbf{H}_{0,k}$, and will assist us in the design of the channel estimation protocol. 

In terms of CSI acquisition, the IRS-assisted system is different from existing popular communication systems, like the conventional MISO system and relay-assisted MISO system, since unlike BS and relay, the IRS has no radio resources of its own to send pilot symbols to help the BS estimate $\mathbf{H}_1$ nor can it receive and process pilot symbols from the users to estimate $\mathbf{h}_{2,k}$s.  This is one of the biggest challenges in the practical design of IRS-assisted systems.  In terms of precoding/beamforming design, the IRS-assisted system model is much more difficult to analyze than existing models, due to the constant-modulus constraints on elements of the reflect beamforming vector $\mathbf{v}$. Although beamforming optimization under unit-modulus constraints has been studied in the context of hybrid digital/analog mmWave architectures \cite{CE_mm, mm23}, such designs are mainly restricted to the BS  side, and are not directly applicable to the joint design of the precoding at the BS and reflect beamforming at the IRS.

\subsection{Channel Model}

The design of IRS-assisted systems also requires the correct modeling of $\mathbf{h}_{2,k}$ and $\mathbf{H}_{1}$. Existing works (e.g. \cite{8741198,LIS_new,WSR, RISn,SWIPT,LIS,phy}) utilize the independent Rayleigh and Rician models to analyze the system performance, which are only practical if the IRS elements are spaced far enough and the environment has rich scattering. In most practical settings, the channels with respect to IRS elements will be spatially correlated which will impact the performance. In this work, we will evaluate the performance of the IRS-assisted system under the correlated Rayleigh channel model for $\mathbf{h}_{2,k}$ and $\mathbf{h}_{d,k}$ given as
\begin{align}
&\mathbf{h}_{2,k}=\sqrt{\beta_{2,k}} \mathbf{R}_{IRS_{k}}^{1/2} \mathbf{z}_k, \\
&\mathbf{h}_{d,k}=\sqrt{\beta_{d,k}} \mathbf{R}_{BS_{k}}^{1/2} \mathbf{z}_{d,k},
\end{align}
where $\mathbf{R}_{IRS_{k}} \in \mathbb{C}^{N\times N}$ and $\mathbf{R}_{BS_{k}} \in \mathbb{C}^{M\times M}$ are the correlation matrices at the IRS and the BS respectively with respect to (w.r.t.) user $k$, with $\text{tr}(\mathbf{R}_{IRS_{k}})=N$ and $\text{tr}(\mathbf{R}_{BS_{k}})=M$. Moreover, $\mathbf{z}_k\sim\mathcal{CN}(\mathbf{0}, \mathbf{I}_N)$ and $\mathbf{z}_{d,k}\sim\mathcal{CN}(\mathbf{0}, \mathbf{I}_M)$ are the fast fading vectors for IRS-to-user $k$ link and BS-to-user $k$ link respectively, and $\beta_{2,k}$ and $\beta_{d,k}$ are the path loss factors for the IRS-to-user $k$ link and BS-to-user $k$ link respectively.  We will adopt the correlation model developed for arrays of discrete antennas in \cite{ourwork, ourworkSC}, assuming that the underlying IRS technology is a reflective antenna array or a reflect-array. It is important to note that  the conventional statistical correlation models for arrays of discrete antennas are not directly applicable if the IRS is realized using a reconfigurable meta-surface. The correct modeling of the spatial correlation for this implementation still requires significant attention from researchers who are conversant in both communication and electromagnetic theory.

The  IRS is envisioned to be installed on a high rise building close to the BS, which will result in a  LoS channel between the BS and the IRS \cite{LIS, LIS_jour}. Since the BS and the IRS have co-located elements, so the channel matrix $\mathbf{H}_{1}$ is likely to  have rank one, i.e. $\mathbf{H}_{1}=\mathbf{a}\mathbf{b}^{H}$, where $\mathbf{a} \in \mathbb{C}^{M\times 1}$ and $\mathbf{b}\in\mathbb{C}^{N\times 1}$ are the array responses at the BS and IRS defined in \cite{LIS_jour}. Under such a setting, the degrees of freedom offered by the overall IRS-assisted link $\mathbf{H}_{0,k}$ will be one and the IRS will only yield performance gains when $K=1$ \cite{LIS_jour}. To benefit from the IRS in a multi-user setting, we must have $\text{rank}(\mathbf{H}_{1})\geq K$. One way to introduce this rank is to have deterministic scattering between the BS and the IRS or place the IRS close to the BS such that the LoS channel could be made of high rank. The high-rank LoS BS-to-IRS channel matrix $\mathbf{H}_1$ for a multi-user setting can be generated as \cite{LIS_jour}
\begin{align}
&[\mathbf{H}_{1}]_{m,n}\hspace{-.01in}=\hspace{-.01in}\sqrt{\beta_1} \exp\hspace{-.04in}\Big( j\frac{2\pi}{\lambda}(m-1) d_{BS}\sin\theta_{LoS_{1,n}} \sin\phi_{LoS_{1,n}}\hspace{-.02in}\nonumber \\
\label{H_1}
&+(n-1) d_{IRS}\sin\theta_{LoS_{2,m}} \sin\phi_{LoS_{2,m}}\hspace{-.04in}\Big),
\end{align}
$m=1,\dots, M$, $n=1,\dots, N$, where $\lambda$ is the carrier wavelength, $\theta_{LoS_{1,n}}$ and $\phi_{LoS_{1,n}}$ represent the elevation and azimuth LoS angles of departure (AoD) respectively at the BS w.r.t IRS element $n$, and $\theta_{LoS_{2,m}}$ and $\phi_{LoS_{2,m}}$ represent the elevation and azimuth LoS angles of arrival (AoA) respectively at the IRS. Moreover $\beta_1$ is the path loss factor for the BS-to-IRS link, $d_{BS}$ is the inter-antenna separation at the BS and $d_{IRS}$ is the inter-element separation at the IRS.

 \section{Channel Estimation Protocol}

Channel estimation is necessary to compute the precoding vectors at the BS and the reflect beamforming vector $\mathbf{v}$ at the IRS.  The real difficulty is in the estimation of $\mathbf{H}_{1}$ and $\mathbf{h}_{2,k}$s as the IRS has no radio resources and signal processing capability to send pilot symbols to the BS to enable the estimation of $\mathbf{H}_{1}$ or to receive pilot symbols from users and estimate $\mathbf{h}_{2,k}$. Recently a few papers have proposed LS estimates for the IRS-assisted channels assuming a single-user IRS-assisted MISO system in \cite{LS} and \cite{LS1}. More specifically, \cite{LS} proposes an ON/OFF channel estimation protocol, where first the direct channel is estimated by keeping all IRS elements OFF and then the IRS-assisted channels $\mathbf{h}_{0,n,k}$, $n=1,\dots, N$, are estimated one-by-one by switching one element of the IRS ON at a time. In \cite{LS1}, LS channel estimates are derived keeping all the IRS elements active throughout the channel estimation phase with an optimal IRS phase shift matrix given as the DFT matrix. The idea was extended in \cite{esto} to an OFDM system and in \cite{est_new} to an IRS-assisted system with discrete phase shifts while focusing on a single-user scenario. In parallel to these works, a few channel estimation algorithms exploiting the sparsity of the cascaded channel matrix  $\mathbf{H}_{0,k}$ have also been recently proposed as discussed in the introduction. 

In this section, we will outline our channel estimation protocol where the BS computes the MMSE estimates of the IRS-assisted channel vectors based on the received pilot sequences from  users over multiple sub-phases, where in each sub-phase the IRS applies an optimal reflect beamforming vector $\mathbf{v}$.  MMSE estimator significantly outperforms the LS estimator since it is based on the Bayesian estimation technique which achieves the minimum MSE between the true and estimated channel by exploiting prior knowledge of the channel's large scale fading statistics \cite{est}. These statistics stay constant over several coherence intervals and can be accurately learned and tracked at the BS as discussed later in this section. After deriving the MMSE estimates, we will  analytically compare the normalized MSE of both the LS and MMSE estimates. Simulation results are also provided to compare the MSE and bit error rate (BER) performance of the proposed protocol with existing methods. The important symbols used in this section are summarized in Table \ref{table2} for readers' convenience. 

\begin{table}[!t]
\centering
\normalsize
\caption{Important symbols defining the channel estimation protocol.}
\begin{tabular}{|l|l|}
\hline
  \textbf{Symbol} & \textbf{Definition} \\ 
\hline
$\tau$ sec & Length of coherence interval. \\
$\tau_C$ sec &Length of channel estimation phase. \\
$\tau_D$ sec & Length of transmission phase. \\
$S$& Number of channel estimation \\
& sub-phases. \\
$\tau_S$ sec& Duration of each sub-phase. \\
$T_S$& Length in symbols of pilot \\
& sequence sent by each user. \\
$\tilde{\tau}$& Duration of each pilot symbol. \\
$P_C$ & Tx power of each user. \\
$\sigma^2$ & Received noise variance at BS. \\
$\mathbf{x}_{p,k}\in \mathbf{C}^{T_S\times 1}$ & Pilot sequence of user k. \\
$\mathbf{V}^{tr}\in \mathbb{C}^{S\times N+1}$ & Matrix of IRS reflect beamforming \\
& vectors $\mathbf{v}_s$ in $s=1,.., S$ sub-phases.\\
$\mathbf{v}^{tr}_{i}$& Column vector $i$ of $\mathbf{V}^{tr}$. \\
$\mathbf{Y}_s^{tr}\in \mathbb{C}^{M\times T_S}$& Received training signal at the \\
& BS in sub-phase $s$. \\
$\mathbf{N}_s^{tr}\in \mathbb{C}^{M\times T_S}$& Received noise at BS in sub-phase $s$.\\
$\mathbf{r}_{s,k}^{tr}\in \mathbb{C}^{M\times 1}$& Observation vector for user $k$ in \\
& sub-phase $s$ given as $\mathbf{r}_{s,k}^{tr}=\mathbf{Y}_s^{tr}\frac{\mathbf{x}_{p,k}}{P_C\tau_S}$. \\
$\mathbf{n}_{s,k}^{tr}\in \mathbb{C}^{M\times 1}$ & $\mathbf{n}_{s,k}^{tr}=\mathbf{N}_s^{tr}\mathbf{x}_{p,k}$.\\
$\mathbf{r}^{tr}_k\in \mathbb{C}^{MS \times 1}$ & Concatenation of all $\mathbf{r}_{s,k}^{tr}$ given \\
& as $\mathbf{r}^{tr}_k=[\mathbf{r}^{tr^T}_{1,k}, \dots, \mathbf{r}^{tr^T}_{S,k}]^T$.\\
$\mathbf{n}^{tr}_k\in \mathbb{C}^{MS \times 1}$ & $[\mathbf{n}^{tr^T}_{1,k}, \dots, \mathbf{n}^{tr^T}_{S,k}]^T$.\\
 $\bar{\mathbf{h}}_k\in \mathbb{C}^{M(N+1)\times 1}$& Concatenation of $\mathbf{h}_{d,k}$ and $\mathbf{h}_{0,n,k}$s\\
& as $\bar{\mathbf{h}}_k=[\mathbf{h}_{d,k}^T, \mathbf{h}_{0,1,K}^T, \dots, \mathbf{h}_{0,N,k}^T]^T$. \\
$\tilde{\mathbf{r}}^{tr}_k\in \mathbb{C}^{M(N+1)\times 1}$&Observation vector after processing \\
& $\mathbf{r}^{tr}_k$ with pseudo-inverse of $\bar{\mathbf{V}}^{tr}$. \\
$\hat{\mathbf{h}}_{d,k}\in \mathbb{C}^{M\times 1}$ &MMSE estimates of  $\mathbf{h}_{d,k}$.\\
$\hat{\mathbf{h}}_{0,n,k}\in \mathbb{C}^{M\times 1}$& MMSE estimate of $\mathbf{h}_{0,n,k}$. \\
$\hat{\mathbf{h}}_k\in \mathbb{C}^{M\times 1}$ & MMSE estimate of overall channel\\
&  $\mathbf{h}_k$ given as $\hat{\mathbf{h}}_k=\hat{\mathbf{h}}_{d,k}+ \hat{\mathbf{H}}_{0,k} \mathbf{v}$.\\
$\hat{\mathbf{h}}^{LS}_{d,k},\hat{\mathbf{h}}^{LS}_{0,n,k}$ &LS estimates of  $\mathbf{h}_{d,k}$ and $\mathbf{h}_{0,n,k}$. \\
$\tilde{\mathbf{h}}_{d,k}\in \mathbb{C}^{M\times 1}$ &Error in estimate of  $\mathbf{h}_{d,k}$. \\
$\tilde{\mathbf{h}}_{0,n,k}\in \mathbb{C}^{M\times 1}$ &Error in estimate of  $\mathbf{h}_{0,n,k}$. \\
$\boldsymbol{\Psi}_{d,k} \in \mathbb{C}^{M\times M}$& Covariance matrix for $\hat{\mathbf{h}}_{d,k}$.\\
$\boldsymbol{\Psi}_{n,k}\in \mathbb{C}^{M\times M}$&Covariance matrix for $\hat{\mathbf{h}}_{0,n,k}$.\\
$\tilde{\boldsymbol{\Psi}}_{d,k}\in \mathbb{C}^{M\times M}$& Covariance matrix for $\tilde{\mathbf{h}}_{d,k}$.\\
$\tilde{\boldsymbol{\Psi}}_{n,k}\in \mathbb{C}^{M\times M}$& Covariance matrix for $\tilde{\mathbf{h}}_{0,n,k}$.\\
$r_{n,k}$& $\mathbf{R}_{IRS_k}(n,n)$.\\
$\text{NMSE}(\hat{\mathbf{h}})$& Normalized mean squared error in \\
& estimate $\hat{\mathbf{h}}$.\\
$c$& Constant defined as $\frac{\sigma^2}{SP_C \tau_S}$. \\
\hline
\end{tabular}
\label{table2}
\end{table}

\subsection{Proposed MMSE-DFT Channel Estimation Protocol}

Given the passive nature of the IRS, we exploit channel reciprocity under the TDD protocol in estimating the downlink channels using the received uplink pilot signals from the users.  For this purpose, we divide the channel coherence period of $\tau$ seconds (sec) into an uplink training phase of $\tau_{C}$ sec and a downlink transmission phase of $\tau_{D}$ sec. Throughout the uplink training phase, the users transmit mutually orthogonal pilot symbols. Since the IRS has no radio resources to send or receive and process pilot symbols, the BS has to estimate all the channels.  To this end, note that $\textbf{H}_{1}$ and $\textbf{h}_{2,k}$ have been cascaded as $\textbf{H}_{0,k}\in \mathbb{C}^{M\times N}$ in (\ref{ch1}), where $\textbf{H}_{0,k}=[\textbf{h}_{0,1,k}, \dots, \textbf{h}_{0,N,K}]$ is a matrix of $N$ column vectors. Since the estimation of $\mathbf{h}_{2,k}$ separately is extremely difficult due to the passive nature of IRS elements, we will focus on the MMSE estimation of the cascaded IRS-assisted channels $\textbf{h}_{0,n,k}$, $n=1,\dots, N$ and the direct channel $\textbf{h}_{d,k}$ for all $k=1,\dots, K$ users at the BS.

In the considered channel estimation protocol, the total channel estimation period of $\tau_C$ sec is divided into $S$ sub-phases\footnote{We will see later that $S\geq N+1$ to obtain the LS and MMSE estimates under the proposed protocol.}, each of length $\tau_S=\frac{\tau_C}{S}$ sec. The IRS applies the reflect beamforming vector $\mathbf{v}_s =[v_{s,1}, \dots, v_{s,N}]^T \in \mathbb{C}^{N \times 1}$ throughout sub-phase $s$, $s=1,\dots, S$, where $v_{s,n}=\alpha_{s,n} \exp(j\theta_{s,n})$. In each sub-phase,  the users transmit $T_S=\frac{\tau_S}{\tilde{\tau}}$ pilot symbols, where $\tilde{\tau}$ is the duration of each symbol. Users transmit $S$ copies of orthogonal pilot sequences across the $S$ sub-phases,  where the pilot sequence of user $k$ is denoted as $\mathbf{x}_{p,k}=[x_{p,k,1}, \dots, x_{p,k,T_S}]^T \in \mathbb{C}^{T_S \times 1}$, such that $\mathbf{x}_{p,k}^H\mathbf{x}_{p,l}=0$, for $k\neq l$, $k,l=1,\dots, K$ and $\mathbf{x}_{p,k}^H\mathbf{x}_{p,k}=P_C T_S \tilde{\tau}=P_C \tau_S$ Joules, where $P_C$  is the transmit power of user. The received training signal, $\mathbf{Y}^{tr}_s\in \mathbb{C}^{M\times T_S}$ in sub-phase $s$ is given as
\begin{align}
&\mathbf{Y}^{tr}_s=\sum_{k=1}^K (\mathbf{h}_{d,k}+\mathbf{H}_{0,k} \mathbf{v}_s) \mathbf{x}_{p,k}^H+ \mathbf{N}_s^{tr},\hspace{.01in} s=1,\dots, S,
\end{align}
where  $\mathbf{N}^{tr}_s \in \mathbb{C}^{M\times T_S}$ is the matrix of noise vectors at the BS, with each column distributed independently as $\mathcal{CN}(\textbf{0},\sigma^2\textbf{I}_{M})$.  To get the observation vector with respect to each user, the BS correlates the received training signal with the pilot sequence of user $k$ to obtain the observation vector, $\mathbf{r}^{tr}_{s,k}\in \mathbb{C}^{M\times 1}$, for user $k$ in sub-phase $s$ as
\begin{align}
\label{LSS}
&\mathbf{r}^{tr}_{s,k}=(\mathbf{h}_{d,k}+\mathbf{H}_{0,k} \mathbf{v}_s)+\frac{\mathbf{n}^{tr}_{s,k}}{P_C \tau_S}, \hspace{.02in} k=1,\dots, K,
\end{align}
where $\mathbf{n}^{tr}_{s,k}=\mathbf{N}_s^{tr} \mathbf{x}_{p,k}$. Let $\mathbf{r}^{tr}_k=[\mathbf{r}^{tr^T}_{1,k}, \dots, \mathbf{r}^{tr^T}_{S,k}]^T \in \mathbb{C}^{MS \times 1}$, $\bar{\mathbf{h}}_k=[\mathbf{h}_{d,k}^T, \mathbf{h}_{0,1,K}^T, \dots, \mathbf{h}_{0,N,k}^T]^T\in \mathbb{C}^{M(N+1)\times 1}$ and $\mathbf{n}_k^{tr}=[\mathbf{n}^{tr^T}_{1,k}, \dots, \mathbf{n}^{tr^T}_{S,k}]^T \in \mathbb{C}^{MS \times 1}$. Collecting the observation vectors in \eqref{LSS} across $S$ training sub-phases, we   obtain
\begin{align}
\label{LSS1}
&\mathbf{r}^{tr}_k=(\mathbf{V}^{tr}\otimes \mathbf{I}_M) \bar{\mathbf{h}}_k+\frac{\mathbf{n}^{tr}_{k}}{P_C\tau_S}, \hspace{.02in} k=1,\dots, K,
\end{align}
where,
\begin{align}
\label{Vtr}
&\mathbf{V}^{tr}=\begin{bmatrix} 1&\mathbf{v}_{1}^T \\
		\vdots & \vdots \\
    1 &  \mathbf{v}_{S}^T
  \end{bmatrix} \in \mathbb{C}^{S\times N+1}.
	\end{align}
The received observation vector in \eqref{LSS1} is processed at the BS with the left pseudo-inverse of $\bar{\mathbf{V}}^{tr}=\mathbf{V}^{tr}\otimes \mathbf{I}_M \in \mathbf{C}^{MS\times M(N+1)}$, provided that $S\geq N+1$,\footnote{The full column rank condition, i.e. $S\geq N+1$, is needed for the left pseudo-inverse of $\bar{\mathbf{V}}^{tr}$ to exist.} as 
	\begin{align}
	\label{LSS2}
	&\tilde{\mathbf{r}}^{tr}_k=(\bar{\mathbf{V}}^{tr^H} \bar{\mathbf{V}}^{tr})^{-1}\bar{\mathbf{V}}^{tr^H}\mathbf{r}^{tr}_k.
	\end{align}
	Performing the pseudo-inverse operation in \eqref{LSS2} will result in
	\begin{align}
	\label{LSS22}
	&\tilde{\mathbf{r}}^{tr}_k=\underbrace{\bar{\mathbf{h}}_k}_{\text{True channels}}+\underbrace{(\bar{\mathbf{V}}^{tr^H} \bar{\mathbf{V}}^{tr})^{-1}\bar{\mathbf{V}}^{tr^H} \frac{\mathbf{n}^{tr}_{k}}{P_C \tau_S}}_{\text{Noise vector $\tilde{\mathbf{n}}_k^{tr} \in \mathbb{C}^{M(N+1)\times 1}$}}  \hspace{.02in} k=1,\dots, K,
	\end{align}
	which is the function of the true channel vectors $\mathbf{h}_{d,k}$ and $\mathbf{h}_{0,n,k}$, $n=1,\dots, N$ collected in $\bar{\mathbf{h}}_k$ and the noise $\tilde{\mathbf{n}}_k^{tr}$ in the received observation vector. 	 The remaining task before proceeding to the derivation of the MMSE estimates is to design $\mathbf{V}^{tr}$. The appropriate design criteria is to minimize the variances of the elements of the noise vector $\tilde{\mathbf{n}}^{tr}_k$, while keeping the noise across the estimation of different channel vectors uncorrelated. The covariance matrix of the noise $\tilde{\mathbf{n}}^{tr}_k$ denoted as $\mathbf{C}_{\tilde{\mathbf{n}}^{tr}_k}=\mathbb{E}[\tilde{\mathbf{n}}_k^{tr}\tilde{\mathbf{n}}_k^{tr^H}] \in \mathbb{C}^{M(N+1)\times M(N+1)}$ is given as  
	\begin{align}
	&\mathbf{C}_{\tilde{\mathbf{n}}^{tr}_k}=(\bar{\mathbf{V}}^{tr^H} \bar{\mathbf{V}}^{tr})^{-1}\bar{\mathbf{V}}^{tr^H} \frac{\mathbb{E}\left[\mathbf{n}^{tr}_{k} \mathbf{n}^{tr^H}_{k}\right]}{(P_C\tau_S)^2}\bar{\mathbf{V}}^{tr}(\bar{\mathbf{V}}^{tr^H} \bar{\mathbf{V}}^{tr})^{-1}, \\
	\label{C_n}
	&=\frac{\sigma^2 P_C \tau_S}{(P_C\tau_S)^2}(\bar{\mathbf{V}}^{tr^H} \bar{\mathbf{V}}^{tr})^{-1}=\frac{\sigma^2}{P_C \tau_S}(\mathbf{V}^{tr^H} \mathbf{V}^{tr})^{-1}\otimes \mathbf{I}_M.
	\end{align}

To ensure uncorrelated noise across the estimated channels, $\mathbf{C}_{\tilde{\mathbf{n}}^{tr}_k}$ should be a scaled identity matrix and therefore $\mathbf{V}^{tr}$ should have orthogonal columns. Furthermore, we will aim to achieve the same noise variance in the estimation of all channels, which will require equally scaled orthogonal columns of $\mathbf{V}^{tr}$ i.e. $(\mathbf{V}^{tr^H} \mathbf{V}^{tr})^{-1}=\zeta \mathbf{I}_{N+1}$. Minimizing the variance of the noise is then equivalent to minimizing $\zeta$ with the constraints that 1) $\mathbf{V}^{tr}$ has the structure in \eqref{Vtr}, 2) $v_{s,n}=\alpha_{s,n}\exp(j\theta_{s,n})$, 3) $\alpha_{s,n}\in[0,1]$, 4) $\theta_{s,n} \in [0,2\pi]$, and 5) $(\mathbf{V}^{tr^H} \mathbf{V}^{tr})^{-1}=\zeta \mathbf{I}_{N+1}$. To this end, note that the last constraint implies $\zeta=\frac{N+1}{\text{tr}(\mathbf{V}^{tr^H} \mathbf{V}^{tr})}=\frac{N+1}{\sum_{n=1}^{N+1} \sum_{s=1}^S |[\mathbf{V}^{tr}]_{s,n}|^2}$. The maximum value of $|[\mathbf{V}^{tr}]_{s,n}|$ under the third constraint is $1$. Therefore a lower bound on $\zeta$ can be obtained as 
\begin{align}
\label{LB}
&\zeta\geq \frac{1}{S}.
\end{align}
Under the outlined constraints on $\mathbf{V}^{tr}$, a possible optimal design that attains the lower bound in \eqref{LB} is the $N+1$ leading columns of a $S\times S$ DFT matrix given as \cite{LS1}
\begin{align}
\label{des}
&[\mathbf{V}^{tr}]_{s,n}=\exp\left(-j\frac{2\pi(s-1)(n-1)}{S}\right),
\end{align}
$s=1,\dots, S$, $n=1,\dots, N+1$. Under the DFT design, we have $(\mathbf{V}^{tr^H} \mathbf{V}^{tr})^{-1}=\frac{1}{S}\mathbf{I}_{N+1}$ and therefore $\zeta=\frac{1}{S}$. This choice for $\mathbf{V}^{tr}$ indeed attains the  lower bound in \eqref{LB} while meeting all constraints. 

We now derive the MMSE estimates based on the received observation vector $\tilde{\mathbf{r}}^{tr}_{k}$ in \eqref{LSS22}, which can be simplified under the DFT design in \eqref{des} as 
\begin{align}
\label{LSS222}
&\tilde{\mathbf{r}}^{tr}_k=\bar{\mathbf{h}}_k+\frac{1}{S}\bar{\mathbf{V}}^{tr^H} \frac{\mathbf{n}^{tr}_{k}}{P_C \tau_S}.
\end{align}
 We can write \eqref{LSS222} as $\tilde{\mathbf{r}}^{tr}_{k}=[\tilde{\mathbf{r}}^{tr^T}_{1,k} \tilde{\mathbf{r}}^{tr^T}_{2,k}, \dots,\tilde{\mathbf{r}}^{tr^T}_{N+1,k}]^T$, where $\tilde{\mathbf{r}}^{tr}_{i,k} \in \mathbb{C}^{M\times 1}$ is given as $\tilde{\mathbf{r}}^{tr}_{k}([M(i-1)+1:Mi])$, $i=1,\dots, N+1$. To derive the MMSE-DFT estimate of $\mathbf{h}_{d,k}$, we exploit the relationship between $\tilde{\mathbf{r}}^{tr}_{1,k}$ and $\mathbf{h}_{d,k}$ given as
	\begin{align}
	\label{RR1}
	&\tilde{\mathbf{r}}^{tr}_{1,k}=\mathbf{h}_{d,k}+\frac{1}{S} (\mathbf{v}^{tr}_1 \otimes \mathbf{I}_M)^H \frac{\mathbf{n}^{tr}_{k}}{P_C\tau_S}, \hspace{.02in} k=1,\dots, K. 
	\end{align}
	where $\mathbf{v}^{tr}_1$ is the first $S\times 1$  column of $\mathbf{V}^{tr}$. Based on the observation vector in \eqref{RR1}, the BS can compute the estimate of $\mathbf{h}_{d,k}$ and the result is stated in the following lemma.

\begin{lemma}\label{L1}
The MMSE estimate $\hat{\mathbf{h}}_{d,k}$  of $\mathbf{h}_{d,k}$ is given as
\begin{align}
\label{h_d_est}
&\hat{\mathbf{h}}_{d,k}=\beta_{d,k} \mathbf{R}_{BS_k}\mathbf{Q}_{d,k} \tilde{\mathbf{r}}^{tr}_{1,k},
\end{align}
which is distributed as $\hat{\mathbf{h}}_{d,k}\sim \mathcal{CN}(\mathbf{0}, \mathbf{\Psi}_{d,k})$, where
\begin{align}
&\mathbf{\Psi}_{d,k}=\beta_{d,k}^2 \mathbf{R}_{BS_k} \mathbf{Q}_{d,k}  \mathbf{R}_{BS_k}.
\end{align}
and $\mathbf{Q}_{d,k}=\left(\beta_{d,k} \mathbf{R}_{BS_k}+\frac{\sigma^2}{S P_C \tau_S} \mathbf{I}_M \right)^{-1}$.
\end{lemma}
\begin{IEEEproof}
The proof is provided in Appendix \ref{Sec:L1}. 
\end{IEEEproof}

Invoking the orthogonality property of the MMSE estimate \cite{orth}, we can decompose the channel $\mathbf{h}_{d,k}$ as $\mathbf{h}_{d,k}= \hat{\mathbf{h}}_{d,k} + \tilde{\mathbf{h}}_{d,k}$, where $\tilde{\mathbf{h}}_{d,k}\sim \mathcal{CN}(\mathbf{0},\tilde{\mathbf{\Psi}}_{d,k})$ is the uncorrelated estimation error (which is also statistically independent of $\hat{\mathbf{h}}_{d,k}$ due to the joint  Gaussianity of both vectors) and $\tilde{\mathbf{\Psi}}_{d,k}=\beta_{d,k}\mathbf{R}_{BS_k}-\mathbf{\Psi}_{d,k}$.

We now find the MMSE-DFT estimate of $\mathbf{h}_{0,n,k}$, $n=1,\dots, N$ using the received observation vector $\tilde{\mathbf{r}}^{tr}_{n+1,k}$, which is given using \eqref{LSS222} as
	\begin{align}
	\label{RR2}
	&\tilde{\mathbf{r}}^{tr}_{n+1,k}=\mathbf{h}_{0,n,k}+\frac{1}{S} (\mathbf{v}^{tr}_{n+1}\otimes \mathbf{I}_M)^H \frac{\mathbf{n}^{tr}_{k}}{P_C \tau_S},
	\end{align}
	where $\mathbf{v}^{tr}_{n+1}$ is the $(n+1)^{th}$ column vector  of $\mathbf{V}^{tr}$. Based on this observation vector, the BS can compute the estimate of $\mathbf{h}_{0,n,k}$ and the result is stated in the following lemma.

\begin{lemma} \label{L2} The MMSE estimate $\hat{\mathbf{h}}_{0,n,k}$ of $\mathbf{h}_{0,n,k}$ is given as
\begin{align}
\label{h_irs_est}
&\hat{\mathbf{h}}_{0,n,k}= r_{n,k} \beta_{2,k} \mathbf{h}_{1,n} \mathbf{h}_{1,n}^H  \mathbf{Q}_{n,k}\tilde{\mathbf{r}}^{tr}_{n+1,k}, 
\end{align}
for $ n=1,\dots, N$, $k=1,\dots, K$, which is distributed as $\hat{\mathbf{h}}_{0,n,k}\sim \mathcal{CN}(\mathbf{0}, \mathbf{\Psi}_{n,k})$, where
\begin{align}
&\mathbf{\Psi}_{n,k}= r_{n,k} r_{n,k}^* \beta^2_{2,k} \mathbf{h}_{1,n} \mathbf{h}_{1,n}^H  \mathbf{Q}_{n,k} \mathbf{h}_{1,n} \mathbf{h}_{1,n}^H  ,
\end{align}
and
\begin{align}
\label{er1}
&\mathbf{Q}_{n,k}=\left(r_{n,k} \beta_{2,k} \mathbf{h}_{1,n} \mathbf{h}_{1,n}^H  +\frac{\sigma^2}{SP_C \tau_S}\mathbf{I}_M\right)^{-1}. 
\end{align}
 Also $r_{n,k}$ is the $(n,n)^{th}$ entry of the matrix $\mathbf{R}_{IRS_k}$ and $\mathbf{h}_{1,n}$ is the $n^{th}$ column of $\mathbf{H}_1$.
\end{lemma}
\begin{IEEEproof}
The proof is provided in Appendix \ref{Sec:L2}. 
\end{IEEEproof}

Invoking the orthogonality property of the MMSE estimate, we can decompose  $\mathbf{h}_{0,n,k}$ as $\mathbf{h}_{0,n,k}= \hat{\mathbf{h}}_{0,n,k} + \tilde{\mathbf{h}}_{0,n,k}$, where $\tilde{\mathbf{h}}_{0,n,k}\sim \mathcal{CN}(\mathbf{0},\tilde{\boldsymbol{\Psi}}_{n,k})$ is the uncorrelated estimation error, where $\tilde{\boldsymbol{\Psi}}_{n,k}=\beta_{2,k} r_{n,k} \mathbf{h}_{1,n} \mathbf{h}_{1,n}^H  -\mathbf{\Psi}_{n,k}$. Under the proposed design in \eqref{des}, the MMSE estimates do not depend on the cross-correlation between IRS elements, so knowledge of $\mathbf{R}_{IRS}$ is not required at the BS\footnote{The diagonal elements of the correlation matrix of a correlated Rayleigh channel vector equal unity so $r_{n,k}=1$, $\forall n$.}. 

To calculate the MMSE estimates, the BS will require knowledge of the correlation matrices $\mathbf{R}_{BS_k}$, $k=1,\dots, K$, and the LoS BS-to-IRS channel vectors $\mathbf{h}_{1,n}$, $n=1,\dots, N$. The LoS channel vectors are deterministic which depend only on the LoS angles between the BS and IRS. These angles need to be calculated only once at the BS using knowledge of the IRS location, which is fixed. The correlation matrices vary very slowly as compared to the fast fading process and stay constant over many coherence intervals. As discussed in several works, they can be calculated based on knowledge of only the users' AoAs (which depend on their locations) and angular spread in the environment, both of which can be accurately learned and tracked at the BS  \cite{usergroups, ourworkTWC}\footnote{Even for nomadic users, the AoA and angular spread which determine the channel correlation evolve in time much slower than the actual channel fading process, and can be considered ``locally constant". Algorithms for covariance tracking are well known and widely investigated, and could be employed here to track the slow variations of the users'   channel covariance matrices \cite{usergroups}. However, the covariance tracking aspect of the system is out of the scope of this work.}. In fact, second-order channel statistics are generally assumed to be perfectly known at the BS in massive MIMO literature \cite{massiveMIMObook}.

Unlike LS estimates, the MMSE estimates depend on the distribution of $\mathbf{H}_1$, $\mathbf{h}_{2,k}$ and $\mathbf{h}_{d,k}$. The derived results can be easily generalized to other channel fading models. For example, the MMSE estimates under independent Rayleigh fading $\mathbf{h}_{2,k}$s and $\mathbf{h}_{d,k}$s can be obtained by setting $\mathbf{R}_{BS}=\mathbf{I}_M$. The estimates when $\mathbf{H}_1$ is not fixed but rather follows a fading model can be be similarly developed. After obtaining the MMSE estimates, the BS uses them to design the precoder $\mathbf{G}^*$, power allocation matrix $\mathbf{P}^*$ as well as the reflect beamforming vector $\mathbf{v}^*$ in (\ref{ch1}) based on the performance criteria of interest. The BS then provides information on the required IRS phase shifts vector $\mathbf{v}^*$  for downlink transmission to the IRS micro-controller.  Wireless backhaul links in mmWave and THz bands are suitable candidates for the BS to communicate with the IRS controller under strict latency requirements \cite{SRE}.

\subsection{NMSE Comparison with Least Squares Estimation}

The LS estimates  are obtained by correlating the received training signal $\mathbf{Y}^{tr}_s$ with the pilot sequence of user $k$ as shown in \eqref{LSS} and applying the pseudo-inverse of $\bar{\mathbf{V}}^{tr}$ on the resulting observation vector as done in \eqref{LSS2}  \cite{LS1}. Under the DFT design for $\mathbf{V}^{tr}$ in \eqref{des}, the LS estimates are given as
\begin{align}
\label{h_d_ls}
	&\hat{\mathbf{h}}^{LS}_{d,k}=\mathbf{h}_{d,k}+\frac{1}{S} (\mathbf{v}^{tr}_1 \otimes \mathbf{I}_M)^H \frac{\mathbf{n}^{tr}_{k}}{P_C\tau_S},  \\
	\label{h_irs_ls}
&\hat{\mathbf{h}}^{LS}_{0,n,k}=\mathbf{h}_{0,n,k}+\frac{1}{S} (\mathbf{v}^{tr}_{n+1}\otimes \mathbf{I}_M)^H \frac{\mathbf{n}^{tr}_{k}}{P_C \tau_S},
\end{align}
where $\mathbf{v}^{tr}_{n+1}$ is the $(n+1)^{th}$ column of $\mathbf{V}^{tr}$. 

We develop analytical expressions for the normalized MSE (NMSE) in the LS and MMSE estimates of direct and IRS-assisted channel vectors. The NMSE is defined as
\begin{align}
&\text{NMSE}(\hat{\mathbf{h}}_{d,k})\hspace{-.01in}=\hspace{-.01in}\frac{\text{tr}(\mathbb{E}[(\hat{\mathbf{h}}_{d,k}-\mathbf{h}_{d,k})(\hat{\mathbf{h}}_{d,k}-\mathbf{h}_{d,k})^H])}{\text{tr}(\mathbb{E}[\mathbf{h}_{d,k}\mathbf{h}_{d,k}^H])}, \\
&\text{NMSE}(\hat{\mathbf{h}}_{0,n,k})\hspace{-.03in}=\hspace{-.03in}\frac{\text{tr}\hspace{-.01in}(\mathbb{E}[(\hat{\mathbf{h}}_{0,n,k}-\mathbf{h}_{0,n,k})(\hat{\mathbf{h}}_{0,n,k}-\mathbf{h}_{0,n,k})^H])}{\text{tr}(\mathbb{E}[\mathbf{h}_{0,n,k}\mathbf{h}_{0,n,k}^H])}.
\end{align}

 To enable an analytical comparison, we set $\mathbf{R}_{BS_k}=\mathbf{I}_M$, $k=1,\dots, K$\footnote{This assumption does not affect the NMSE in LS estimates. Under the MMSE-DFT protocol, the NMSE in the estimation of IRS-assisted channels is independent of the structure of the correlation matrix $\mathbf{R}_{IRS_k}$ as discussed in Lemma 2. Only the NMSE in the MMSE estimation of direct channel is affected by $\mathbf{R}_{BS_k}$ and this effect will be studied through simulations.}. Noting that $\text{tr}(\mathbb{E}[\mathbf{h}_{d,k}\mathbf{h}_{d,k}^H])=\beta_{d,k} \text{tr}(\mathbf{R}_{BS_k})=M\beta_{d,k}$, the NMSE in the LS-DFT estimate of $\mathbf{h}_{d,k}$ is given as
\begin{align}
&\text{NMSE}(\hat{\mathbf{h}}_{d,k}^{LS})=\frac{\text{tr}\left((\mathbf{v}^{tr}_1 \otimes \mathbf{I}_M)^H \mathbb{E}\left[\mathbf{n}^{tr}_{k}\mathbf{n}^{tr^H}_{k}\right] (\mathbf{v}^{tr}_1 \otimes \mathbf{I}_M)\right)}{M\beta_{d,k} S^2(P_C \tau_S)^2}, \nonumber \\
&=\frac{1}{M\beta_{d,k}}\frac{\sigma^2 P_C \tau_S}{S^2 (P_C \tau_S)^2}\text{tr}\left((\mathbf{v}^{tr}_1 \otimes \mathbf{I}_M)^H (\mathbf{v}^{tr}_1 \otimes \mathbf{I}_M)\right)\\
\label{exxpp3}
&=\frac{\sigma^2}{\beta_{d,k} S P_C \tau_S}.
\end{align}
The result follows from using $\mathbb{E}\left[\mathbf{n}^{tr}_{k}\mathbf{n}^{tr^H}_{k}\right]=\sigma^2P_C \tau_S \mathbf{I}_{MS}$ as proved in \eqref{nn} and that $\text{tr}((\mathbf{v}^{tr}_1 \otimes \mathbf{I}_M)^H(\mathbf{v}^{tr}_1 \otimes \mathbf{I}_M))=\text{tr}(\mathbf{v}^{tr^H}_1\mathbf{v}^{tr}_1 \otimes \mathbf{I}_M)=MS$. The expression reveals that the NMSE in the LS estimate increases linearly as $\sigma^2$ grows large or $\beta_{d,k}$, $P_C$, $\tau_S$ grow small. This result can also be derived directly as the trace of the first $M\times M$ block diagonal matrix of $\mathbf{C}_{\tilde{\mathbf{n}}^{tr}_k}$ in \eqref{C_n}.

The NMSE in the MMSE-DFT estimate of $\mathbf{h}_{d,k}$ in Lemma \ref{L1} can be computed as $\text{NMSE}(\hat{\mathbf{h}}_{d,k})=\frac{1}{M\beta_{d,k}}\text{tr}(\tilde{\boldsymbol{\Psi}}_{d,k})$ resulting in
\begin{align}
\label{expp1}
&\text{NMSE}(\hat{\mathbf{h}}_{d,k})=\frac{1}{M\beta_{d,k}}\Big(\beta_{d,k}\text{tr}\left(\mathbf{I}_M \right) -\beta_{d,k}^2 \text{tr} \Big(\beta_{d,k}\mathbf{I}_M\nonumber \\
&+\frac{\sigma^2}{S P_C \tau_S} \mathbf{I}_M \Big)^{-1}\Big), \\
\label{exxx}
&=\frac{1}{M\beta_{d,k}}\frac{M \beta_{d,k} \frac{\sigma^2}{S P_C \tau_S}}{\left(\beta_{d,k}+\frac{\sigma^2}{S P_C \tau_S}\right)}=\frac{\frac{\sigma^2}{S P_C \tau_S}}{\beta_{d,k}+\frac{\sigma^2}{S P_C \tau_S}}.
\end{align} 
We observe that the NMSE in the MMSE estimate approaches $1$ as $\sigma^2$ grows large or $\beta_{d,k}$, $P_C$, $\tau_S$ grow small. The NMSE value of $1$ signifies that the error in the channel estimate has the same power as the true channel itself. Any beamforming transmission under estimates having NMSE values of $1$ or beyond will correspond to isotropic transmission, i.e. as if the BS and IRS beamform with no CSI at all \cite{LS}. However, as compared to the LS estimate, the NMSE in MMSE-DFT estimate will increase to $1$ much slowly (i.e. when the noise becomes very high or $\beta_{d,k}$ becomes very small) as can be seen by comparing \eqref{exxpp3} and \eqref{exxx}. This implies that MMSE-DFT estimates will be more accurate even at  low values of training signal-to-noise ratio (SNR). Finally denoting $c=\frac{\sigma^2}{S P_C \tau_S}$ we have
\begin{align}
&\text{NMSE}(\hat{\mathbf{h}}^{LS}_{d,k})-\text{NMSE}(\hat{\mathbf{h}}_{d,k})=\frac{c}{\beta_{d,k}} - \frac{c}{\beta_{d,k}+c}\nonumber \\
&=\frac{1}{\beta_{d,k}}\frac{c^2}{(\beta_{d,k}+c)}\geq 0,
\end{align}
since $c$ and $\beta_{d,k}$ are non-negative. Therefore, the MMSE-DFT estimate of the direct channel will always outperform the LS-DFT estimate for any value of $\sigma^2$, $P_c$, $S$, $\tau_S$ and $\beta_{d,k}$.

Next we compute the NMSE in the LS-DFT estimates of $\mathbf{h}_{0,n,k}$ in a similar manner as \eqref{exxpp3}. Noting that $\text{tr}(\mathbb{E}[\mathbf{h}_{0,n,k}\mathbf{h}_{0,n,k}^H])=M\beta_1\beta_{2,k}=M\beta_k$, we obtain
\begin{align}
&\text{NMSE}(\hat{\mathbf{h}}_{0,n,k}^{LS})\hspace{-.04in}=\hspace{-.04in}\frac{\text{tr}((\mathbf{v}^{tr}_{n+1} \otimes \mathbf{I}_M)^H \mathbb{E}[\mathbf{n}^{tr}_{k}\mathbf{n}^{tr^H}_{k}](\mathbf{v}^{tr}_{n+1} \otimes \mathbf{I}_M))}{M\beta_kS^2(P_C \tau_S)^2}\\
\label{expp2}
&=\frac{1}{M\beta_k}\frac{ \sigma^2 P_C \tau_S}{S^2 (P_C \tau_S)^2}\text{tr}(\mathbf{v}^{tr^H}_{n+1}\mathbf{v}^{tr}_{n+1} \otimes \mathbf{I}_M)= \frac{1}{\beta_k} \frac{\sigma^2}{S P_C \tau_S}.
\end{align}
The NMSE in the LS estimation of each $\mathbf{h}_{0,n,k}$ is the same as the NMSE in the LS estimation of the direct channel in \eqref{exxpp3}. 

The NMSE in the MMSE-DFT estimates of $\mathbf{h}_{0,n,k}$ in Lemma \ref{L2} can be computed as $\frac{1}{M\beta_k}\text{tr}(\tilde{\boldsymbol{\Psi}}_{n,k})$ resulting in
\begin{align}
&\text{NMSE}(\hat{\mathbf{h}}_{0,n,k})=\frac{1}{M\beta_k}\Big(\beta_{2,k} \text{tr}(\mathbf{h}_{1,n} \mathbf{h}_{1,n}^H) -\beta_{2,k}^2 \text{tr}\Big(\mathbf{h}_{1,n} \mathbf{h}_{1,n}^H\nonumber \\
& \left(\beta_{2,k} \mathbf{h}_{1,n} \mathbf{h}_{1,n}^H +\frac{\sigma^2}{S P_C \tau_S}\mathbf{I}_M\right)^{-1} \mathbf{h}_{1,n} \mathbf{h}_{1,n}^H\Big)\Big), \nonumber \\
\label{ins}
&=\frac{1}{M\beta_k}\left(M\beta_k-\frac{\beta_k^2 M^2}{\frac{\sigma^2}{S P_C \tau_S}}+\frac{\beta_k^3 M^3}{(\frac{\sigma^2}{S P_C \tau_S})^2+M\frac{\sigma^2}{S P_C \tau_S}\beta_k}\right), \\
\label{exx}
&=\frac{1}{M\beta_k}\left(\frac{M\beta_k \frac{\sigma^2}{S P_C \tau_S}}{M\beta_k+\frac{\sigma^2}{S P_C \tau_S}}\right)=\frac{\frac{\sigma^2}{S P_C \tau_S}}{M\beta_k+\frac{\sigma^2}{S P_C \tau_S}}.
\end{align}
where \eqref{ins} follows from applying the Sherman–Morrison formula on the inverse term and noting that $\text{tr}(\mathbf{h}_{1,n} \mathbf{h}_{1,n}^H)=\beta_1 M$ under the definitions in Sec. II-B. 

Denoting $c=\frac{\sigma^2}{S P_C \tau_S}$ and using straightforward calculation we can show that
\begin{align}
&\text{NMSE}(\hat{\mathbf{h}}_{0,n,k}^{LS})-\text{NMSE}(\hat{\mathbf{h}}_{0,n,k})=\frac{c}{\beta_k}-\frac{ c}{M\beta_k+c}\nonumber \\
&=\frac{c^2+c\beta_{k} (M-1)}{\beta_k(c+M\beta_k)}\geq 0,
\end{align}
since $c \geq 0$, $\beta_k\geq 0$ and $M\geq 1$. Therefore the NMSE in the MMSE-DFT estimate of $\mathbf{h}_{0,n,k}$ will always be lower than the NMSE in the LS-DFT estimate for any value of noise, power, sub-phase duration and path loss factor. Also $\text{NMSE}(\hat{\mathbf{h}}_{0,n,k})$ approaches $1$ as $c$ grows large or $\beta_k$ grows small.

\subsection{Performance Evaluation of the Proposed Protocol}

\begin{figure*}[!t]
\begin{subfigure}[t]{.48\textwidth}
\tikzset{every picture/.style={scale=.95}, every node/.style={scale=.8}}
%
%
\definecolor{mycolor1}{rgb}{1.00000,0.00000,1.00000}%
\definecolor{mycolor2}{rgb}{0.00000,0.49804,0.00000}%
\definecolor{mycolor3}{rgb}{0.92941,0.69412,0.12549}%
	 	\definecolor{mycolor4}{rgb}{0.52, 0.52, 0.51}
	\definecolor{mycolor5}{rgb}{0.72, 0.45, 0.2}
\begin{tikzpicture}

\begin{axis}[%
width=.95\columnwidth,
height=.85\columnwidth,
scale only axis,
xmode=log,
xmin=5e-07,
xmax=0.05,
xminorticks=true,
xlabel style={at={(.5,-0.03)},font=\color{white!15!black}},
xlabel={$\text{Noise variance }\sigma{}^\text{2}\text{ (J)}$},
ymode=log,
ymin=1e-04,
ymax=1e+4,
yminorticks=true,
ylabel style={at={(-.05,0.5)},font=\color{white!15!black}},
ylabel={NMSE},
axis background/.style={fill=white},
xmajorgrids,
xminorgrids,
ymajorgrids,
yminorgrids,
legend style={at={(5e-7,1.027)}, anchor=north west, legend cell align=left,align=left,draw=white!15!black, /tikz/column 2/.style={
                column sep=5pt,
            }},
]

\addplot [color=blue, line width=1.35pt,mark size=2.2pt,mark=o,mark options={solid}]
  table[row sep=crcr]{%
5e-07	0.000913153897721869\\
1e-06	0.00184609711560569\\
5e-06	0.00926842305845908\\
1e-05	0.0178687187097176\\
5e-05	0.0857415235213299\\
0.0001	0.15348353308952\\
0.0005	0.489439634841084\\
0.001	0.644271280156953\\
0.005	0.91701251435247\\
0.01	0.941989470058917\\
0.05	0.986010667681169\\
};
\addlegendentry{MMSE-DFT}

\addplot [color=mycolor2, line width=1.0pt,mark size=2.0pt,mark=x,mark options={solid}]
  table[row sep=crcr]{%
5e-07	0.000908265213442325\\
1e-06	0.00181488203266788\\
5e-06	0.00900900900900901\\
1e-05	0.0178571428571429\\
5e-05	0.0833333333333333\\
0.0001	0.153846153846154\\
0.0005	0.476190476190476\\
0.001	0.645161290322581\\
0.005	0.900900900900901\\
0.01	0.947867298578199\\
0.05	0.989119683481701\\
};
\addlegendentry{MMSE-DFT Th. \eqref{exxx}}

\addplot [color=mycolor1, line width=1.35pt,mark size=2pt,mark=square,mark options={solid}]
  table[row sep=crcr]{%
5e-07	0.000914181907462083\\
1e-06	0.00185173400303849\\
5e-06	0.009344646584439\\
1e-05	0.018293848434533\\
5e-05	0.0932854638997813\\
0.0001	0.183378109480078\\
0.0005	0.955672006952576\\
0.001	1.83500442793136\\
0.005	9.24417409939511\\
0.01	18.8302429098837\\
0.05	89.7816895910912\\
};
\addlegendentry{LS-DFT}

\addplot [color=black, line width=1pt,mark size=2pt,mark=triangle,mark options={solid}]
  table[row sep=crcr]{%
5e-07	0.000909090909090909\\
1e-06	0.00181818181818182\\
5e-06	0.00909090909090909\\
1e-05	0.0181818181818182\\
5e-05	0.0909090909090909\\
0.0001	0.181818181818182\\
0.0005	0.909090909090909\\
0.001	1.81818181818182\\
0.005	9.09090909090909\\
0.01	18.1818181818182\\
0.05	90.9090909090909\\
};
\addlegendentry{LS-DFT Th. \eqref{exxpp3}}

\addplot [color=mycolor4, dashed, line width=1pt,mark size=2pt,mark=pentagon,mark options={solid}]
  table[row sep=crcr]{%
5e-07	0.000909563953235263\\
1e-06	0.00174784307325126\\
5e-06	0.00780379719103767\\
1e-05	0.0142577685162664\\
5e-05	0.051304483826077\\
0.0001	0.0809165271913416\\
0.0005	0.230316631178038\\
0.001	0.356452852506205\\
0.005	0.678771765764542\\
0.01	0.859621692127249\\
0.05	0.992121255696073\\
};
\addlegendentry{MMSE-DFT (Corr.)}

\addplot [color=mycolor5, dashed, line width=.85pt,mark size=1.75pt,mark=triangle, mark options={solid, rotate=270, mycolor5}]
  table[row sep=crcr]{%
5e-07	0.000922527679206635\\
1e-06	0.0017998965756198\\
5e-06	0.00896788065885701\\
1e-05	0.0177944876418871\\
5e-05	0.0909174097514928\\
0.0001	0.184085369036565\\
0.0005	0.90600486372447\\
0.001	1.76941661130344\\
0.005	9.15449320282464\\
0.01	18.6131453922204\\
0.05	94.1300574760424\\
};
\addlegendentry{LS-DFT (Corr.)}

\addplot [color=red, line width=1.0pt, mark size=2.0pt, mark=asterisk, mark options={solid, red}]
  table[row sep=crcr]{%
5e-07	0.0109995338471803\\
1e-06	0.0191078243206767\\
5e-06	0.0942592293934699\\
1e-05	0.171653875570059\\
5e-05	0.481433816682558\\
0.0001	0.602612698908088\\
0.0005	0.934457300726459\\
0.001	0.884035368938685\\
0.005	0.972183758520154\\
0.01	1.0379373813187\\
0.05	1.01301281800574\\
};
\addlegendentry{MMSE ON/OFF}

\addplot [color=mycolor3, line width=1.0pt, mark size=2.0pt, mark=star, mark options={solid, mycolor3}]
  table[row sep=crcr]{%
5e-07	0.0110803635939518\\
1e-06	0.0192573381082145\\
5e-06	0.105584878971697\\
1e-05	0.205198981372483\\
5e-05	1.02462609543148\\
0.0001	1.70438635484415\\
0.0005	10.0720740635193\\
0.001	21.4568447384982\\
0.005	110.41429695116\\
0.01	182.105151658948\\
0.05	1095.01943892857\\
};
\addlegendentry{LS ON/OFF}

\end{axis}
\end{tikzpicture}%
\caption{NMSE in the estimation of $\mathbf{h}_{d,k}$.}
\label{Fig1_est}
\end{subfigure}
\hspace{.05cm}
\begin{subfigure}[t]{.48\textwidth}
\tikzset{every picture/.style={scale=.95}, every node/.style={scale=.8}}
%
%
\definecolor{mycolor1}{rgb}{1.00000,0.00000,1.00000}%
\definecolor{mycolor2}{rgb}{0.00000,0.49804,0.00000}%
\definecolor{mycolor3}{rgb}{0.92941,0.69412,0.12549}%
	 	\definecolor{mycolor4}{rgb}{0.52, 0.52, 0.51}
	\definecolor{mycolor5}{rgb}{0.72, 0.45, 0.2}

\begin{tikzpicture}

\begin{axis}[%
width=.95\columnwidth,
height=.85\columnwidth,
scale only axis,
xmode=log,
xmin=5e-07,
xmax=.05,
xminorticks=true,
xlabel style={at={(.5,-0.03)},font=\color{white!15!black}},
xlabel={$\text{Noise variance }\sigma{}^\text{2}\text{ (J)}$},
ymode=log,
ymin=1e-04,
ymax=1e+4,
yminorticks=true,
ylabel style={at={(-.05,0.5)},font=\color{white!15!black}},
ylabel={NMSE},
axis background/.style={fill=white},
xmajorgrids,
xminorgrids,
ymajorgrids,
yminorgrids,
legend style={at={(5e-7,1.027)}, anchor=north west, legend cell align=left,align=left,draw=white!15!black, /tikz/column 2/.style={
                column sep=5pt,
            }},
						]

\addplot [color=blue, line width=1.5pt,mark size=2.2pt,mark=o,mark options={solid}]
  table[row sep=crcr]{%
5e-07	0.00022902140391587\\
1e-06	0.000446529976706775\\
5e-06	0.00229041294965168\\
1e-05	0.00447401714873743\\
5e-05	0.0225419757307469\\
0.0001	0.0436698950235311\\
0.0005	0.183683816183303\\
0.001	0.302519866668551\\
0.005	0.701964718322593\\
0.01	0.814318494704893\\
0.05	0.956267646742323\\
};
\addlegendentry{MMSE-DFT}

\addplot [color=mycolor2, line width=1.0pt,mark size=2.0pt,mark=x,mark options={solid}]
  table[row sep=crcr]{%
5e-07	0.000227221086116793\\
1e-06	0.000454338936846884\\
5e-06	0.00226757369614513\\
1e-05	0.00452488687782804\\
5e-05	0.0222222222222222\\
0.0001	0.0434782608695653\\
0.0005	0.185185185185186\\
0.001	0.3125\\
0.005	0.694444444444452\\
0.01	0.819672131147531\\
0.05	0.957854406130264\\
};
\addlegendentry{MMSE-DFT Th. \eqref{exx}}

\addplot [color=mycolor1, line width=1.5pt,mark size=2.0pt,mark=square,mark options={solid}]
  table[row sep=crcr]{%
5e-07	0.000916995895385511\\
1e-06	0.00180347541358252\\
5e-06	0.00906001255230904\\
1e-05	0.0182116128602454\\
5e-05	0.0905488852801516\\
0.0001	0.179402033445227\\
0.0005	0.89449303663276\\
0.001	1.80414538826404\\
0.005	9.04737712168145\\
0.01	18.2299424684776\\
0.05	91.1285475143017\\
};
\addlegendentry{LS-DFT}

\addplot [color=black, line width=1pt,mark size=2pt,mark=triangle,mark options={solid}]
  table[row sep=crcr]{%
5e-07	0.000909090909090909\\
1e-06	0.00181818181818182\\
5e-06	0.00909090909090909\\
1e-05	0.0181818181818182\\
5e-05	0.0909090909090909\\
0.0001	0.181818181818182\\
0.0005	0.909090909090909\\
0.001	1.81818181818182\\
0.005	9.09090909090909\\
0.01	18.1818181818182\\
0.05	90.9090909090909\\
};
\addlegendentry{LS-DFT Th. \eqref{expp2}}

\addplot [color=mycolor4, dashed, line width=1pt,mark size=2pt,mark=pentagon,mark options={solid}]
  table[row sep=crcr]{%
5e-07	0.000230393758186297\\
1e-06	0.0004519880080595\\
5e-06	0.00219688641648266\\
1e-05	0.00457159473763846\\
5e-05	0.0220868719386701\\
0.0001	0.0432805487789176\\
0.0005	0.188115292700727\\
0.001	0.309956389278437\\
0.005	0.704917516039559\\
0.01	0.795191215930315\\
0.05	0.951387523820257\\
};
\addlegendentry{MMSE-DFT (Corr.)}

\addplot [color=mycolor5, dashed, line width=.85pt,mark size=1.75pt,mark=triangle, mark options={solid, rotate=270, mycolor5}]
  table[row sep=crcr]{%
5e-07	0.00092483072803912\\
1e-06	0.00182811612120628\\
5e-06	0.00902650702623669\\
1e-05	0.018048296892459\\
5e-05	0.0907063319749246\\
0.0001	0.181141494380446\\
0.0005	0.913009687877462\\
0.001	1.81057856805997\\
0.005	9.17601168443558\\
0.01	18.1412898865052\\
0.05	90.9037537324196\\
};
\addlegendentry{LS-DFT (Corr.)}

\addplot [color=red, line width=1.0pt, mark size=2.0pt, mark=asterisk, mark options={solid, red}]
  table[row sep=crcr]{%
5e-07	0.00544947411151575\\
1e-06	0.0119652239392242\\
5e-06	0.0444928330697225\\
1e-05	0.0965135108559008\\
5e-05	0.265472179972676\\
0.0001	0.397824022636052\\
0.0005	0.756966703052943\\
0.001	0.848482890301741\\
0.005	0.959505546057865\\
0.01	0.926089555548564\\
0.05	1.05265226753104\\
};
\addlegendentry{MMSE ON/OFF}

\addplot [color=mycolor3, line width=1.0pt, mark size=2.0pt, mark=star, mark options={solid, mycolor3}]
  table[row sep=crcr]{%
5e-07	0.01947795282062\\
1e-06	0.0425469744539112\\
5e-06	0.197740542206657\\
1e-05	0.41492800830797\\
5e-05	2.03803933963539\\
0.0001	4.34243311592111\\
0.0005	20.5391991570389\\
0.001	42.4203397610334\\
0.005	200.447153389857\\
0.01	404.947448779512\\
0.05	1975.74511906777\\
};
\addlegendentry{LS ON/OFF}


\end{axis}
\end{tikzpicture}%
\caption{NMSE in the estimation of $\mathbf{h}_{0,n,k}$.}
\label{Fig2_est}
\end{subfigure}
\caption{NMSE comparison between MMSE-DFT and LS-DFT estimates against $\sigma^2$ for $M=4$, $N=10$ under independent Rayleigh fading and correlated (Corr.) Rayleigh channels.}
\label{Fig_est}
\end{figure*}
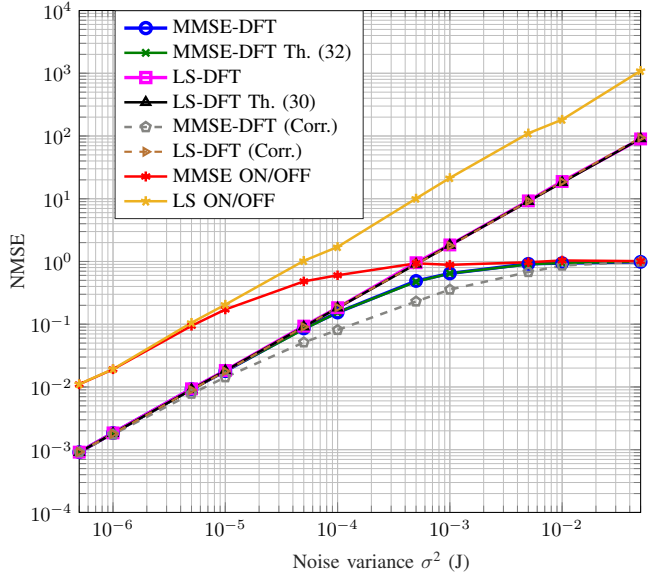
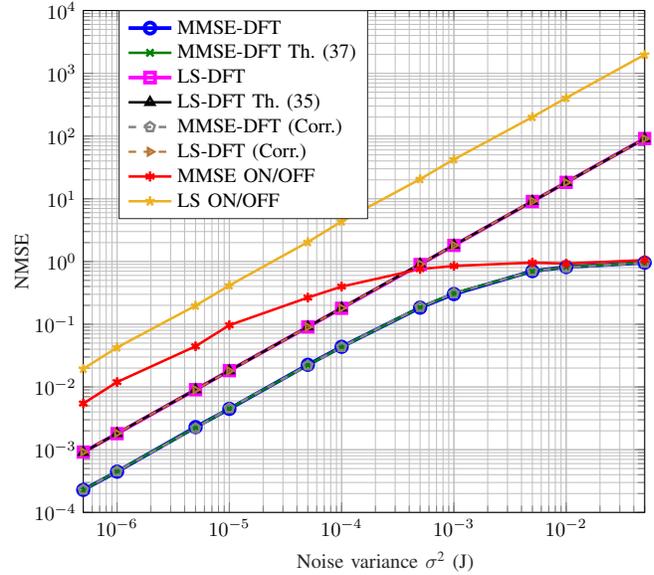

The NMSE in the LS-DFT and the MMSE-DFT estimates of the direct and IRS-assisted channels are compared in Fig. \ref{Fig_est} versus the noise variance $\sigma^2$. Fig. \ref{Fig1_est} shows the Monte-Carlo simulated $\text{NMSE}(\hat{\mathbf{h}}_{d,k})$  as well as the theoretical (Th.) expressions in \eqref{exxpp3} and \eqref{exxx} for LS-DFT and MMSE-DFT estimates respectively. Fig. \ref{Fig2_est} shows the simulated quantity $\frac{1}{N}\sum_{n=1}^N \text{NMSE}(\hat{\mathbf{h}}_{0,n,k})$ as well as the theoretical expressions  in \eqref{expp2} and \eqref{exx} for LS-DFT and  MMSE-DFT estimates respectively. The parameter values are set as $M=4$, $N=10$, $P_C=1$, $T_S=K=1$, $\tilde{\tau}=50\mu$s, $\tau_S=T_S\tilde{\tau}$ and $S=N+1$. The simulated NMSE matches the theoretical expressions perfectly. Moreover, the MMSE-DFT estimates achieve a lower NMSE than the LS-DFT estimates especially for moderate to high values of $\sigma^2$ (i.e. low SNR regime).  We observe that the NMSE in the MMSE-DFT and LS-DFT estimates of $\mathbf{h}_{d,k}$ becomes the same for very low values of noise while the NMSE in the MMSE-DFT estimates of $\mathbf{h}_{0,n,k}$s is always lower as compared to that in LS-DFT estimates. The NMSE in the MMSE estimates approaches $1$ for both the direct channel and the IRS-assisted channels as the noise variance increases, while the NMSE in the LS estimates grows even beyond $1$. However, as we discussed earlier, the NMSE value of $1$ implies that the estimation error has the same power as the actual channel being estimated. For NMSE values of $1$ and beyond under any estimation protocol, the performance of the IRS-assisted system  will correspond to isotropic transmission, i.e. transmission without any CSI, which actually provides the worst bound on the performance under estimation errors \cite{LS}. However, the NMSE under LS-DFT protocol grows to one much quicker than MMSE-DFT protocol, making LS-DFT more prone to estimation errors.

We also plot the NMSE for the correlated (Corr.) scenario where $[\mathbf{R}_{BS_k}]_{m,m'}=\eta^{|m-m'|}$  and $[\mathbf{R}_{IRS_k}]_{n,n'}=\eta^{|n-n'|}$ and $\eta$ is set as $0.95$. The NMSE in the LS-DFT estimates is unaffected and the NMSE in the MMSE-DFT estimates of $\mathbf{h}_{0,n,k}$s is also unaffected by the structure of correlation. The NMSE in the MMSE-DFT estimate of the direct channel $\mathbf{h}_{d,k}$ actually reduces with the introduction of correlation. 

We also compare the results against the LS-ON/OFF protocol in \cite{LS}, which sets $S=N+1$ and uses $\mathbf{V}^{tr}=\begin{bmatrix} 
    1 & \mathbf{0}_{N}^T  \\
		\mathbf{1}_N& \mathbf{I}_N 
  \end{bmatrix} \in \mathbb{C}^{N+1\times N+1}$. The drawbacks of this approach is that the cascaded channel is only estimated one-by-one such that the noise variance in each element of the received observation vector given in \eqref{C_n} is  $\frac{\sigma^2}{P_C \tau_S}$ instead of $\frac{\sigma^2}{S P_C \tau_S}$, and the error in the estimation of $\mathbf{h}_{d,k}$ is propagated to the estimation of $\mathbf{h}_{0,n,k}$s. The NMSE in the LS estimates of $\mathbf{h}_{d,k}$ and $\mathbf{h}_{0,n,k}$ under ON/OFF protocol can be straightforwardly calculated to be $\frac{\sigma^2}{\beta_{d,k} P_C \tau_S }$ and $\frac{2\sigma^2}{\beta_{d,k}P_C \tau_S}$ respectively. Compared to \eqref{exxpp3} and \eqref{expp2}, we see a factor of $S$ and $2S$ increase respectively in the NMSE in $\hat{\mathbf{h}}^{LS}_{d,k}$ and $\hat{\mathbf{h}}^{LS}_{0,n,k}$ under ON/OFF protocol, which can also be observed by comparing the  LS-DFT and LS-ON/OFF curves in Fig. \ref{Fig_est}. 
	
	Furthermore, the MMSE estimates under ON/OFF protocol can be derived in a similar manner as done in this work (details have been skipped for brevity in writing). The NMSE in the MMSE-ON/OFF estimates can be derived as $\frac{\frac{\sigma^2}{P_c \tau_S}}{\beta_{d,k}+\frac{\sigma^2}{P_c \tau_S}}$ for $\mathbf{h}_{d,k}$ and $\frac{1}{\left(1+\frac{M \beta_k \left(1+\frac{\sigma^2}{P_c \tau_S}\right)}{\left(\frac{\sigma^2}{P_c \tau_S}\right)^2+\frac{\sigma^2}{P_c \tau_S}(\beta_{d,k}+1)}\right)}$ for $\mathbf{h}_{0,n,k}$s. Compared to \eqref{exxx} and \eqref{exx}, we can see that the MMSE-ON/OFF protocol causes a factor of $S$ and $S(1+\beta_{d,k})$ increase in the NMSE in  $\hat{\mathbf{h}}_{d,k}$ and $\hat{\mathbf{h}}_{0,n,k}$ respectively in the low noise regime. In the high noise regime, the NMSE in MMSE-ON/OFF estimates and MMSE-DFT estimates becomes close. These results can also be observed by comparing the  MMSE-DFT and MMSE-ON/OFF curves in Fig. \ref{Fig_est}. 
	
\begin{figure*}[!t]
\begin{subfigure}[t]{.48\textwidth}
\tikzset{every picture/.style={scale=.95}, every node/.style={scale=.8}}
%
%
\definecolor{mycolor1}{rgb}{0.00000,0.49804,0.00000}%
\definecolor{mycolor2}{rgb}{1.00000,0.00000,1.00000}%
\begin{tikzpicture}

\begin{axis}[%
width=.95\columnwidth,
height=.85\columnwidth,
scale only axis,
xmin=0,
xmax=1,
xlabel style={font=\color{white!15!black}},
xlabel={$\beta_{d,k}$},
ymode=log,
ymin=0.1,
ymax=18.4709755509772,
yminorticks=true,
ylabel style={font=\color{white!15!black}},
ylabel={NMSE},
axis background/.style={fill=white},
xmajorgrids,
ymajorgrids,
yminorgrids,
legend style={at={(1,1.03)}, anchor=north east, legend cell align=left,align=left,draw=white!15!black, /tikz/column 2/.style={
                column sep=5pt,
            }},]
\addplot [color=blue, line width=1pt, mark size=2.0pt, mark=o, mark options={solid, blue}]
  table[row sep=crcr]{%
0.05	0.929403273652807\\
0.1	0.884162905256498\\
0.15	0.863148364121979\\
0.2	0.818199191561248\\
0.25	0.780003484912198\\
0.3	0.739735099690007\\
0.35	0.768917849397478\\
0.4	0.697093176220531\\
0.45	0.671284042153159\\
0.5	0.640409479370211\\
0.55	0.60459785340822\\
0.6	0.59904815523924\\
0.65	0.586279826831476\\
0.7	0.558612930404967\\
0.75	0.553532544061949\\
0.8	0.522457111834903\\
0.85	0.525963301320237\\
0.9	0.514905236792377\\
0.95	0.487110364722463\\
1	0.472455944362346\\
};
\addlegendentry{MMSE-DFT}

\addplot [color=mycolor1, line width=1pt, mark size=1.75pt, mark=x, mark options={solid, mycolor1}]
  table[row sep=crcr]{%
0.05	0.947867298578199\\
0.1	0.900900900900901\\
0.15	0.858369098712446\\
0.2	0.819672131147541\\
0.25	0.784313725490196\\
0.3	0.75187969924812\\
0.35	0.722021660649819\\
0.4	0.694444444444444\\
0.45	0.668896321070234\\
0.5	0.645161290322581\\
0.55	0.623052959501558\\
0.6	0.602409638554217\\
0.65	0.583090379008746\\
0.7	0.564971751412429\\
0.75	0.547945205479452\\
0.8	0.531914893617021\\
0.85	0.516795865633075\\
0.9	0.50251256281407\\
0.95	0.488997555012225\\
1	0.476190476190476\\
};
\addlegendentry{MMSE-DFT Th. \eqref{exxx}}

\addplot [color=mycolor2, line width=1pt, mark size=2pt, mark=square, mark options={solid, mycolor2}]
  table[row sep=crcr]{%
0.05	17.4609755509772\\
0.1	9.05647708928203\\
0.15	6.04280324597233\\
0.2	4.55161117978598\\
0.25	3.59598414198592\\
0.3	3.05075519620549\\
0.35	2.64706986667683\\
0.4	2.27424126378682\\
0.45	2.02161245554017\\
0.5	1.80928650741041\\
0.55	1.66379359557815\\
0.6	1.50440205252026\\
0.65	1.42265819074863\\
0.7	1.2648005906703\\
0.75	1.21586558856748\\
0.8	1.11106928488205\\
0.85	1.09872993427142\\
0.9	1.02696079241276\\
0.95	0.974413650528252\\
1	0.912001966873308\\
};
\addlegendentry{LS-DFT}

\addplot [color=black, line width=1pt, mark size=1.75pt, mark=triangle, mark options={solid, black}]
  table[row sep=crcr]{%
0.05	17.1818181818182\\
0.1	9.09090909090909\\
0.15	6.06060606060606\\
0.2	4.54545454545454\\
0.25	3.63636363636364\\
0.3	3.03030303030303\\
0.35	2.5974025974026\\
0.4	2.27272727272727\\
0.45	2.02020202020202\\
0.5	1.81818181818182\\
0.55	1.65289256198347\\
0.6	1.51515151515152\\
0.65	1.3986013986014\\
0.7	1.2987012987013\\
0.75	1.21212121212121\\
0.8	1.13636363636364\\
0.85	1.06951871657754\\
0.9	1.01010101010101\\
0.95	0.956937799043062\\
1	0.909090909090909\\
};
\addlegendentry{LS-DFT Th. \eqref{exxpp3}}

\addplot [color=blue, dashed, line width=1pt, mark size=2.0pt, mark=o, mark options={solid, blue}]
  table[row sep=crcr]{%
0.05	0.921649862340428\\
0.1	0.835303663007024\\
0.15	0.749611445390845\\
0.2	0.711396669279406\\
0.25	0.628294204355048\\
0.3	0.593353270788445\\
0.35	0.55807181281386\\
0.4	0.534590864051794\\
0.45	0.494349450974949\\
0.5	0.47167274083977\\
0.55	0.449113115254343\\
0.6	0.420376671819517\\
0.65	0.405563504424339\\
0.7	0.398582290769839\\
0.75	0.373762824377813\\
0.8	0.365210865586349\\
0.85	0.356399114498526\\
0.9	0.332920766066118\\
0.95	0.326932034418347\\
1	0.315768437903972\\
};

\addplot [color=mycolor1, dashed, line width=1pt, mark size=1.750pt, mark=x, mark options={solid, mycolor1}]
  table[row sep=crcr]{%
0.05	0.900900900900901\\
0.1	0.819672131147541\\
0.15	0.75187969924812\\
0.2	0.694444444444444\\
0.25	0.645161290322581\\
0.3	0.602409638554217\\
0.35	0.564971751412429\\
0.4	0.531914893617021\\
0.45	0.50251256281407\\
0.5	0.476190476190476\\
0.55	0.452488687782806\\
0.6	0.431034482758621\\
0.65	0.411522633744856\\
0.7	0.393700787401575\\
0.75	0.377358490566038\\
0.8	0.36231884057971\\
0.85	0.348432055749129\\
0.9	0.335570469798658\\
0.95	0.323624595469256\\
1	0.3125\\
};

\addplot [color=mycolor2, dashed, line width=1pt, mark size=2pt, mark=square, mark options={solid, mycolor2}]
  table[row sep=crcr]{%
0.05	9.03499617382047\\
0.1	4.54962663713891\\
0.15	3.01026091950118\\
0.2	2.26469979260378\\
0.25	1.7950063069835\\
0.3	1.47019563729555\\
0.35	1.304439146875\\
0.4	1.14207927202027\\
0.45	1.00497245931903\\
0.5	0.917346476397931\\
0.55	0.820491303867893\\
0.6	0.730900022628351\\
0.65	0.704624328050318\\
0.7	0.66858281896759\\
0.75	0.606862461902868\\
0.8	0.564030111681728\\
0.85	0.540252718266027\\
0.9	0.498015386136616\\
0.95	0.480133510396511\\
1	0.456262962493899\\
};

\addplot [color=black, dashed, line width=1pt, mark size=1.75pt, mark=triangle, mark options={solid, black}]
  table[row sep=crcr]{%
0.05	9.09090909090909\\
0.1	4.54545454545454\\
0.15	3.03030303030303\\
0.2	2.27272727272727\\
0.25	1.81818181818182\\
0.3	1.51515151515152\\
0.35	1.2987012987013\\
0.4	1.13636363636364\\
0.45	1.01010101010101\\
0.5	0.909090909090909\\
0.55	0.826446280991735\\
0.6	0.757575757575758\\
0.65	0.699300699300699\\
0.7	0.649350649350649\\
0.75	0.606060606060606\\
0.8	0.568181818181818\\
0.85	0.53475935828877\\
0.9	0.505050505050505\\
0.95	0.478468899521531\\
1	0.454545454545455\\
};

\addplot [color=blue, dotted, line width=1pt, mark size=2.0pt, mark=o, mark options={solid, blue}]
  table[row sep=crcr]{%
0.05	0.815831607820158\\
0.1	0.674903546731055\\
0.15	0.585577177258822\\
0.2	0.533177576923502\\
0.25	0.481434248476187\\
0.3	0.431428138705617\\
0.35	0.395938275399229\\
0.4	0.363940193565824\\
0.45	0.330393977141959\\
0.5	0.306136642441036\\
0.55	0.292611040775027\\
0.6	0.27127832026041\\
0.65	0.256607433367131\\
0.7	0.245131395407776\\
0.75	0.228479436756117\\
0.8	0.218662021924785\\
0.85	0.215635363210237\\
0.9	0.205104905207342\\
0.95	0.190685319034595\\
1	0.183788462216525\\
};

\addplot [color=mycolor1, dotted, line width=1pt, mark size=1.750pt, mark=x, mark options={solid, mycolor1}]
  table[row sep=crcr]{%
0.05	0.819672131147541\\
0.1	0.694444444444444\\
0.15	0.602409638554217\\
0.2	0.531914893617021\\
0.25	0.476190476190476\\
0.3	0.431034482758621\\
0.35	0.393700787401575\\
0.4	0.36231884057971\\
0.45	0.335570469798658\\
0.5	0.3125\\
0.55	0.292397660818713\\
0.6	0.274725274725275\\
0.65	0.259067357512953\\
0.7	0.245098039215686\\
0.75	0.232558139534884\\
0.8	0.221238938053097\\
0.85	0.210970464135021\\
0.9	0.201612903225806\\
0.95	0.193050193050193\\
1	0.185185185185185\\
};

\addplot [color=mycolor2, dotted, line width=1pt, mark size=2pt, mark=square, mark options={solid, mycolor2}]
  table[row sep=crcr]{%
0.05	4.59820074080551\\
0.1	2.30270487094401\\
0.15	1.51924602593232\\
0.2	1.1413485438926\\
0.25	0.899554589829826\\
0.3	0.76014381482377\\
0.35	0.650994423867796\\
0.4	0.568721121144363\\
0.45	0.497089415409219\\
0.5	0.455796958826524\\
0.55	0.415022003869079\\
0.6	0.372662063279651\\
0.65	0.349153925032678\\
0.7	0.319608116672955\\
0.75	0.302113798514702\\
0.8	0.284534547643889\\
0.85	0.274746896313458\\
0.9	0.259783951034114\\
0.95	0.233401726869783\\
1	0.225018060457986\\
};

\addplot [color=black, dotted, line width=1pt, mark size=1.75pt, mark=triangle, mark options={solid, black}]
  table[row sep=crcr]{%
0.05	4.54545454545454\\
0.1	2.27272727272727\\
0.15	1.51515151515151\\
0.2	1.13636363636364\\
0.25	0.909090909090909\\
0.3	0.757575757575758\\
0.35	0.649350649350649\\
0.4	0.568181818181818\\
0.45	0.505050505050505\\
0.5	0.454545454545455\\
0.55	0.413223140495868\\
0.6	0.378787878787879\\
0.65	0.34965034965035\\
0.7	0.324675324675325\\
0.75	0.303030303030303\\
0.8	0.284090909090909\\
0.85	0.267379679144385\\
0.9	0.252525252525252\\
0.95	0.239234449760766\\
1	0.227272727272727\\
};
\node at (axis cs: 0,.29) [anchor = west] {Solid: $S=(N+1)$};
\node at (axis cs: 0,.2) [anchor = west] {Dashed: $S=2(N+1)$};
\node at (axis cs: 0,.14) [anchor = west] {Dotted: $S=4(N+1)$};

\end{axis}
\end{tikzpicture}%
\caption{NMSE in the estimation of $\mathbf{h}_{d,k}$ against $\beta_{d,k}$.}
\label{Fig1PL_est}
\end{subfigure}
\hspace{.05cm}
\begin{subfigure}[t]{.48\textwidth}
\tikzset{every picture/.style={scale=.95}, every node/.style={scale=.8}}
%
%
\definecolor{mycolor1}{rgb}{0.00000,0.49804,0.00000}%
\definecolor{mycolor2}{rgb}{1.00000,0.00000,1.00000}%
\begin{tikzpicture}

\begin{axis}[%
width=.95\columnwidth,
height=.85\columnwidth,
scale only axis,
xmin=0,
xmax=1,
xlabel style={font=\color{white!15!black}},
xlabel={$\beta_k$},
ymode=log,
ymin=.04,
ymax=18.2818181818182,
yminorticks=true,
ylabel style={font=\color{white!15!black}},
ylabel={NMSE},
axis background/.style={fill=white},
xmajorgrids,
ymajorgrids,
yminorgrids,
legend style={at={(1,1.03)}, anchor=north east, legend cell align=left,align=left,draw=white!15!black, /tikz/column 2/.style={
                column sep=5pt,
            }},]
\addplot [color=blue, line width=1pt, mark size=2.0pt, mark=o, mark options={solid, blue}]
  table[row sep=crcr]{%
0.05	0.828128352523601\\
0.1	0.691847087065525\\
0.15	0.595139843548118\\
0.2	0.538493919369746\\
0.25	0.481722451236932\\
0.3	0.432280575773519\\
0.35	0.390764130464029\\
0.4	0.365256291231727\\
0.45	0.335247224891363\\
0.5	0.315998791898895\\
0.55	0.295130864503557\\
0.6	0.273838984144325\\
0.65	0.262429298060077\\
0.7	0.244060113543765\\
0.75	0.229305945038211\\
0.8	0.214626534653436\\
0.85	0.212374878862724\\
0.9	0.204152731438915\\
0.95	0.192899302754973\\
1	0.186145866733837\\
};
\addlegendentry{MMSE-DFT}

\addplot [color=mycolor1, line width=1pt, mark size=1.75pt, mark=x, mark options={solid, mycolor1}]
  table[row sep=crcr]{%
0.05	0.819672131147526\\
0.1	0.694444444444433\\
0.15	0.602409638554209\\
0.2	0.531914893617015\\
0.25	0.476190476190481\\
0.3	0.431034482758623\\
0.35	0.393700787401569\\
0.4	0.362318840579713\\
0.45	0.335570469798654\\
0.5	0.3125\\
0.55	0.292397660818712\\
0.6	0.274725274725273\\
0.65	0.259067357512952\\
0.7	0.245098039215682\\
0.75	0.232558139534882\\
0.8	0.2212389380531\\
0.85	0.210970464135019\\
0.9	0.20161290322581\\
0.95	0.193050193050195\\
1	0.185185185185188\\
};
\addlegendentry{MMSE-DFT Th. \eqref{exx}}

\addplot [color=mycolor2, line width=1pt, mark size=2pt, mark=square, mark options={solid, mycolor2}]
  table[row sep=crcr]{%
0.05	17.1577413143366\\
0.1	9.13951970981278\\
0.15	6.031645242162\\
0.2	4.5332010531962\\
0.25	3.65113902499944\\
0.3	3.0366482539281\\
0.35	2.59710201563243\\
0.4	2.28099928381809\\
0.45	2.01393569404435\\
0.5	1.83105090715234\\
0.55	1.65792894619237\\
0.6	1.51414087948704\\
0.65	1.40932145053982\\
0.7	1.29326716470563\\
0.75	1.2106962057222\\
0.8	1.13293481717545\\
0.85	1.06776317896626\\
0.9	1.01373245344893\\
0.95	0.959828100899342\\
1	0.916031509003869\\
};
\addlegendentry{LS-DFT}

\addplot [color=black, line width=1pt, mark size=1.75pt, mark=triangle, mark options={solid, black}]
  table[row sep=crcr]{%
0.05	17.1818181818182\\
0.1	9.09090909090909\\
0.15	6.06060606060606\\
0.2	4.54545454545454\\
0.25	3.63636363636364\\
0.3	3.03030303030303\\
0.35	2.5974025974026\\
0.4	2.27272727272727\\
0.45	2.02020202020202\\
0.5	1.81818181818182\\
0.55	1.65289256198347\\
0.6	1.51515151515152\\
0.65	1.3986013986014\\
0.7	1.2987012987013\\
0.75	1.21212121212121\\
0.8	1.13636363636364\\
0.85	1.06951871657754\\
0.9	1.01010101010101\\
0.95	0.956937799043062\\
1	0.909090909090909\\
};
\addlegendentry{LS-DFT Th. \eqref{expp2}}

\addplot [color=blue, dashed, line width=1pt, mark size=2.0pt, mark=o, mark options={solid, blue}]
  table[row sep=crcr]{%
0.05	0.709453878255944\\
0.1	0.53588258843023\\
0.15	0.431613845061835\\
0.2	0.363482620147643\\
0.25	0.315837591605738\\
0.3	0.273649495052112\\
0.35	0.246731775353246\\
0.4	0.219923666533621\\
0.45	0.204053997877462\\
0.5	0.185540808972804\\
0.55	0.171934364275452\\
0.6	0.159677518813477\\
0.65	0.149245867091414\\
0.7	0.142825285995439\\
0.75	0.131055816507884\\
0.8	0.125404585362494\\
0.85	0.119508295197118\\
0.9	0.11031101299554\\
0.95	0.105215809512636\\
1	0.100427977593714\\
};

\addplot [color=mycolor1, dashed, line width=1pt, mark size=1.75pt, mark=x, mark options={solid, mycolor1}]
  table[row sep=crcr]{%
0.05	0.694444444444433\\
0.1	0.531914893617015\\
0.15	0.431034482758623\\
0.2	0.362318840579713\\
0.25	0.3125\\
0.3	0.274725274725273\\
0.35	0.245098039215682\\
0.4	0.2212389380531\\
0.45	0.20161290322581\\
0.5	0.185185185185188\\
0.55	0.171232876712326\\
0.6	0.159235668789811\\
0.65	0.148809523809523\\
0.7	0.139664804469271\\
0.75	0.13157894736842\\
0.8	0.124378109452738\\
0.85	0.117924528301886\\
0.9	0.112107623318387\\
0.95	0.106837606837608\\
1	0.102040816326532\\
};

\addplot [color=mycolor2, dashed, line width=1pt, mark size=2pt, mark=square, mark options={solid, mycolor2}]
  table[row sep=crcr]{%
0.05	9.05294667086495\\
0.1	4.55515532175535\\
0.15	3.01954715627012\\
0.2	2.28442675007533\\
0.25	1.82282299454997\\
0.3	1.5076079345703\\
0.35	1.30186875949905\\
0.4	1.1347692449895\\
0.45	1.01469480678069\\
0.5	0.908008371520714\\
0.55	0.826837675519386\\
0.6	0.755663296416579\\
0.65	0.71046570257978\\
0.7	0.652002944746512\\
0.75	0.604752278309919\\
0.8	0.573997769430925\\
0.85	0.534618980764829\\
0.9	0.505157832454156\\
0.95	0.479980755584126\\
1	0.45458593041697\\
};

\addplot [color=black, dashed, line width=1pt, mark size=1.75pt, mark=triangle, mark options={solid, black}]
  table[row sep=crcr]{%
0.05	9.09090909090909\\
0.1	4.54545454545454\\
0.15	3.03030303030303\\
0.2	2.27272727272727\\
0.25	1.81818181818182\\
0.3	1.51515151515152\\
0.35	1.2987012987013\\
0.4	1.13636363636364\\
0.45	1.01010101010101\\
0.5	0.909090909090909\\
0.55	0.826446280991735\\
0.6	0.757575757575758\\
0.65	0.699300699300699\\
0.7	0.649350649350649\\
0.75	0.606060606060606\\
0.8	0.568181818181818\\
0.85	0.53475935828877\\
0.9	0.505050505050505\\
0.95	0.478468899521531\\
1	0.454545454545455\\
};

\addplot [color=blue, dotted, line width=1pt, mark size=2.0pt, mark=o, mark options={solid, blue}]
  table[row sep=crcr]{%
0.05	0.535769489161926\\
0.1	0.370129866626219\\
0.15	0.274709417544751\\
0.2	0.220998944517516\\
0.25	0.183239361602141\\
0.3	0.160518033485198\\
0.35	0.140604167785555\\
0.4	0.125459440229628\\
0.45	0.112395628188499\\
0.5	0.101014841328353\\
0.55	0.0939473006261732\\
0.6	0.0857935331770881\\
0.65	0.0808706166098883\\
0.7	0.075682198874298\\
0.75	0.0709787173559616\\
0.8	0.0660101264851536\\
0.85	0.0612436796322715\\
0.9	0.0587548997478472\\
0.95	0.0560198273888906\\
1	0.0544674339676842\\
};

\addplot [color=mycolor1, dotted, line width=1pt, mark size=1.75pt, mark=x, mark options={solid, mycolor1}]
  table[row sep=crcr]{%
0.05	0.531914893617015\\
0.1	0.362318840579713\\
0.15	0.274725274725273\\
0.2	0.2212389380531\\
0.25	0.185185185185188\\
0.3	0.159235668789811\\
0.35	0.139664804469272\\
0.4	0.124378109452738\\
0.45	0.112107623318387\\
0.5	0.102040816326532\\
0.55	0.0936329588014966\\
0.6	0.0865051903114174\\
0.65	0.0803858520900313\\
0.7	0.0750750750750763\\
0.75	0.0704225352112667\\
0.8	0.0663129973474794\\
0.85	0.0626566416040095\\
0.9	0.0593824228028511\\
0.95	0.0564334085778774\\
1	0.0537634408602156\\
};

\addplot [color=mycolor2, dotted, line width=1pt, mark size=2pt, mark=square, mark options={solid, mycolor2}]
  table[row sep=crcr]{%
0.05	4.54805472120662\\
0.1	2.29141972487246\\
0.15	1.5086484542597\\
0.2	1.1235313266586\\
0.25	0.909735367611485\\
0.3	0.758271000197827\\
0.35	0.648970977709537\\
0.4	0.569184448481948\\
0.45	0.50299035114569\\
0.5	0.452146606829183\\
0.55	0.413682975895276\\
0.6	0.375404626831649\\
0.65	0.350772935314085\\
0.7	0.322688591775864\\
0.75	0.304586141171428\\
0.8	0.282808547279896\\
0.85	0.264853619468421\\
0.9	0.25197326453416\\
0.95	0.239488126310023\\
1	0.228467989533383\\
};

\addplot [color=black, dotted, line width=1pt, mark size=1.75pt, mark=triangle, mark options={solid, black}]
  table[row sep=crcr]{%
0.05	4.54545454545454\\
0.1	2.27272727272727\\
0.15	1.51515151515151\\
0.2	1.13636363636364\\
0.25	0.909090909090909\\
0.3	0.757575757575758\\
0.35	0.649350649350649\\
0.4	0.568181818181818\\
0.45	0.505050505050505\\
0.5	0.454545454545455\\
0.55	0.413223140495868\\
0.6	0.378787878787879\\
0.65	0.34965034965035\\
0.7	0.324675324675325\\
0.75	0.303030303030303\\
0.8	0.284090909090909\\
0.85	0.267379679144385\\
0.9	0.252525252525252\\
0.95	0.239234449760766\\
1	0.227272727272727\\
};
\node at (axis cs: 0,.1) [anchor = west] {Solid: $S=(N+1)$};
\node at (axis cs: 0,.066) [anchor = west] {Dashed: $S=2(N+1)$};
\node at (axis cs: 0,.045) [anchor = west] {Dotted: $S=4(N+1)$};
\end{axis}
\end{tikzpicture}%
\caption{NMSE in the estimation of $\mathbf{h}_{0,n,k}$ against $\beta_k$.}
\label{Fig2PL_est}
\end{subfigure}
\caption{NMSE comparison between MMSE-DFT and LS-DFT estimates against path loss.}
\label{FigPL_est}
\end{figure*}
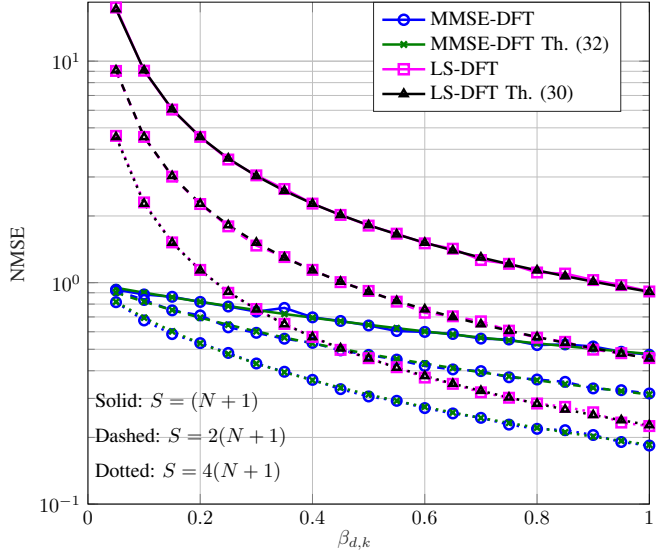
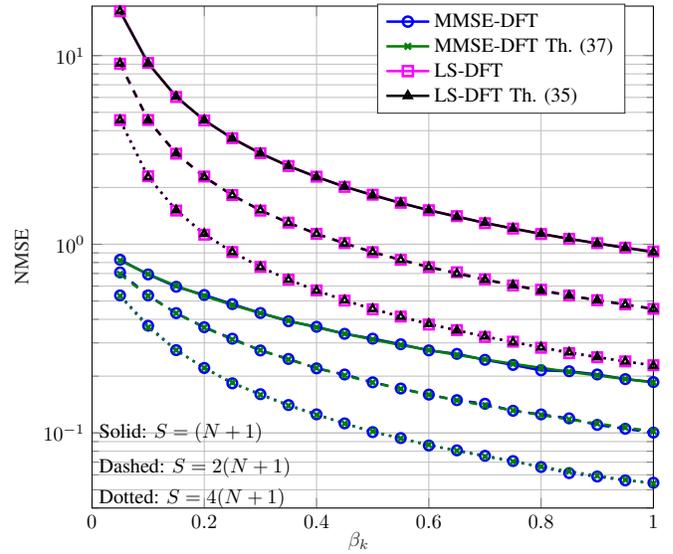

In Fig. \ref{Fig1PL_est}, we compare the NMSE in the estimation of $\mathbf{h}_{d,k}$ against $\beta_{d,k}$ and in Fig. \ref{Fig2PL_est}, we compare the NMSE in the estimation of $\mathbf{h}_{0,n,k}$ (we plot $\frac{1}{N}\sum_{n=1}^N \text{NMSE}(\hat{\mathbf{h}}_{0,n,k})$) against $\beta_k$ under both MMSE-DFT and LS-DFT protocols. The value for $\sigma^2$ is set as $5 \times 10^{-4}$J in these results. The match between the theoretical expressions of the NMSE derived in this section and the simulated values is perfect. The NMSE in MMSE-DFT estimates is always lower than that in LS-DFT estimates. We also show the effect of increasing the number of sub-phases $S$ beyond $N+1$. As evident in \eqref{exxpp3} and \eqref{expp2} there is a factor of $S$ decrease in the NMSE in LS estimates over the entire range of $\beta_{d,k}$ and $\beta_k$. The NMSE in MMSE estimates decreases by a factor of less than $S$ in the low path loss (high SNR) regime while it approaches $1$ in the high path loss regime irrespective of the value of $S$. However, the MMSE-DFT estimates are seen to outperform LS-DFT estimates for the considered values of $S$, with the performance gap becoming small as $S$ increases. It is important to note that although we see a significant NMSE improvement  by increasing the number of sub-phases $S$, there will also be a  rate loss due to channel training as $S$ increases. This is because the time left for downlink transmission reduces with $S$ under the relation $\tau_D=\tau-S\tau_S$. Therefore, the system will suffer a rate loss factor of $1-\frac{S \tau_S}{\tau}$ during downlink transmission, rendering the IRS-assisted system performance sensitive to the value of $S$ and the quality of estimates.  This trade-off will be studied in the simulation results in Sec. V.

To gain further insights into how these NMSE values are related to the system performance, we numerically study the bit error rate (BER) achieved by an IRS-assisted system with $M=4$ antennas and $N=10$ reflecting elements serving a single-antenna user. For a single-user, it is well-known that the optimal precoding strategy at the BS is maximum ratio transmission (MRT), i.e. the precoding vector is set as $\mathbf{g}_k=\frac{\hat{\mathbf{h}}_k}{||\hat{\mathbf{h}}_k||}$, where $\hat{\mathbf{h}}_k=\hat{\mathbf{h}}_{d,k}+\hat{\mathbf{H}}_{0,k}\mathbf{v}$. The estimates $\hat{\mathbf{h}}_{d,k}$ and $\hat{\mathbf{h}}_{0,n,k}$, $n=1,\dots, N$, are given by \eqref{h_d_est} and \eqref{h_irs_est} respectively under MMSE-DFT protocol, while under the LS-DFT protocol, they are given by \eqref{h_d_ls} and \eqref{h_irs_ls} respectively. A close to optimal design for $\mathbf{v}$ that maximizes the received signal power at the user is proposed in \cite{LS} as $\mathbf{v}=\exp(j \angle(\hat{\mathbf{H}}_{0,k}^H\hat{\mathbf{h}}_{d,k}))$. 

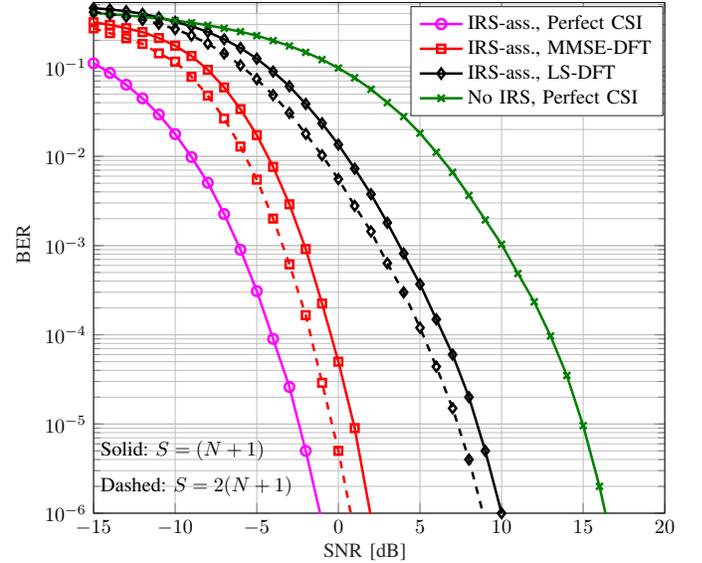
\begin{figure}[t]
\tikzset{every picture/.style={scale=.95}, every node/.style={scale=.8}}
%
%
\definecolor{mycolor1}{rgb}{0.00000,0.49804,0.00000}%
\definecolor{mycolor2}{rgb}{1.00000,0.00000,1.00000}%
\begin{tikzpicture}

\begin{axis}[%
width=.95\columnwidth,
height=.85\columnwidth,
scale only axis,
xmin=-15,
xmax=20,
xlabel style={at={(0.5,-0.04)},font=\color{white!15!black}},
xlabel={SNR [dB]},
ymode=log,
ymin=1e-06,
ymax=0.53,
yminorticks=true,
ylabel style={font=\color{white!15!black}},
ylabel={BER},
axis background/.style={fill=white},
xmajorgrids,
ymajorgrids,
yminorgrids,
legend style={at={(1.05,.99)}, anchor=north east, legend cell align=left,align=left,draw=white!15!black, /tikz/column 2/.style={
                column sep=5pt,
            }},
						]
\addplot [color=mycolor2, line width=1pt,mark size=2pt, mark=o, mark options={solid, mycolor2}]
  table[row sep=crcr]{%
-15	0.110525\\
-14	0.086145\\
-13	0.063473\\
-12	0.04467\\
-11	0.029486\\
-10	0.017767\\
-9	0.009823\\
-8	0.005061\\
-7	0.002259\\
-6	0.000898\\
-5	0.000309\\
-4	9e-05\\
-3	2.6e-05\\
-2	5e-06\\
-1	8e-07\\
0	0\\
1	0\\
2	0\\
3	0\\
4	0\\
5	0\\
6	0\\
7	0\\
8	0\\
9	0\\
10	0\\
11	0\\
12	0\\
13	0\\
14	0\\
15	0\\
16	0\\
17	0\\
18	0\\
19	0\\
20	0\\
};
\addlegendentry{IRS-ass., Perfect CSI}

\addplot [color=red, line width=1pt,mark size=1.6pt, mark=square, mark options={solid, red}]
  table[row sep=crcr]{%
-15	0.318622\\
-14	0.298333\\
-13	0.273561\\
-12	0.24704755\\
-11	0.2121\\
-10	0.174053\\
-9	0.133313\\
-8	0.09335\\
-7	0.059234\\
-6	0.033899\\
-5	0.017307\\
-4	0.007631\\
-3	0.002901\\
-2	0.000915\\
-1	0.000224\\
0	5e-05\\
1	9e-06\\
2	9e-07\\
3	0\\
4	0\\
5	0\\
6	0\\
7	0\\
8	0\\
9	0\\
10	0\\
11	0\\
12	0\\
13	0\\
14	0\\
15	0\\
16	0\\
17	0\\
18	0\\
19	0\\
20	0\\
};
\addlegendentry{IRS-ass., MMSE-DFT}

\addplot [color=black, line width=1pt,mark size=2pt, mark=diamond, mark options={solid, black}]
  table[row sep=crcr]{%
-15	0.458888\\
-14	0.442962\\
-13	0.424033\\
-12	0.3958336\\
-11	0.3615729\\
-10	0.322653\\
-9	0.287669\\
-8	0.248154\\
-7	0.20559\\
-6	0.164148\\
-5	0.124472\\
-4	0.089174\\
-3	0.060996\\
-2	0.038698\\
-1	0.023498\\
0	0.013549\\
1	0.007284\\
2	0.003767\\
3	0.001804\\
4	0.000814\\
5	0.000369\\
6	0.000149\\
7	6e-05\\
8	2e-05\\
9	5e-06\\
10	1e-06\\
11	0\\
12	0\\
13	0\\
14	0\\
15	0\\
16	0\\
17	0\\
18	0\\
19	0\\
20	0\\
};
\addlegendentry{IRS-ass., LS-DFT}

\addplot [color=mycolor1, line width=1pt,mark size=2pt, mark=x, mark options={solid, mycolor1}]
  table[row sep=crcr]{%
-15	0.405116\\
-14	0.392674\\
-13	0.379832\\
-12	0.366103\\
-11	0.351138\\
-10	0.333045\\
-9	0.315203\\
-8	0.294719\\
-7	0.272385\\
-6	0.249805\\
-5	0.225711\\
-4	0.198414\\
-3	0.173254\\
-2	0.146653\\
-1	0.121575\\
0	0.097734\\
1	0.075898\\
2	0.056549\\
3	0.04004\\
4	0.027911\\
5	0.018186\\
6	0.011123\\
7	0.00662\\
8	0.003647\\
9	0.001937\\
10	0.001032\\
11	0.000487\\
12	0.000234\\
13	9.7e-5\\
14	3.5e-05\\
15	9.6e-06\\
16	2e-06\\
17	3e-07\\
18	0\\
19	0\\
20	0\\
};
\addlegendentry{No IRS, Perfect CSI}


\addplot [color=red, dashed, line width=1pt,mark size=1.6pt, mark=square, mark options={solid, red}]
  table[row sep=crcr]{%
-15	0.272358\\
-14	0.240463\\
-13	0.2105084\\
-12	0.181965\\
-11	0.142841\\
-10	0.115334\\
-9	0.077997\\
-8	0.047946\\
-7	0.026536\\
-6	0.012889\\
-5	0.00549\\
-4	0.002009\\
-3	0.000616\\
-2	0.000166\\
-1	2.9e-05\\
0	5e-06\\
1	6e-07\\
2	0\\
3	0\\
4	0\\
5	0\\
6	0\\
7	0\\
8	0\\
9	0\\
10	0\\
11	0\\
12	0\\
13	0\\
14	0\\
15	0\\
16	0\\
17	0\\
18	0\\
19	0\\
20	0\\
};

\addplot [color=black, dashed, line width=1pt,mark size=2pt, mark=diamond, mark options={solid, black}]
  table[row sep=crcr]{%
-15	0.414113\\
-14	0.393058\\
-13	0.368373\\
-12	0.338711\\
-11	0.306054\\
-10	0.267112\\
-9	0.225985\\
-8	0.18348\\
-7	0.142217\\
-6	0.104824\\
-5	0.073259\\
-4	0.04856\\
-3	0.030678\\
-2	0.017871\\
-1	0.010276\\
0	0.005565\\
1	0.002789\\
2	0.00144\\
3	0.000633\\
4	0.000299\\
5	0.00012\\
6	4.4e-05\\
7	1.5e-05\\
8	4e-06\\
9	8e-07\\
10	0\\
11	0\\
12	0\\
13	0\\
14	0\\
15	0\\
16	0\\
17	0\\
18	0\\
19	0\\
20	0\\
};

\node at (axis cs: -15,5e-6) [anchor = west] {Solid: $S=(N+1)$};
\node at (axis cs: -15,2e-6) [anchor = west] {Dashed: $S=2(N+1)$};

\end{axis}
\end{tikzpicture}%
\caption{BER performance of an IRS-assisted (IRS-ass.) MISO system under the proposed channel estimation protocol.}
\label{BER}
\end{figure}

Under these designs for precoding at the BS and reflect beamforming at the IRS, we plot  in Fig. \ref{BER} the BER achieved by the IRS-assisted system under binary phase-shift keying (BPSK) signaling, against  the SNR defined as the ratio of the transmit power to the noise variance. The BER curves under perfect CSI and imperfect CSI with MMSE-DFT estimation as well as LS-DFT estimation are shown. We also plot the BER achieved by a conventional MISO system having $4$ antennas at the BS and no IRS. As expected, the BER decreases with increasing SNR while it approaches the maximum value of $0.5$ for very low values of SNR. We observe that the IRS-assisted system achieves a significantly better BER performance as compared to the conventional system without IRS, with the BER for the former decreasing to $10^{-6}$ at SNR level of near $0$\rm{dB}, similar to the observation made in \cite{basar}. In fact, the SNR gap between the IRS-assisted system and the conventional system  to achieve  the BER rate of $10^{-6}$ is around $17$\rm{dB}, which shows that the IRS is capable of improving the reliability of the underlying communication channel by manipulating the propagation of radio waves in the environment. This superior BER performance is explained in \cite{basar} using the analytical result that the received signal power at the user scales quadratically as $N^2$ with the number of IRS elements $N$, whereas in the conventional MISO system it scales linearly with the number of BS antennas $M$. As a result the IRS provides approximately a factor of $N^2$ improvement in the received signal power\footnote{This improvement is compromised to some extent by the double path loss effect in the IRS-assisted link, which suffers from the product of the path loss in BS-to-IRS and IRS-to-user links. In generating this simulation result, we set the path loss factor for each link as $0.25$ and still observe the positive effect of $N^2$ gain to dominate over the negative effect of double path loss in the IRS-assisted link resulting in significant BER improvement.}, because of which even when the SNR is relatively low, the BER achieved by the IRS-assisted system is quite low.

Under channel estimation errors in an IRS-assisted system, the BER performance of the MMSE-DFT protocol is significantly better than the LS-DFT protocol, with an SNR gap of almost $8$\rm{dB} to achieve the BER of $10^{-6}$. This is in accordance with the insights drawn earlier from the NMSE analysis where we showed the MMSE-DFT estimates to always achieve a lower NMSE. Further, we note that the BER under LS-DFT protocol approaches the maximum value at SNR level of $-15$\rm{dB} whereas under MMSE-DFT protocol, it will reach the maximum BER slower (in fact it does not reach the maximum value for the SNR range considered in the figure). This can also be confirmed from Fig. \ref{Fig1_est} and \ref{Fig2_est}, where we see that the NMSE values in MMSE-DFT estimates approach $1$ much slower (at higher values of noise) than the LS-DFT estimates.  Finally, we see that the BER decreases with increasing number of sub-phases $S$ for both protocols. This is due to the decrease in NMSE with increasing $S$ as observed earlier in Fig. \ref{Fig1PL_est}  and Fig. \ref{Fig2PL_est}.

It is important to remark here that both ON/OFF and DFT protocols require long channel training times when $N$ is very large since the number of sub-phases $S$ has to be greater than $N+1$. As an extension, the scenario where IRS elements that experience strong correlation and therefore similar channels are grouped together can be studied. The number of sub-phases needed can then be reduced to the number of groups instead of the number of IRS elements. However this will also reduce the degrees of freedom offered by the IRS for performance improvement since elements in the same group will apply the same reflection coefficient. We  stress that MMSE estimates yield convenient analytical expressions unlike the algorithms in \cite{CE_MU,   cas} and can be extended under future channel estimation protocols that reduce training overhead. 

\section{Joint Active and Passive Beamforming Design}

In this section, we design the precoding vectors and power allocation at the BS and the phase shifts vector at the IRS. The amplitude reflection coefficients $\alpha_n$, $\forall n$ are assumed to be unity as done in almost all existing works, motivated by the recent advances in the design and development of lossless metasurfaces \cite{MS_loss, MS_loss2}. Similar to channel estimation, we assume that all the design computations take place at the BS since the IRS has no signal processing capability. The BS then informs the IRS controller about the required optimal reflect beamforming vector $\mathbf{v}^*$ through a backhaul link, and the controller triggers the elements of the IRS to apply the required phase-shifts.

The performance metric employed is the max-min  rate, which provides a good balance between system throughput and user fairness. The rate of user $k$ is defined as $R_k=\log_2(1+\gamma_k)$, where $\gamma_k$ is the SINR of user $k$ given as
\begin{align}
\label{SINR}
& \gamma_{k}=\frac{\frac{p_{k}}{K} |\textbf{h}_{k}^{H} \textbf{g}_{k}|^{2}}{\sum_{i\neq k} \frac{p_{i}}{K} |\textbf{h}_{k}^{H} \textbf{g}_{i}|^{2}+ \sigma_n^2}, 
\end{align}
where $\textbf{h}_{k}= \mathbf{h}_{d,k}+\mathbf{H}_{0,k}\mathbf{v}$ is the overall channel from BS to user $k$ as defined in \eqref{ch1}. Since logarithm is a monotonically increasing function so max-min rate problem is equivalent to solving the max-min SINR problem.

\subsection{Problem Formulation}

The BS utilizes the information it has on the direct and the IRS-assisted channels to find the optimal precoding vectors $\mathbf{G}^*=[\mathbf{g}_1, \dots, \mathbf{g}_K]$,   allocated powers $\textbf{p}^*=[p_{1},\dots, p_{K}]^T$, and the IRS reflect beamforming vector $\mathbf{v}^*$ as the solution of the following max-min SINR problem. 
\begin{subequations}
 \begin{alignat}{2} \textit{(P1)} \hspace{.35in}
&\!\max_{\mathbf{p}, \mathbf{G}, \mathbf{v}} \hspace{.1in} \!\min_{k}        &\qquad& \gamma_{k} \label{P1}\\
&\text{subject to} &      & \frac{1}{K}\mathbf{1}_{K}^{T}\mathbf{p} \leq P_{max}, \label{constraint0} \\
&& & ||\mathbf{g}_{k}||=1, \forall k, \label{constraint1}\\
&&      & |v_{n}|=1, \hspace{.08in} n=1,\dots, N,\label{constraint3}
\end{alignat}
\end{subequations}
where $v_n=\exp(j\theta_n)$ is the $n^{th}$ element of $\mathbf{v}$. Note that the constraints in \eqref{constraint0} and (\ref{constraint1})  meet the constraint in (\ref{p_cons}).   We would like to highlight that with the exception of \cite{LIS_jour}, the max-min SINR problem has not been dealt with in the context of IRS-assisted systems. In contrast to \cite{LIS_jour} which focuses on the problem formulation and solution under perfect CSI in the asymptotic regime where $M$, $N$ and $K$ grow infinitely large, we focus on the exact problem in \textit{(P1)} and deal with both perfect and imperfect CSI. 

Due to the non-convex nature of the problem in which the precoding vectors, allocated powers and phase shifts are coupled, we will adopt an AO technique, where the precoding vectors and power allocation at the BS are optimized iteratively with the phase shifts at the IRS, until convergence is achieved. For fixed $\mathbf{v}$, we have the following sub-problem
\begin{subequations}
 \begin{alignat}{2} \textit{(P2)} \hspace{.35in}
&\!\max_{\mathbf{p}, \mathbf{G}} \hspace{.1in} \!\min_{k}        &\qquad& \gamma_{k} \label{P2}\\
&\text{subject to} &      & \frac{1}{K}\mathbf{1}_{K}^{T}\mathbf{p} \leq P_{max}, \\
&& &  ||\mathbf{g}_{k}||=1, \forall k.
\end{alignat}
\end{subequations}
It was shown in \cite{LIS_jour} that the optimal linear precoder (OLP) that solves (\textit{P2}) optimally with respect to $\mathbf{G}$ and $\mathbf{p}$ takes the form
\begin{align}
\label{G_opt}
\textbf{g}_{k}^*=\frac{\left(\sum_{i\neq k} \frac{q_{i}^*}{K}\textbf{h}_{i}\textbf{h}_{i}^{H}+\sigma_n^2\textbf{I}_{M}\right)^{-1}\textbf{h}_{k}}{||\left(\sum_{i\neq k} \frac{q_{i}^*}{K}\textbf{h}_{i}\textbf{h}_{i}^{H}+\sigma_n^2\textbf{I}_{M}\right)^{-1}\textbf{h}_{k}||},
\end{align}
where $q_{k}^*$s are obtained as the unique positive solution of the following fixed-point equations
\begin{align}
\label{q}
q_{k}^*=\frac{\tau^*}{\frac{1}{K}\textbf{h}_{k}^H\left(\sum_{i\neq k} \frac{q_{i}^*}{K}\textbf{h}_{i}\textbf{h}_{i}^{H}+\sigma_n^2\textbf{I}_{M}\right)^{-1}\textbf{h}_{k}},
\end{align}
with $\tau^*=\frac{KP_{max}}{\sum_{k=1}^{K}\left(\frac{1}{K}\textbf{h}_{k}^H\left(\sum_{i\neq k} \frac{q_{i}^*}{K}\textbf{h}_{i}\textbf{h}_{i}^{H}+\sigma_n^2 \textbf{I}_{M}\right)^{-1}\textbf{h}_{k}\right)^{-1}}$. The optimal powers $p_{k}^*$s are obtained as
\begin{align}
\label{P_opt}
&\textbf{p}^*=\left(\textbf{I}_{K}-\tau^*\textbf{D}\textbf{F}  \right)^{-1}\tau^*\sigma_n^2\textbf{D}\textbf{1}_{K},
\end{align}
where $\textbf{D}=\text{diag}\left(\frac{1}{\frac{1}{K}|\textbf{h}_{1}^{H}\textbf{g}^*_{1}|^{2}}, \dots, \frac{1}{\frac{1}{K}|\textbf{h}_{K}^{H}\textbf{g}^*_{K}|^{2}} \right)$ and $[\textbf{F}]_{k,i}=\frac{1}{K}|\textbf{h}_{k}^{H}\textbf{g}^*_{i}|^{2}$, if $k\neq i$ and $0$ otherwise.

On the other hand, for fixed ${\mathbf{g}}_k$s and ${p}_k$s, \textit{(P1)} is reduced to
\begin{subequations}
 \begin{alignat}{2} \textit{(P3)} \hspace{.35in}
&\!\max_{ \mathbf{v}} \hspace{.1in} \!\min_{k}        &\qquad& \gamma_{k} \label{P3}\\
&\text{subject to} &      &|v_{n}|=1, \hspace{.08in} n=1,\dots, N. \label{C3}
\end{alignat}
\end{subequations}
We will propose a solution for \textit{(P3)} in the next subsection. The proposed AO algorithm will then solve problem \textit{(P1)} by solving problems \textit{(P2)} and \textit{(P3)} alternatively. The extension to imperfect CSI is summarized in Section IV-C. The AO technique has been utilized in \cite{LIS} to solve the transmit power minimization problem and in \cite{8741198} for energy efficiency maximization problem. However, the sub-problems constituting the AO algorithm in this work are different.

\subsection{Problem Solution}

The optimal solution for the precoding vectors and allocated powers in \textit{(P2)} are already provided in \eqref{G_opt} and \eqref{P_opt} respectively.  Here, we develop a solution for the design of reflect beamforming vector in \textit{(P3)}, which is a non-convex problem. However, we observe that the numerator and denominator of $\gamma_k$ in \eqref{SINR} which is the objective function in \eqref{P3} can be transformed into quadratic forms. To see this note that the terms $|\mathbf{h}_k^H \mathbf{g}_i|^2$ in \eqref{SINR} can be written as
\begin{align}
&|\mathbf{h}_k^H \mathbf{g}_i|^2=\mathbf{v}^H \mathbf{a}_{k,i}\mathbf{a}_{k,i}^H \mathbf{v}  +\mathbf{v}^H \mathbf{a}_{k,i} b_{k,i}^* + b_{k,i}\mathbf{a}_{k,i}^H \mathbf{v}+ b_{k,i}b_{k,i}^*,
\end{align}
where $\mathbf{a}_{k,i}=\mathbf{H}_{0,k}^H \mathbf{g}_{i}$ and $b_{k,i}=\mathbf{h}_{d,k}^H \mathbf{g}_i$. By introducing an auxiliary variable $t$, \textit{(P3)} can be reformulated in terms of quadratic forms as
\begin{subequations}
 \begin{alignat}{2} \textit{(P4)} \hspace{.35in}
&\!\max_{ \bar{\mathbf{v}}} \hspace{.1in} \!\min_{k}        &\qquad& \frac{\frac{p_k}{K} (\bar{\mathbf{v}}^H \mathbf{R}_{k,k} \bar{\mathbf{v}} + |b_{k,k}|^2)}{\sum_{i\neq k}^{K} \frac{p_i}{K} (\bar{\mathbf{v}}^H \mathbf{R}_{k,i} \bar{\mathbf{v}} + |b_{k,i}|^2)+\sigma_n^2} \label{P5}\\
&\text{subject to} &      &|\bar{v}_{n}|^2=1, \hspace{.08in} n=1,\dots, N+1,
\end{alignat}
\end{subequations}
where $\mathbf{R}_{k,i}=   \begin{bmatrix}
    \mathbf{a}_{k,i}\mathbf{a}_{k,i}^H & \mathbf{a}_{k,i} b_{k,i}^*  \\
    b_{k,i} \mathbf{a}_{k,i}^H & 0 
  \end{bmatrix} $ and $\bar{\mathbf{v}}=\begin{bmatrix} \mathbf{v} \\ t \end{bmatrix}$. 
	
	However the problem \textit{(P4)} is NP-hard in general \cite{semidefiniterelax}. Note that $\bar{\mathbf{v}}^H \mathbf{R}_{k,i} \bar{\mathbf{v}}=\text{tr}(\mathbf{R}_{k,i} \bar{\mathbf{v}}\bar{\mathbf{v}}^H)$. Therefore, we can reformulate \textit{(P4)} by defining $\bar{\mathbf{V}}=\bar{\mathbf{v}}\bar{\mathbf{v}}^H$, which needs to satisfy $\bar{\mathbf{V}} \succeq 0$ and $\text{rank}(\bar{\mathbf{V}}) = 1$. Since the rank-one constraint is non-convex, we apply semi-definite relaxation to relax this constraint by letting $\bar{\mathbf{V}}$ be a positive semi-definite matrix of arbitrary rank. The semi-definite relaxed problem is given as
\begin{subequations}
 \begin{alignat}{2} \textit{(P5)} \hspace{.35in}
&\!\max_{ \bar{\mathbf{V}}} \hspace{.1in} \!\min_{k}        &\qquad& \frac{\frac{p_k}{K} (\text{tr}(\mathbf{R}_{k,k} \bar{\mathbf{V}}) + |b_{k,k}|^2)}{\sum_{i\neq k}^{K} \frac{p_i}{K} (\text{tr} \mathbf{R}_{k,i} \bar{\mathbf{V}} + |b_{k,i}|^2)+\sigma_n^2} \label{P6}\\
&\text{subject to} &      &\bar{\mathbf{V}} \succeq 0, \\
&& & \bar{\mathbf{V}}_{n,n}=1, \hspace{.08in} n=1,\dots, N+1. \label{C5}
\end{alignat}
\end{subequations}
Problem \textit{(P5)} is efficiently solved using fractional programming, which provides tools to maximize the minimum of ratios in which the numerator is a concave function, the denominator is a convex function, and the constraint set is convex \cite{fractional, dinkelbach} . An efficient method to do so is the generalized Dinkelbach's algorithm, outlined in Appendix A of \cite{dinkelbach}, which is guaranteed to converge to the global solution of the max-min fractional problem with limited complexity. The objective function in (\ref{P6}) considers a set of ratios of two functions, where we denote the numerator by $n_{k}(\bar{\mathbf{V}})$ and the denominator by $d_{k}(\bar{\mathbf{V}})$, $k=1, \dots, K$. By exploiting the fact that $\text{tr}(\textbf{A} \textbf{B}) = vec(\textbf{A}^{T})^{T} vec(\textbf{B})$, we write   $n_{k}(\bar{\mathbf{V}})$ and $d_{k}(\bar{\mathbf{V}})$ as
\begin{align}
\label{f}
&n_{k}(\bar{\mathbf{V}})=\frac{p_k}{K} (vec(\mathbf{R}_{k,k}^T)^T vec(\bar{\mathbf{V}})+ |b_{k,k}|^2), \\
\label{g}
&d_{k}(\bar{\mathbf{V}})=\sum_{i\neq k}^{K} \frac{p_i}{K} (vec(\mathbf{R}_{k,i}^T)^T vec(\bar{\mathbf{V}}) + |b_{k,i}|^2)+\sigma_n^2.
\end{align}

It can be seen from (\ref{f}) and \eqref{g} that $n_{k}(\bar{\mathbf{V}})$ and $d_{k}(\bar{\mathbf{V}})$ are  linear functions of $\bar{\mathbf{V}}$. Problem  \textit{(P5)} therefore considers a set of ratios $\{\frac{n_{k}(\bar{\mathbf{V}} )} {d_{k}(\bar{\mathbf{V}} )}\}_{k=1}^{K}$, where each ratio has an affine numerator $n_{k}(\bar{\mathbf{V}})$, affine denominator $d_{k}(\bar{\mathbf{V}})$ and convex constraints and can therefore be solved optimally using the generalized Dinkelbach's algorithm \cite{dinkelbach}. Once the optimal $\bar{\mathbf{V}}^{*}$ is obtained, the corresponding  vector $\bar{\mathbf{v}}$ that solves \textit{ (P4)} needs to be extracted. If the resulting matrix $\bar{\mathbf{V}}^{*}$ turns out to have rank-one, the optimal solution $\bar{\mathbf{v}}^*$ can be obtained as  \vspace{-.05in}
\begin{align}
\label{11}
&\bar{\mathbf{v}}^*=\mathbf{u}_{max}(\bar{\mathbf{V}}^{*}),
\end{align}
 where $\mathbf{u}_{max}(\mathbf{A})$ is the eigenvector corresponding to maximum eigenvalue of $\mathbf{A}$. If the rank turns out to be greater than one, then Gaussian randomization can be applied to find  $\bar{\mathbf{v}}^*$ by using the eigenvalue decomposition $\bar{\mathbf{V}}^*=\mathbf{U}\boldsymbol{\Lambda}\mathbf{U}^H$ and computing $\bar{\mathbf{v}}_l=\mathbf{U} \boldsymbol{\Lambda}^{1/2} \mathbf{r}_l$, where $\mathbf{r}_l\sim \mathcal{CN}(\mathbf{0}, \mathbf{I}_{N+1})$ for $l=1,\dots,L$. The solution $\bar{\mathbf{v}}^*$ can then be found as \vspace{-.05in}
\begin{align}
&l^*=\underset{l}{\text{max }} \underset{k}{\text{min}} \frac{\frac{p_k}{K} (\bar{\mathbf{v}}_l^H \mathbf{R}_{k,k} \bar{\mathbf{v}}_l + |b_{k,k}|^2)}{\sum_{i\neq k}^{K} \frac{p_i}{K} (\bar{\mathbf{v}}_l^H \mathbf{R}_{k,i} \bar{\mathbf{v}}_l + |b_{k,i}|^2)+\sigma_n^2}, \\
\label{22}
&\bar{\mathbf{v}}^*=\bar{\mathbf{v}}_{l^*}.
\end{align}
With a sufficiently large number of randomizations $L$, we can guarantee a very accurate approximation of the optimal objective value of \textit{(P4)} \cite{LIS_new, semidefiniterelax}. In our extensive simulations, we have always observed the optimal solution of Problem \textit{(P5)} to have rank-one and therefore $\bar{\mathbf{v}}^*$ in \eqref{11} is indeed optimal for \textit{(P4)}. The same observation was reported in some other works including \cite{ak, article1}.  

Finally, the solution to \textit{(P3)} can be recovered by accounting for the constraint that the last element of $\bar{\mathbf{v}}^*$ (which is $t$) should equal one and the first $N$ elements of $\bar{\mathbf{v}}^*$ need to satisfy the constraint \eqref{C3}. The resulting solution as outlined in \cite{LIS, LIS_new} is $\mathbf{v}^*=\exp\left(j \angle\left(\left[\frac{\bar{\mathbf{v}}^*}{\bar{\mathbf{v}}^*_{N+1}}\right]_{(1:N)}\right)\right)$, where $[\mathbf{x}]_{(1:N)}$ denotes the vector of first $N$ elements of $\mathbf{x}$ and $\bar{\mathbf{v}}^*_{N+1}$ is the last entry of $\bar{\mathbf{v}}^*$.  The Dinkelbach's procedure to solve \textit{(P3)} as well as the overall AO algorithm to solve \textit{(P1)} is outlined in \textbf{Algorithm 1}. 

The convergence of \textbf{Algorithm 1} is ensured by the noting that the objective value of \textit{(P1)}, i.e. $\underset{k}{\text{min}} \frac{\frac{p_{k}}{K} |\textbf{h}_{k}^{H} \textbf{g}_{k}|^{2}}{\sum_{i\neq k} \frac{p_{i}}{K} |\textbf{h}_{k}^{H} \textbf{g}_{i}|^{2}+ \sigma_n^2}$, is upper-bounded due to the constraint set in \textit{(P1)} and is non-decreasing over the iterations by applying \textbf{Algorithm 1}. To see this, denote the objective value of \textit{(P1)} based on a solution $(\mathbf{G}^*, \mathbf{p}^*, \mathbf{v}^*)$ as $f(\mathbf{G}^*, \mathbf{p}^*, \mathbf{v}^*)$. Let $(\mathbf{G}^{r^*}, \mathbf{p}^{r^*}, \mathbf{v}^{r^*})$ and $(\mathbf{G}^{r+1^*}, \mathbf{p}^{r+1^*}, \mathbf{v}^{r+1^*})$ be the solutions to (P2) in the $r^{th}$ and $(r + 1)^{th}$ iterations, respectively in step $5$ of the algorithm. It then follows that $f(\mathbf{G}^{r+1^*}, \mathbf{p}^{r+1^*}, \mathbf{v}^{r+1^*}) \geq f(\mathbf{G}^{r^*}, \mathbf{p}^{r^*}, \mathbf{v}^{r+1^*}) \geq f(\mathbf{G}^{r^*}, \mathbf{p}^{r^*}, \mathbf{v}^{r^*})$, where first inequality holds since for given $\mathbf{v}^{r+1^*}$ in step $5$ of \textbf{Algorithm 1}, $\mathbf{G}^{r+1^*}$, $\mathbf{p}^{r+1^*}$ is the optimal solution to problem \textit{(P2)}, and second inequality holds because $\mathbf{v}^{r+1^*}$ increases the objective value of \textit{(P3)} for given $\mathbf{G}^{r^*}$, $\mathbf{p}^{r^*}$ in step $14$. However, no global optimality claim can be made since \textit{(P1)} is not jointly convex with respect to $\mathbf{G}$, $\mathbf{P}$ and $\mathbf{v}$.

\begin{algorithm}[!t]
\caption{Alternating Optimization Algorithm}\label{alg:euclid}
\begin{algorithmic}[1]
\State Input: $\epsilon>0$, $\epsilon_1>0$, $\sigma^2_n$, $\mathbf{h}_{d,k}$ and $\mathbf{h}_{0,n,k}$ $\forall n$ $\forall k$.
\State Set the iteration number $r = 1$ and initialize the phase shifts vector $\mathbf{v}^r$.
\Repeat 
\Procedure{Solution to \textit{(P2)} for given $\mathbf{v}^r$}{Output: $\textbf{g}^{r^*}_k$, $p^{r^*}_k$, $k=1,\dots, K$}
\State Compute $\textbf{g}^{r^*}_k$ and $p^{r^*}_k$, $\forall k$, as the solution to \eqref{G_opt} and \eqref{P_opt}.
\EndProcedure
\Procedure{Solution to \textit{(P3)} for given $\mathbf{g}^{r^*}_k$, $p^{r^*}_k$}{Output: $\mathbf{v}^{(r+1)^*}$}
\State Initialize $\lambda=0$;
\Repeat 
\State $\bar{\textbf{V}}^{*} = \underset{\bar{\textbf{V}} \in \mathbb{C}^{N+1 \times N+1}}{\text{max}} \{\underset{1 \leq k \leq K}{\text{min}} [n_{k}(\bar{\textbf{V}})-\lambda d_{k}(\bar{\textbf{V}})] \}$, where $n_{k}(\bar{\textbf{V}})$ and $d_{k}(\bar{\textbf{V}})$ are given by (\ref{f}) and (\ref{g}) respectively, subject to  $\bar{\textbf{V}} \succeq 0$ and $\bar{\textbf{V}}_{n,n}=1$, $n=1,\dots, N+1$;
\State $F= \text{ min}_{1 \leq k \leq K}  \{n_{k}(\bar{\textbf{V}}^*)-\lambda d_{k}(\bar{\textbf{V}}^{*})\}$;
\State $\lambda = \text{ min}_{1 \leq k \leq K} \{n_{k}(\bar{\textbf{V}}^{*})/d_{k}(\bar{\textbf{V}}^{*})\} $;
\Until $F < \epsilon_1$.
\State $\bar{\mathbf{v}}^*$ computed using \eqref{11} or \eqref{22};
\State $\textbf{v}^{(r+1)^*}=\exp\left(j \angle\left(\left[\frac{\bar{\mathbf{v}}^*}{\bar{\mathbf{v}}^*_{N+1}}\right]_{(1:N)}\right)\right)$;
\State $r=r+1$;
\EndProcedure
\Until the fractional increase in $\underset{k}{\text{min }} \gamma_k$ is below $\epsilon$.
\end{algorithmic}
\end{algorithm}

\subsection{Imperfect CSI Scenario}

When only imperfect CSI is available at the BS, the BS can implement the AO algorithm by using $\underset{\mathbf{p}, \mathbf{G}, \mathbf{v}}{\text{max}} \underset{k}{\text{min }} \hat{\gamma}_k$ as the objective function in \textit{(P1)}, where
\begin{align}
\label{SINR_imp}
&\hat{\gamma}_k =\frac{\frac{p_{k}}{K} |\hat{\textbf{h}}_{k}^{H} \textbf{g}_{k}|^{2}}{\sum_{i\neq k} \frac{p_{i}}{K} |\hat{\textbf{h}}_{k}^{H} \textbf{g}_{i}|^{2}+ \sigma_n^2},
\end{align}
where $\hat{\mathbf{h}}_k=\hat{\mathbf{h}}_{d,k}+\hat{\mathbf{H}}_{0,k}\mathbf{v}$ with $\hat{\mathbf{h}}_{d,k}$ and $\hat{\mathbf{H}}_{0,k}$ being the MMSE estimates defined in \eqref{h_d_est} and \eqref{h_irs_est} respectively. The BS can not compute the true SINR values in \eqref{SINR} since it only has the estimates of $\mathbf{h}_k$'s available. As a consequence the solutions for \textit{(P2)} and \textit{(P3)} will be optimal in terms of the estimated minimum  SINR in \eqref{SINR_imp} instead of the true minimum SINR in \eqref{SINR}. Finding the optimal solution to \textit{(P1)} under imperfect CSI using the true minimum SINR  as an objective function is extremely difficult with no optimal solution in the literature. Therefore, replacing $\mathbf{h}_{d,k}$s and $\mathbf{H}_{0,k}$s with their estimates is a reasonable approach to tackle this problem and is similar to what is done in  \cite{LS, THz} that deal with the design of IRS-assisted system under CSI errors\footnote{As an extension, maximizing the expected minimum SINR where the expectation is performed with respect to the distribution of the sample space, which is dominated by knowledge of channel estimates and distribution of channel estimation error, can be considered as an objective function to make the algorithm robust to CSI errors. This will yield a stochastic optimization problem with two sub-problems both of which are non-convex. Moreover, the objective function will contain the expectation operator, and the probability density function of the sample space is usually very complicated with no closed-form expression as well. Therefore, designing an algorithm to solve such a stochastic problem is a really challenging task and has been left for future work.}.

Solving  \textit{(P2)} with $\text{max } \text{min } \hat{\gamma}_k$ as the objective function for fixed $\mathbf{v}$ will result in
\begin{align}
\label{G_opt_imp}
&\textbf{g}_{k}^*=\frac{\left(\sum_{i\neq k} \frac{\hat{q}^*_{i}}{K}\hat{\textbf{h}}_{i}\hat{\textbf{h}}_{i}^{H}+\sigma_n^2 \textbf{I}_{M}\right)^{-1}\hat{\textbf{h}}_{k}}{||\left(\sum_{i\neq k} \frac{\hat{q}_{i}^*}{K}\hat{\textbf{h}}_{i}\hat{\textbf{h}}_{i}^{H}+\sigma_n^2 \textbf{I}_{M}\right)^{-1}\hat{\textbf{h}}_{k}||},
\end{align}
where $\hat{q}^*_{k}$s are obtained as the unique positive solution to $\hat{q}^*_{k}=\frac{\hat{\tau}^*}{\frac{1}{K}\hat{\textbf{h}}_{k}^H\left(\sum_{i\neq k} \frac{\hat{q}^*_{i}}{K}\hat{\textbf{h}}_{i}\hat{\textbf{h}}_{i}^{H}+\sigma_n^2 \textbf{I}_{M}\right)^{-1}\hat{\textbf{h}}_{k}}$ with  $\hat{\tau}^*=\frac{KP_{max}}{\sum_{k=1}^{K}\left(\frac{1}{K}\hat{\textbf{h}}_{k}^H\left(\sum_{i\neq k} \frac{\hat{q}^*_{i}}{K}\hat{\textbf{h}}_{i}\hat{\textbf{h}}_{i}^{H}+\sigma_n^2 \textbf{I}_{M}\right)^{-1}\hat{\textbf{h}}_{k}\right)^{-1}}$. The allocated powers $p^*_k$ are given as
\begin{align}
\label{P_opt_imp}
&\textbf{p}^*=[p_1^*, \dots, p_K^*]^T=\left(\textbf{I}_{K}-\hat{\tau}^*\hat{\textbf{D}}\hat{\textbf{F}}  \right)^{-1}\hat{\tau}^* \sigma_n^2 \hat{\textbf{D}}\textbf{1}_{K},
\end{align}
where $\hat{\textbf{D}}=\text{diag}\left(\frac{1}{\frac{1}{K}|\hat{\textbf{h}}_{1}^{H}\textbf{g}^*_{1}|^{2}}, \dots, \frac{1}{\frac{1}{K}|\hat{\textbf{h}}_{K}^{H}\textbf{g}^*_{K}|^{2}} \right)$ and $[\hat{\textbf{F}}]_{k,i}=\frac{1}{K}|\hat{\textbf{h}}_{k}^{H}\textbf{g}^*_{i}|^{2}$, if $k\neq i$ and $0$ otherwise.

The optimization with respect to $\mathbf{v}$ in \textit{(P3)} using $\text{max } \text{min } \hat{\gamma}_k$ as the objective function can be performed by expressing the numerator and denominator of \eqref{SINR_imp} in terms of quadratic forms, with the difference being that  $\mathbf{h}_{d,k}$s and $\mathbf{H}_{0,k}$s will be replaced with their estimates in the definitions of $\mathbf{a}_{k,i}$ and $b_{k,i}$ in \eqref{P5}. The resulting problem can be relaxed using semi-definite relaxation and then solved using the Dinkelbach's algorithm. 

The overall AO algorithm will be the same as \textbf{Algorithm 1}, with the difference being that the input channel vectors $\mathbf{h}_{d,k}$ and $\mathbf{h}_{0,n,k}$s in step 1 will be replaced by their estimates $\hat{\mathbf{h}}_{d,k}$ and $\hat{\mathbf{h}}_{0,n,k}$s in \eqref{h_d_est} and \eqref{h_irs_est} respectively and the stopping criteria in step 16 will be applied on $\underset{k}{\text{min }} \hat{\gamma}_k$ where $\hat{\gamma}_k$ is defined in \eqref{SINR_imp}. The algorithm will therefore alternate between the computation of $\mathbf{g}^*_k$s and $p_k^*$s in \eqref{G_opt_imp} and \eqref{P_opt_imp} respectively for fixed $\mathbf{v}$ and the computation of $\mathbf{v}^*$ for fixed $\mathbf{g}_k$s and $p_k$s, until convergence is reached, which happens when the fractional increase in  $\underset{k}{\text{min }} \hat{\gamma}_k$  is below a threshold value.  We would stress that the performance of the proposed design is shown in terms of the true minimum SINR in the simulation results and not the estimated minimum SINR.

\begin{table}[!b]
\centering
\normalsize
\caption{Simulation parameters.}
\begin{tabular}{|l|l|}
\hline
  \textbf{Parameter} & \textbf{Value} \\ 
\hline
\textbf{Array parameters:} & \\
\hline
Carrier frequency & $2.5$ GHz \\
 BS, IRS configuration & Uniform linear array (ULA)\\
$d_{BS}$, $d_{IRS}$& $0.5\lambda$\\
Tx power budget ($P_{max}$)& $5$ W \\
Noise level & $-80$\rm{dBm} \\
 \hline
\textbf{Path Loss:} & \\
\hline
Model & $\frac{10^{-C/10}}{d^{\alpha}}$ \\
$C$ (Fixed loss at $d=1$m)  & $26$\rm{dB} ($\beta_{1}$), $28$\rm{dB} ($\beta_{2,k},\beta_{d,k}$)\\
$\alpha$ (Path loss exponent) & $2.2$ ($\beta_{1}$), $3.67$ ($\beta_{2,k},\beta_{d,k}$)\\
\hline
\textbf{Channel Estimation:} & \\
\hline
 $\tau$ & $.05$s \\
$\tau_S$& $50K$ $\mu s$\\
$\tau_C$  &$S \tau_S$\\
$P_C$ & $1$ W\\
\hline
\textbf{Correlation Model:} & \\
\hline
$\mathbf{R}_{BS_{k}}, \mathbf{R}_{IRS_{k}}$ & Generated using [\cite{LIS_jour} Sec. V] \\
\hline
\textbf{Algorithm 1:} & \\
\hline
$\epsilon$, $\epsilon_1$ & $10^{-4}$ \\
$\mathbf{v}^1$& CoM scheme \cite{LIS_jour} \\
\hline
\end{tabular}
\label{T1}
\end{table}

\section{Simulation Results}

\begin{figure}[!t]
\centering
\includegraphics[scale=.46]{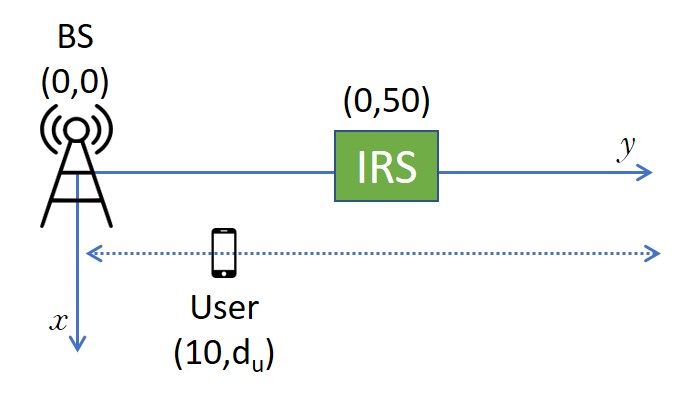}
\caption{IRS-assisted single-user MISO system.  The BS and IRS are marked with their $(x,y)$ coordinates.}
\label{SU1_sim}
\end{figure}

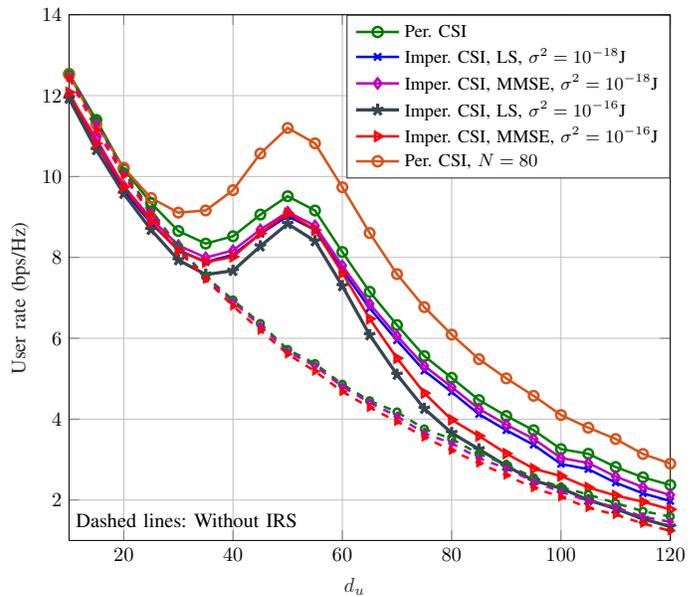
\begin{figure}[t]
\tikzset{every picture/.style={scale=.95}, every node/.style={scale=.8}}
%
%
\definecolor{mycolor1}{rgb}{0.00000,0.49804,0.00000}%
\definecolor{mycolor2}{rgb}{0.74902,0.00000,0.74902}%
\definecolor{mycolor3}{rgb}{0.85098,0.32549,0.09804}%
 	\definecolor{mycolor4}{rgb}{0.21, 0.27, 0.31}
\begin{tikzpicture}

\begin{axis}[%
width=1\columnwidth,
height=.875\columnwidth,
scale only axis,
xmin=10,
xmax=120,
xlabel style={font=\normalsize\color{white!15!black}},
xlabel={$d_u$},
ymin=1,
ymax=14,
ylabel style={at={(axis cs: 14,7)},font=\normalsize\color{white!15!black}},
ylabel={User rate (bps/Hz)},
axis background/.style={fill=white},
xmajorgrids,
ymajorgrids,
legend style={at={(axis cs: 120,14)},anchor=north east,legend cell align=left,align=left,draw=white!15!black, /tikz/column 2/.style={
                column sep=5pt,
            }}
]
\addplot [color=mycolor1, line width=1.0pt, mark size=2.0pt, mark=o, mark options={solid, mycolor1}]
  table[row sep=crcr]{%
10	12.5406305303297\\
15	11.3994767817718\\
20	10.1859503720063\\
25	9.34483985196217\\
30	8.65259376521215\\
35	8.34092073491516\\
40	8.5226982464625\\
45	9.05830658283664\\
50	9.51325702944205\\
55	9.15931512875029\\
60	8.13314411514844\\
65	7.15026436627287\\
70	6.3264458042138\\
75	5.55956801843791\\
80	5.02154665250174\\
85	4.47213425120707\\
90	4.07837055983137\\
95	3.72630029031204\\
100	3.2593033446498\\
105	3.14332354491081\\
110	2.81619711960796\\
115	2.56157400882271\\
120	2.37106364461407\\
};
\addlegendentry{\small Per. CSI}

\addplot [color=blue, line width=1pt, mark size=2pt,  mark=x, mark options={solid, blue}]
  table[row sep=crcr]{%
10	11.9212538176271\\
15	10.8239382094368\\
20	9.65528482816493\\
25	8.84718333509662\\
30	8.18237845444574\\
35	7.88517376466895\\
40	8.06534485511494\\
45	8.58277917494591\\
50	9.0197143495533\\
55	8.68042986676314\\
60	7.69366571889392\\
65	6.74283281538635\\
70	5.94823092427543\\
75	5.19718015826934\\
80	4.67117758681711\\
85	4.1220218864577\\
90	3.7257802261302\\
95	3.36483269341658\\
100	2.88992432858557\\
105	2.76305223131101\\
110	2.42817991094812\\
115	2.1658485900941\\
120	1.97512826702497\\
};
\addlegendentry{\small Imper. CSI, LS, $\sigma^{2}=10^{-18}$\rm{J}}

\addplot [color=mycolor2, line width=1.0pt, mark size=2pt, mark=diamond, mark options={solid, mycolor2}]
  table[row sep=crcr]{%
10	12.0258041774787\\
15	10.9311660042544\\
20	9.76660037035528\\
25	8.95925251008513\\
30	8.29501143491565\\
35	7.99639057136178\\
40	8.1714864657499\\
45	8.68591689295303\\
50	9.12255707086431\\
55	8.78289716685411\\
60	7.79782252775894\\
65	6.85326669820831\\
70	6.05923427516363\\
75	5.31771571373538\\
80	4.79290298833645\\
85	4.25372938491658\\
90	3.86319083789318\\
95	3.51132542802788\\
100	3.03879694234957\\
105	2.91526369965912\\
110	2.58151515305074\\
115	2.32022822341879\\
120	2.12535411152797\\
};
\addlegendentry{\small Imper. CSI, MMSE, $\sigma^{2}=10^{-18}$\rm{J}}

\addplot [color=mycolor4, line width=1.3pt, mark size=2.5pt, mark=star, mark options={solid, mycolor4}]
  table[row sep=crcr]{%
10	11.9287528787835\\
15	10.6602204687472\\
20	9.57779286376544\\
25	8.68544456258139\\
30	7.93506912034131\\
35	7.57529930166525\\
40	7.67064277848532\\
45	8.27689174241353\\
50	8.83170819018325\\
55	8.40378516701631\\
60	7.29359494281436\\
65	6.08048752601461\\
70	5.10161123700082\\
75	4.25936448032001\\
80	3.65954579581349\\
85	3.25644183421905\\
90	2.84390854758593\\
95	2.48156818488257\\
100	2.29179792401596\\
105	2.000553408279329\\
110	1.78439670799259\\
115	1.53262123046876\\
120	1.34214275996422\\
};
\addlegendentry{\small Imper. CSI, LS, $\sigma^2=10^{-16}$\rm{J}}

\addplot [color=red, line width=1.0pt, mark size=2pt, mark=triangle, mark options={solid, rotate=270, red}]
  table[row sep=crcr]{%
10	12.0812917638023\\
15	10.8163276407182\\
20	9.74408880916356\\
25	8.87843205542434\\
30	8.17906325207851\\
35	7.88364327549623\\
40	8.01517389683666\\
45	8.59321201324142\\
50	9.09255443470979\\
55	8.67517930266721\\
60	7.61876176218284\\
65	6.4818500971709\\
70	5.50613580355482\\
75	4.64520495028215\\
80	3.98374318809458\\
85	3.58351257782328\\
90	3.14915380739941\\
95	2.77554433683861\\
100	2.59355496259148\\
105	2.30924085854491\\
110	2.11177151346019\\
115	1.95760673424005\\
120	1.76823804236288\\
};
\addlegendentry{\small Imper. CSI, MMSE, $\sigma^2=10^{-16}$\rm{J}}

\addplot [color=mycolor3, line width=1.0pt, mark size=2pt, mark=o, mark options={solid, mycolor3}]
  table[row sep=crcr]{%
10	12.5217177588537\\
15	11.2919779110278\\
20	10.2174871319516\\
25	9.46873248787327\\
30	9.10703805403576\\
35	9.16127845041163\\
40	9.66251649820147\\
45	10.5669629087439\\
50	11.2012357651345\\
55	10.8208415355699\\
60	9.73443183488402\\
65	8.60302113296702\\
70	7.58701749098601\\
75	6.77007638807243\\
80	6.09166286165065\\
85	5.48292645084991\\
90	5.00858180461231\\
95	4.5770016530829\\
100	4.10215826639419\\
105	3.78845112662978\\
110	3.50982250838061\\
115	3.13811709588789\\
120	2.89908916617098\\
};
\addlegendentry{\small Per. CSI, $N=80$}

\addplot [color=mycolor1, dashed, line width=1.0pt, mark size=1.4pt, mark=o, mark options={solid, mycolor1}]
  table[row sep=crcr]{%
10	12.5294496247771\\
15	11.3762310444771\\
20	10.1178596417022\\
25	9.17414628419435\\
30	8.26539123539033\\
35	7.52787494647389\\
40	6.947334409357\\
45	6.36557652448908\\
50	5.7233409238907\\
55	5.3726829646319\\
60	4.86285140078564\\
65	4.45569768598571\\
70	4.17872598884125\\
75	3.75870596347902\\
80	3.50842279551929\\
85	3.1415032714317\\
90	2.8789523239658\\
95	2.55171759666915\\
100	2.31868663899795\\
105	2.1363839251543\\
110	1.92435679168892\\
115	1.72416238492069\\
120	1.59240773086788\\
};

\addplot [color=mycolor2, dashed, line width=1pt, mark size=1.3pt, mark=diamond, mark options={solid, mycolor2}]
  table[row sep=crcr]{%
10	12.4769125640271\\
15	11.324838101316\\
20	10.06077032011749\\
25	9.12490058474729\\
30	8.21698448649337\\
35	7.48012104484378\\
40	6.90002282248437\\
45	6.308701150676\\
50	5.66679901093553\\
55	5.31632681174887\\
60	4.80655232092326\\
65	4.40933248591449\\
70	4.07229911459975\\
75	3.65199543528446\\
80	3.40153649899354\\
85	3.03415397933723\\
90	2.78110383372404\\
95	2.45894122829964\\
100	2.19939197231433\\
105	1.97732783597598\\
110	1.8240838582952\\
115	1.58345797202486\\
120	1.46125105083746\\
};

\addplot [color=red, dashed, line width=1.0pt, mark size=1.3pt, mark=triangle, mark options={solid, rotate=270, red}]
  table[row sep=crcr]{%
10	12.4417915788738\\
15	11.1583296892159\\
20	10.0215199212508\\
25	9.0599983570302\\
30	8.17201077777506\\
35	7.460321977553112\\
40	6.79796167213817\\
45	6.19442479571987\\
50	5.60893387655277\\
55	5.18422631933289\\
60	4.68159764963905\\
65	4.28790184982133\\
70	3.94095274211312\\
75	3.55284944448608\\
80	3.22624691055952\\
85	2.912271971476901\\
90	2.612737361537664\\
95	2.300167976773584\\
100	2.07632801180237\\
105	1.80819620706469\\
110	1.63805792256673\\
115	1.41954142984159\\
120	1.23534201008731\\
};

\node at (axis cs: 10,1.5) [anchor = west] {\normalsize Dashed lines: Without IRS};

\end{axis}

\end{tikzpicture}%
\caption{Performance of an IRS-assisted single-user MISO system  under perfect (per.) and imperfect (imper.) CSI for $M=4$, $N=40$ and $S=N+1$.}
\label{SU2_sim}
\end{figure}

We utilize the parameter values described in Table \ref{T1} in generating the simulation results. The path loss parameters are computed  at $2.5$ GHz operating frequency for the 3GPP Urban Micro (UMi) scenario from TR36.814 (detailed in Section V of \cite{LIS_jour}). We use the LoS version to generate path loss for $\mathbf{H}_1$ and the non-LOS (NLOS) version to generate path losses for $\mathbf{h}_{2,k}$ and $\mathbf{h}_{d,k}$. Moreover, $5$ \rm{dBi} antennas are considered at the BS and IRS. Note that the IRS is deployed much higher than the BS to avoid the penetration losses and blockages caused by ground structures like buildings. Therefore, we assume a penetration loss of $15$ \rm{dB} in each BS-to-user link, whereas we assume negligible penetration loss in the IRS-to-user links.   

  We first focus on the single-user IRS-assisted system shown in Fig. \ref{SU1_sim} and plot in Fig. \ref{SU2_sim} the rate achieved by the user for varying values of $d_{u}$. Note that for a single-user system, the SINR in \eqref{SINR} is simplified to SNR given as $\gamma_k=p_k |\mathbf{h}_k^H \mathbf{g}_k|^2$ and the user rate is related to the SNR as $R_k=\left( 1-\frac{\tau_C}{\tau} \right) \log_2(1+\gamma_k)$, where the factor $\left( 1-\frac{\tau_C}{\tau} \right)$ accounts for the rate loss due to channel training. 	The results are plotted under the optimized precoding vector $\mathbf{g}_k^*$ and phase-shifts vector $\mathbf{v}^*$\footnote{Note that for a single-user setting, the solution in step 5 of Algorithm 1 for $\mathbf{g}_k^*$ can be simplified to MRT precoding and the solution to \textit{(P3)} can be given as $\mathbf{v}^*=\exp(j \angle(\mathbf{H}_{0,k}^H\mathbf{h}_{d,k}))$. Details have been omitted from this work since similar results have appeared in \cite{LIS}. Moreover, for a single-user, we let $\mathbf{H}_1$ be a rank-one LoS channel as generated in \cite{LIS_jour}}.  For the imperfect CSI case, we plot the results under both LS-DFT and MMSE-DFT estimates derived in Section III. We observe that in an IRS-assisted system, the user farther away from the BS can still be closer to the IRS and receive stronger reflected signals from it resulting in an improvement in the performance as observed for $d_{u}>30$. Consequently, the IRS-assisted system is able to provide a higher QoS to a larger region. For example, under perfect CSI it will cover  $120$m with a rate at least $2.3$\rm{bps/Hz}, whereas the system without the IRS can cover about $95$m to achieve the same rate. Moreover, the users placed close to the IRS, e.g. located in $42<d_u < 70$ range will see gains   ranging from $2$ to $4$ \rm{bps/Hz}. Although the rate decreases due to increasing signal attenuation when $d_u>50$ but it is still better than what would have been achieved without the IRS unless the user is so far away that the path loss becomes dominant over the gain provided by the IRS.

Doubling $N$ at the IRS to $80$, the achieved rate scales by about $2$\rm{bps/Hz} for users close to the IRS, which implies that the SNR scales by around $6$\rm{dB}. This corresponds to the scaling of SNR in the order of $N^2$, corresponding to an array gain of $N$ and the reflect beamforming gain of $N$ as analytically proved in \cite{LIS_new}. However, the gain is negligible for $10<d_{u}<25$ because the BS-to-user direct channel is much stronger than the channel through the IRS. Moreover,  higher coverage is possible with large number of reflecting elements as shown through the higher values of achieved rate for $N=80$ under perfect CSI.

The curves under imperfect CSI show that the IRS-assisted system is more sensitive to channel estimation errors than the conventional MISO (without IRS) system. This is because the IRS-assisted system has to estimate $N+1=41$ channel vectors whereas the direct system only needs to estimate one channel vector. Moreover, the error becomes more significant as the user moves away from the IRS because the channel vectors  become weaker and  more difficult to estimate. The IRS-assisted system designed using MMSE-DFT estimates outperforms the system that relies on LS-DFT estimates especially  for higher channel estimation noise, as discussed in Fig. \ref{Fig_est} as well.

Next we study the minimum user rate performance of a multi-user system under imperfect CSI with the BS placed at $(0,0)$, IRS placed at $(0, 100)$ and users distributed uniformly in the square $(x,y)\in[-30,30]\times[70,130]$. Accounting for the rate loss due to channel training, the net achievable rate of user $k$ is given as 
\begin{align}
&R_k=\left(1-\frac{\tau_{C}}{\tau} \right)\text{log}_{2}(1+\gamma_{k})\nonumber \\
\label{R_net}
&=\left(1-\frac{S\tau_S}{\tau} \right)\text{log}_{2}(1+\gamma_{k}),
\end{align}
where $\gamma_k$ is defined in \eqref{SINR}. Note that the total channel estimation $\tau_C$ sec is related to the number of estimation sub-phases $S$ and the duration of each sub-phase $\tau_S$ sec as $\tau_C=S\tau_S$. In Sec. III we saw that increasing $S$ improves the quality of channel estimates by reducing the NMSE by a factor of approximately $S$. Moreover, under the proposed channel estimation protocol the minimum number of required sub-phases $S$ is $N+1$, to ensure that the left pseudo-inverse of $\bar{\mathbf{V}}^{tr}$ in \eqref{LSS2} exists. At the same time, the total channel estimation time $\tau_C$ increases linearly with $S$, which reduces the time left for downlink transmission causing the rate loss factor of $\left(1-\frac{S\tau_{S}}{\tau} \right)$ that we see in \eqref{R_net}. Therefore, $S$ has the positive effect  of improving the channel estimates quality and the adverse impact of increasing the total channel estimation time and should be selected carefully to strike a balance. The next figure will study this trade-off. 

In Fig. \ref{MU1} we plot the net achievable minimum rate against $S$ for an IRS-assisted system serving $4$ users with $M=8$ antennas at the BS, while optimizing the precoding vectors, power allocation and IRS phase shifts vector  using \textbf{Algorithm 1} with the MMSE channel estimates as the input. For the two considered IRS-assisted MISO systems, we find that $S\approx N+1$ is the optimal number of sub-phases that maximizes the achieved minimum user rate, i.e. $S\approx 9$ is optimal for the system with $N=8$ reflecting elements, while $S\approx 17$ is optimal for the system with $N=16$ reflecting elements. For $S<N+1$, the NMSE in the channel estimates becomes very high since the left pseudo-inverse of $\bar{\mathbf{V}}^{tr}$ utilized in \eqref{LSS2} becomes singular as $\bar{\mathbf{V}}^{tr}$ does not have full column rank\footnote{In fact, we are unable to plot the performance for $S<N$ because the pseudo-inverse of $\bar{\mathbf{V}}^{tr}$ needed to implement \eqref{LSS2} does not exist.}. As a result the rate obtained for $S=N$ is lower than that for $S=N+1$, since the computed pseudo-inverse for $S=N$ is inaccurate. 

 Increasing $S$ above $N+1$ has the positive effect of reduced channel estimation error as shown earlier in Fig. \ref{Fig1PL_est} and Fig. \ref{Fig2PL_est}. However, increasing $S$ also increases the channel training time causing a rate loss factor of $\left(1-\frac{S\tau_{S}}{\tau} \right)$ since the total time left for downlink transmission decreases as $\tau-S\tau_S$.  The decrease in downlink transmission time is linear with increasing $S$ as can be seen from \eqref{R_net}, whereas the impact of improvement in estimation quality is only logarithmic with increasing $S$ since the SINR $\gamma_k$ appears inside the log function in \eqref{R_net}. The negative effect of decrease in downlink transmission time dominates over the positive effect of improvement in channel estimates quality as $S$ increases. Therefore, $S\approx N+1$ is the optimal number of sub-phases for both considered settings.  

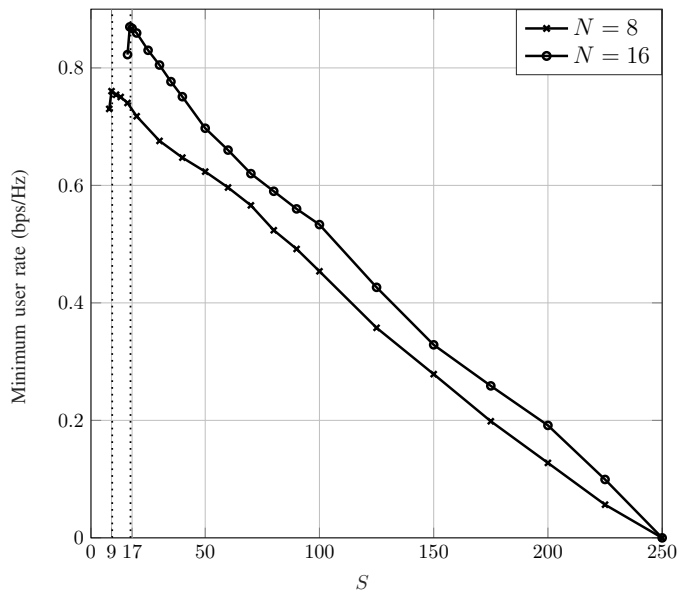
\begin{figure}[t]
\tikzset{every picture/.style={scale=.95}, every node/.style={scale=.8}}
%
%
\definecolor{mycolor1}{rgb}{0.00000,0.44700,0.74100}%
\begin{tikzpicture}

\begin{axis}[%
width=.95\columnwidth,
height=.88\columnwidth,
scale only axis,
xmin=0,
xmax=250,
xlabel style={font=\normalsize\color{white!15!black}},
xlabel={$S$},
ymin=0,
ymax=0.9,
ylabel style={font=\normalsize\color{white!15!black}},
ylabel={Minimum user rate (bps/Hz)},
axis background/.style={fill=white},
extra x ticks={9,18},
extra x tick labels={$9$, $17$},
xmajorgrids,
ymajorgrids,
legend style={at={(axis cs: 250,0.9)},anchor=north east,legend cell align=left,align=left,draw=white!15!black, /tikz/column 2/.style={
                column sep=5pt,
            }}
]
\addplot [color=black, line width=1.0pt,mark size=1.75pt, mark=x, mark options={solid, black}]
  table[row sep=crcr]{%
8	0.730211841128796\\
9	0.760380915857594\\
11	0.754301362876142\\
13	0.750474557101258\\
16	0.740306613614099\\
20	0.717697784187118\\
30	0.67573804897075\\
40	0.647298730307552\\
50	0.623405567893504\\
60	0.596381733893645\\
70	0.565990550526893\\
80	0.523629608181807\\
90	0.49176257328424\\
100	0.453767904165231\\
125 .35757373\\
150	0.27866\\
175 .19847283\\
200	0.127583\\
225 .0564734\\
250	0\\
};
\addlegendentry{\large $N=8$}

\addplot [color=black, line width=1.0pt,mark size=1.5pt, mark=o, mark options={solid, black}]
  table[row sep=crcr]{%
16	0.822641483296072\\
17	0.869860348083887\\
18	0.86729901\\
20	0.859400218310058\\
25	0.83\\
30	0.804873765742851\\
35	0.7766\\
40	0.750970659160087\\
50	0.6973\\
60	0.6601\\
70	0.62\\
80	0.59\\
90	0.56\\
100	0.533271228825328\\
125 .42646272\\
150 .32858393\\
175 .25869392\\
200	0.19143913251152\\
225 .0993647272 \\
250	0\\
};
\addlegendentry{\large $N=16$}

\addplot [color=black, dotted,line width=0.7pt]
  table[row sep=crcr]{%
9.3	0\\
9.3 0.9\\
};
\addplot [color=black, dotted,line width=0.7pt]
  table[row sep=crcr]{%
17.3	0\\
17.3	0.9\\
};

\node[circle,fill,inner sep=1.5pt] at (axis cs:9,-1.4) {};
\node[circle,fill,inner sep=1.5pt] at (axis cs:18,-1.4) {};

\end{axis}

\end{tikzpicture}%
\caption{Number of sub-phases $S$ that maximizes the minimum user rate achieved by the IRS-assisted multi-user MISO system under MMSE-DFT protocol.}
\label{MU1}
\end{figure}

\begin{figure}[t]
\tikzset{every picture/.style={scale=.95}, every node/.style={scale=.8}}
%
%
\definecolor{mycolor1}{rgb}{0.00000,0.49804,0.00000}%
\definecolor{mycolor2}{rgb}{0.74902,0.00000,0.74902}%
\begin{tikzpicture}

\begin{axis}[%
width=.95\columnwidth,
height=.88\columnwidth,
scale only axis,
xmin=8,
xmax=48,
xlabel style={font=\normalsize\color{white!15!black}},
xlabel={$N$},
ymin=0.83,
ymax=2.2,
ylabel style={at={(axis cs: 7.4,1.55)},font=\normalsize\color{white!15!black}},
ylabel={Minimum user rate (bps/Hz)},
axis background/.style={fill=white},
xmajorgrids,
ymajorgrids,
legend style={at={(axis cs: 8,2.2)},anchor=north west,legend cell align=left,align=left,draw=white!15!black, /tikz/column 2/.style={
                column sep=5pt,
            }}
]
\addplot [color=red, line width=1.2pt,mark size=2.5pt, mark=x, mark options={solid, red}]
  table[row sep=crcr]{%
8	1.17035947203653\\
16	1.3114973633652\\
32	1.59667556078273\\
48	1.79241577211594\\
};
\addlegendentry{Per. CSI, $M=12$}

\addplot [color=red,  line width=1.0pt,mark size=2.2pt, mark=o, mark options={solid, red}]
  table[row sep=crcr]{%
8	1.12232497998119\\
16	1.223438478754\\
32	1.38524492656854\\
48	1.48022566993109\\
};
\addlegendentry{Imp. CSI, $M=12$}

\addplot [color=mycolor1,line width=1.0pt,mark size=2pt, mark=square, mark options={solid, mycolor1}]
  table[row sep=crcr]{%
8	1.33493706496252\\
16	1.49686596875183\\
32	1.75740069979966\\
48	2.13920646093385\\
};
\addlegendentry{Per. CSI, $M=15$}

\addplot [color=mycolor1, line width=1.0pt,mark size=2pt, mark=triangle, mark options={solid, mycolor1}]
  table[row sep=crcr]{%
8	1.28552412783746\\
16	1.39409991595269\\
32	1.5244395881222\\
48	1.74975205997337\\
};
\addlegendentry{Imper. CSI, $M=15$}

\addplot [color=mycolor2, line width=0.8pt,mark size=2pt, mark=triangle, mark options={solid, rotate=180, mycolor2}]
  table[row sep=crcr]{%
8	1.54331153032223\\
16	1.55987826508213\\
32	1.52851344248857\\
48	1.5311275180467\\
};
\addlegendentry{Per. CSI, No IRS, $M=20$}

\addplot [color=mycolor2, line width=0.8pt,mark size=2pt, mark=asterisk, mark options={solid}]
  table[row sep=crcr]{%
8	1.53381785233704\\
16	1.55064017768581\\
32	1.51925807579165\\
48	1.52198580476057\\
};
\addlegendentry{Imp. CSI, No IRS, $M=20$}

\addplot [color=red, dashed, line width=1.2pt,mark size=2.5pt, mark=x, mark options={solid, red}]
  table[row sep=crcr]{%
8	1.07952889779757\\
16	1.11926699687943\\
32	1.16526026403422\\
48	1.17706670996068\\
};

\addplot [color=red,  dashed, line width=1.0pt,mark size=2pt, mark=o, mark options={solid, red}]
  table[row sep=crcr]{%
8	1.03904083676585\\
16	1.04202381555526\\
32	1.01048057113093\\
48	0.945719058008384\\
};

\addplot [color=mycolor1, dashed, line width=1.0pt,mark size=1.6pt, mark=square, mark options={solid, mycolor1}]
  table[row sep=crcr]{%
8	1.24837691852963\\
16	1.32450519739852\\
32	1.39550222829437\\
48	1.48505774171789\\
};

\addplot [color=mycolor1, dashed, line width=1.0pt,mark size=2pt, mark=triangle, mark options={solid, mycolor1}]
  table[row sep=crcr]{%
8	1.20218393867698\\
16	1.23354893382957\\
32	1.25844190866888\\
48	1.23997203198015\\
};

\node at (axis cs: 47,.86) [anchor = east] {\normalsize Dashed lines: CoM phases adjustment};
\node at (axis cs: 47,.92) [anchor = east] {\normalsize Solid lines: Proposed Alg. 1};

\end{axis}
\end{tikzpicture}%
\caption{Performance of an  IRS-assisted multi-user system against $N$ under perfect (per.) and imperfect (imper.) CSI.}
\label{MU2}
\end{figure}
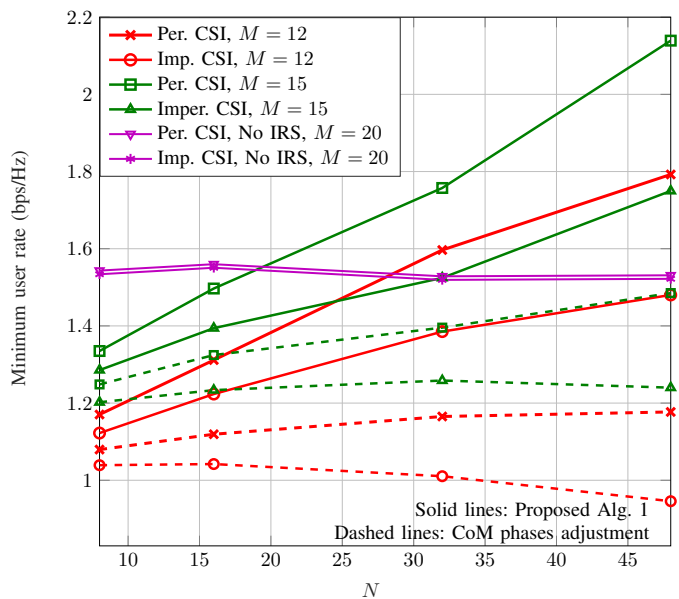

Fig. \ref{MU2} plots the minimum user rate against $N$ for varying number of antennas at the BS in an IRS-assisted system, where the precoding vectors, allocated powers and IRS phases are optimized under \textbf{Algorithm 1} for both perfect CSI and imperfect CSI cases (where for the latter we use the channel estimates as input in step 1 of the algorithm).  The number of sub-phases $S=N+1$ under the MMSE-DFT channel estimation protocol. The performance is compared to that of a conventional large MISO system having $20$ antennas at the BS and no IRS. We show that by appropriately selecting the number of reflecting elements $N$ at the IRS, the IRS-assisted system can perform as well as the large MISO system with a reduced number of antennas at the BS.  Under perfect CSI, the IRS-assisted MISO system with $28$ passive reflecting elements at the IRS and only $12$ active antennas at the BS can achieve the same performance as the considered large MISO system of $20$ antennas.  The same performance can also be achieved with $M=15$ antennas using $N=19$ reflecting elements at the IRS. We also notice that under channel estimation errors, larger array sizes are needed at the IRS to achieve the same performance as the conventional large MISO system. For example, under imperfect CSI an IRS-assisted system with $M=12$ antennas at the BS can achieve nearly the same performance using $N=48$  instead of $N=28$ reflecting elements. Moreover, as the value of $N$ increases the performance gap between perfect and imperfect CSI curves for the IRS-assisted system significantly increases  since the minimum number of required sub-phases $S$ increases linearly in $N$.  This causes a rate loss due to the time spent in channel training. Therefore, accurate and quick CSI acquisition is a critical issue in IRS-assisted communication systems that needs to be addressed to reap the full potential of this technology. However, IRS-assisted communication also has the potential to be an energy-efficient alternative to technologies like massive MISO and network densification by reducing the number of active antennas and RF chains needed at the BS. 

 To test the performance of the proposed \textbf{Algorithm 1}, we consider the benchmark Centre of Means (CoM) scheme from \cite{LIS_jour}, where the IRS phase-shifts are set as the mean of the  LoS angles of all users\footnote{The max-min SINR has not been the subject of any work on IRS-assisted communication systems except \cite{LIS_jour}}. The proposed algorithm is shown to outperform the benchmark scheme considerably. 

\begin{figure}[!t]
\centering
\tikzset{every picture/.style={scale=.95}, every node/.style={scale=.8}}
%
%
\definecolor{mycolor1}{rgb}{0.00000,0.44700,0.74100}%
\definecolor{mycolor2}{rgb}{0.85000,0.32500,0.09800}%
\begin{tikzpicture}

\begin{axis}[%
width=.47\textwidth,
height=.41\textwidth,
scale only axis,
xmin=1,
xmax=15,
xlabel style={font=\color{white!15!black}},
xlabel={Number of iterations},
ymin=0.77,
ymax=0.955,
ylabel style={font=\color{white!15!black}},
ylabel={Minimum user rate (bps/Hz)},
axis background/.style={fill=white},
xmajorgrids,
ymajorgrids,
legend style={at={(0.97,0.03)}, anchor=south east, legend cell align=left,align=left,draw=white!15!black, /tikz/column 2/.style={
                column sep=5pt,
            }}
]

\addplot [color=black, line width=1.0pt,mark size=2pt, mark=x, mark options={solid, black}]
  table[row sep=crcr]{%
2	0.843396329302425\\
3	0.899515739574723\\
4	0.910275409559596\\
5	0.915354330037668\\
6	0.920454796552647\\
7	0.926010784079499\\
8	0.929766053357821\\
9	0.931084298570214\\
10	0.931574429762334\\
11	0.931805098518047\\
12	0.93194047088165\\
13	0.9320360306211\\
14	0.93211366808364\\
15	0.932183032639981\\
};
\addlegendentry{Algorithm 1 (Perfect CSI)}

\addplot [color=black, line width=1.0pt,mark size=2pt, mark=o, mark options={solid, black}]
  table[row sep=crcr]{%
2	0.785282090887264\\
3	0.83667547275108\\
4	0.846860092553733\\
5	0.851865199013081\\
6	0.85688864476675\\
7	0.862037892093024\\
8	0.865388818881526\\
9	0.866727137983349\\
10	0.867175672543169\\
11	0.867382750269051\\
12	0.867503427852728\\
13	0.867589790951313\\
14	0.867661651544168\\
15	0.867727314359532\\
};
\addlegendentry{Algorithm 1 (Imperfect CSI)}

\end{axis}
\end{tikzpicture}%
\caption{Convergence behaviour of the proposed AO algorithm.}
\label{conv}
\end{figure}
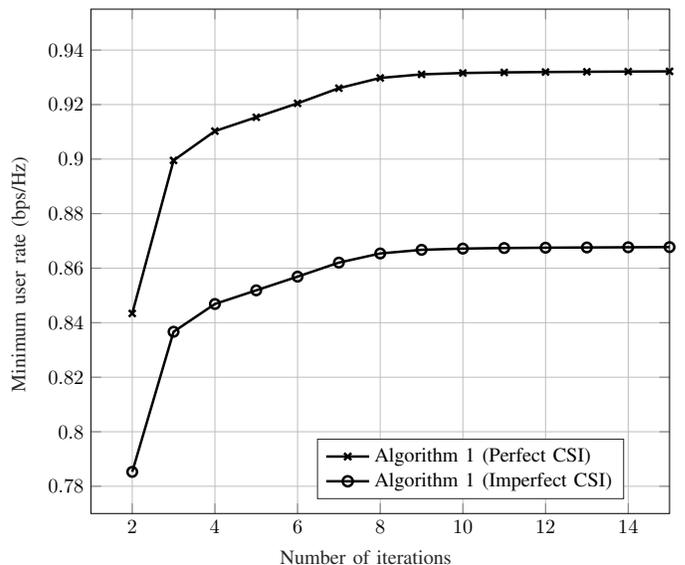

Finally, we show the convergence behaviour of \textbf{Algorithm 1} in Fig. \ref{conv} by setting $M = 8$, $N=16$, $K=4$ and $\epsilon=\epsilon_1=10^{-4}$. The phase shifts are initialized using the CoM scheme. The minimum user rate, computed using the SINR defined in \eqref{SINR}, is plotted against the number of iterations. It is observed that the minimum rate yielded by the proposed algorithm under both perfect and imperfect CSI increases quickly with the number of iterations and the algorithm converges in less than $15$ iterations.

\section{Conclusion}

In this paper, IRS-assisted wireless communication is envisioned to be an important energy-efficient paradigm  for beyond 5G networks, achieving massive MISO like gains with a lower number of active antennas at the BS. The  passive elements constituting the IRS smartly re-configure the signal propagation by introducing phase shifts onto the impinging electromagnetic waves. This paper proposed the MMSE-DFT channel estimation protocol to estimate the direct and IRS-assisted links and compared it with the existing LS based channel estimation protocols. The MMSE estimates were both analytically and numerically shown to achieve a much lower NMSE than the LS estimates. We then proposed an AO algorithm to maximize the minimum SINR, subject to a transmit power constraint and unit-modulus constraints on the IRS elements. The AO algorithm is proved to converge and is shown to yield excellent performance gains in the simulation results that compared the performance of the proposed IRS-assisted system to the conventional MISO system under imperfect CSI. However, the results also highlighted the high sensitivity of the IRS-assisted systems to the quality of the estimates and the rate loss due to channel training.

For future research, it is important to develop low overhead channel estimation protocols where the number of required sub-phases can be reduced to avoid long channel training times. It is also important to make the channel estimation protocols robust in high-speed environments. Another important direction is to study the impact of discrete phase shifts on the performance of the IRS-assisted systems under imperfect CSI. The work can also be extended to multiple IRSs-assisted communication systems as well as IRS-assisted multi-cell systems, where pilot contamination will play a detrimental role in channel estimation. 

\appendix
\subsection{Proof of Lemma \ref{L1}} \label{Sec:L1}
Since both $\tilde{\mathbf{r}}_{1,k}^{tr}$ and $\mathbf{h}_{d,k}$ are jointly Gaussian,  the MMSE estimator is linear. Given the observed training signal, $\tilde{\mathbf{r}}_{1,k}^{tr}$ in \eqref{RR1}, the MMSE estimate of $\mathbf{h}_{d,k}$ is given as 
\begin{align}
\hat{\mathbf{h}}_{d,k}=\mathbf{W}\tilde{\mathbf{r}}_{1,k}^{tr},
\end{align}
 where $\mathbf{W}$ is found as the solution to $ \text{min}_{\mathbf{W}} \hspace{.03in} \text{tr}(\mathbb{E}[(\hat{\mathbf{h}}_{d,k}-\mathbf{h}_{d,k})(\hat{\mathbf{h}}_{d,k}-\mathbf{h}_{d,k})^H])$ and turns out to be 
\begin{align}
\label{e}
&\mathbf{W}=\mathbb{E}[\tilde{\mathbf{r}}_{1,k}^{tr}\mathbf{h}_{d,k}^H](\mathbb{E}[\tilde{\mathbf{r}}_{1,k}^{tr}\tilde{\mathbf{r}}_{1,k}^{trH}])^{-1}.
\end{align}
 Noting that $\mathbf{n}^{tr}_{k}$ and $\mathbf{h}_{d,k}$ are independent random vectors we obtain 
\begin{align}
&\mathbb{E}[\tilde{\mathbf{r}}_{1,k}^{tr}\mathbf{h}_{d,k}^H]=\mathbb{E}\left[\left(\mathbf{h}_{d,k}+\frac{1}{S}(\mathbf{v}_1^{tr}\otimes \mathbf{I}_M)^H \frac{\mathbf{n}^{tr}_{k}}{P_C \tau_S}\right)\mathbf{h}_{d,k}^H\right], \nonumber \\
\label{1_e}
&=\mathbb{E}[\mathbf{h}_{d,k}\mathbf{h}_{d,k}^H]=\beta_{d,k}\mathbf{R}_{BS_k},  
\end{align}
and $\mathbb{E}[\tilde{\mathbf{r}}_{1,k}^{tr}\tilde{\mathbf{r}}_{1,k}^{trH}]=$
\begin{align}
&\mathbb{E}[\mathbf{h}_{d,k}\mathbf{h}_{d,k}^H]+\frac{(\mathbf{v}_1^{tr}\otimes \mathbf{I}_M)^H \mathbb{E}\left[\mathbf{n}^{tr}_{k} \mathbf{n}^{tr^H}_{k}\right](\mathbf{v}_1^{tr}\otimes \mathbf{I}_M)}{S^2 (P_C \tau_S)^2} \\
\label{2_ee}
&=\beta_{d,k}\mathbf{R}_{BS_k}+  \frac{1}{S^2}\frac{\sigma^2 P_C \tau_S}{(P_C\tau_S)^2}(\mathbf{v}_1^{tr}\otimes \mathbf{I}_M)^H \mathbf{I}_{MS} (\mathbf{v}_1^{tr}\otimes \mathbf{I}_M) , \\
&=\beta_{d,k}\mathbf{R}_{BS_k}+  \frac{1}{S^2}\frac{\sigma^2 }{P_C\tau_S} (\mathbf{v}_1^{tr^H}\mathbf{v}_1^{tr}\otimes \mathbf{I}_M)\\
\label{2_e}
&= \beta_{d,k}\mathbf{R}_{BS_k}+  \frac{1}{S}\frac{\sigma^2}{P_C\tau_S} \mathbf{I}_M,
\end{align}
where \eqref{2_ee} follows by noting that $\mathbb{E}\left[\mathbf{n}^{tr}_{k}\mathbf{n}^{tr^H}_{k}\right]=$
\begin{align}
\label{nn}
&\mathbb{E}\left[\mathbf{n}^{tr}_{s,k}\mathbf{n}^{tr^H}_{s,k}\otimes \mathbf{I}_S\right]=\mathbb{E}\left[\mathbf{N}^{tr}_{s}\mathbf{x}_{p,k} \mathbf{x}_{p,k}^H\mathbf{N}^{tr^H}_{s}\right]\otimes \mathbf{I}_S, \\
&=\sigma^2 \mathbf{I}_M \text{tr}(\mathbf{x}_{p,k} \mathbf{x}_{p,k}^H)\otimes \mathbf{I}_S=\sigma^2 P_C \tau_S\mathbf{I}_{MS},
\end{align}
 and \eqref{2_e} follows from $\mathbf{v}_1^{tr^H}\mathbf{v}_1^{tr}=S$ under the DFT design for $\mathbf{V}^{tr}$.

Therefore using \eqref{1_e} and \eqref{2_e} in \eqref{e} we obtain
\begin{align}
&\hat{\mathbf{h}}_{d,k}=\beta_{d,k}\mathbf{R}_{BS_k} \left(\beta_{d,k}\mathbf{R}_{BS_k}+ \frac{\sigma^2 \mathbf{I}_M}{S P_C \tau_S}  \right)^{-1} \tilde{\mathbf{r}}_{1,k}^{tr}.
\end{align}

Moreover it is clear that $\hat{\mathbf{h}}_{d,k}$ is a complex Gaussian vector, the covariance matrix for which can be computed as
\begin{align}
&\mathbb{E}[\hat{\mathbf{h}}_{d,k} \hat{\mathbf{h}}_{d,k}^H]=\beta_{d,k}\mathbf{R}_{BS_k} (\mathbb{E}[\tilde{\mathbf{r}}_{1,k}^{tr}\tilde{\mathbf{r}}_{1,k}^{trH}])^{-1} \mathbb{E}[\tilde{\mathbf{r}}_{1,k}^{tr}\tilde{\mathbf{r}}_{1,k}^{trH}] \nonumber \\
&\left(\mathbb{E}[\tilde{\mathbf{r}}_{1,k}^{tr}\tilde{\mathbf{r}}_{1,k}^{trH}] \right)^{-1} \beta_{d,k}\mathbf{R}_{BS_k}^H\nonumber \\ 
&= \beta^2_{d,k} \mathbf{R}_{BS_k} \left(\beta_{d,k}\mathbf{R}_{BS_k}+ \frac{\sigma^2 \mathbf{I}_M}{S P_S \tau_S}  \right)^{-1} \mathbf{R}_{BS_k}^H.
\end{align}
This completes the proof of Lemma 1.

\subsection{Proof of Lemma \ref{L2}} \label{Sec:L2}

Given the observed training signal, $\tilde{\mathbf{r}}_{n+1,k}^{tr}$ in \eqref{RR2}, we can write the MMSE estimate of $\mathbf{h}_{0,n,k}$ as 
\begin{align}
\label{ee}
\hat{\mathbf{h}}_{0,n,k}=\mathbf{W}\tilde{\mathbf{r}}_{n+1,k}^{tr},
\end{align}
 where $\mathbf{W}=\mathbb{E}[\tilde{\mathbf{r}}_{n+1,k}^{tr}\mathbf{h}_{0,n,k}^H](\mathbb{E}[\tilde{\mathbf{r}}_{n+1,k}^{tr}\tilde{\mathbf{r}}_{n+1,k}^{trH}])^{-1}$.  Noting that $\mathbf{n}^{tr}_{k}$ and $\mathbf{h}_{0,n,k}$ are independent random vectors we obtain
\begin{align}
&\mathbb{E}[\tilde{\mathbf{r}}_{n+1,k}^{tr}\mathbf{h}_{0,n,k}^H]=\mathbb{E}\left[\left(\mathbf{h}_{0,n,k}+\frac{(\mathbf{v}_{n+1}^{tr}\otimes \mathbf{I}_M)^H  \mathbf{n}_{k}^{tr}}{S P_C \tau_S}\right)\mathbf{h}_{0,n,k}^H\right], \nonumber \\
&=\mathbb{E}\left[\mathbf{h}_{0,n,k}\mathbf{h}_{0,n,k}^H\right]=\mathbf{h}_{1,n}\mathbf{h}_{1,n}^H\mathbb{E}[\mathbf{h}_{2,k}(n)\mathbf{h}_{2,k}(n)^*], \\
\label{11_e}
&= r_{n,k} \beta_{2,k}\mathbf{h}_{1,n}\mathbf{h}_{1,n}^H,
\end{align}
where $\mathbf{h}_{1,n}$ is the $n^{th}$ column of $\mathbf{H}_1$ and $r_{n,k}$ is element $(n,n)$ of $\mathbf{R}_{IRS_k}$. Next we obtain the expression of $\mathbb{E}[\tilde{\mathbf{r}}_{n+1,k}^{tr}\tilde{\mathbf{r}}_{n+1,k}^{trH}]=$
\begin{align}
&\mathbb{E}[\mathbf{h}_{0,n,k}\mathbf{h}_{0,n,k}^H]+\frac{(\mathbf{v}_{n+1}^{tr}\otimes \mathbf{I}_M)^H \mathbb{E}[\mathbf{n}^{tr}_{k}\mathbf{n}^{tr^H}_{k} ](\mathbf{v}_{n+1}^{tr}\otimes \mathbf{I}_M)}{S^2 (P_C \tau_S)^2}, \\
\label{22_e}
&=r_{n,k} \beta_{2,k}\mathbf{h}_{1,n}\mathbf{h}_{1,n}^H +\frac{\sigma^2 \mathbf{I}_M}{S P_C \tau_S}.
\end{align}
where $\mathbb{E}[\mathbf{n}^{tr}_{k}\mathbf{n}^{tr^H}_{k} ]$ is computed using similar steps as done in \eqref{nn}. The expression in \eqref{22_e} then follows from realizing that $\mathbf{v}_{n+1}^{tr^H}\mathbf{v}_{n+1}^{tr}=S$ under the proposed DFT design for $\mathbf{V}^{tr}$. Using \eqref{11_e} and \eqref{22_e} in \eqref{ee} we obtain
\begin{align}
&\hat{\mathbf{h}}_{0,n,k}=r_{n,k} \beta_{2,k} \mathbf{h}_{1,n}\mathbf{h}_{1,n}^H  \Big(r_{n,k} \beta_{2,k} \mathbf{h}_{1,n}\mathbf{h}_{1,n}^H \nonumber \\
&+\frac{\sigma^2 \mathbf{I}_M}{S P_C \tau_S}\Big)^{-1} \tilde{\mathbf{r}}_{n+1,k}^{tr}.
\end{align}
Moreover it is clear that $\hat{\mathbf{h}}_{0,n,k}$ is a complex Gaussian vector, the covariance matrix $\Psi_{n,k}=\mathbb{E}[\hat{\mathbf{h}}_{0,n,k} \hat{\mathbf{h}}_{0,n,k}^H]$ for which can be straightforwardly computed.

This completes the proof of Lemma 2.

\bibliographystyle{IEEEtran}
\bibliography{bib}

\begin{IEEEbiography}[{\includegraphics[width=1 in, height=1.35 in,clip,keepaspectratio ]{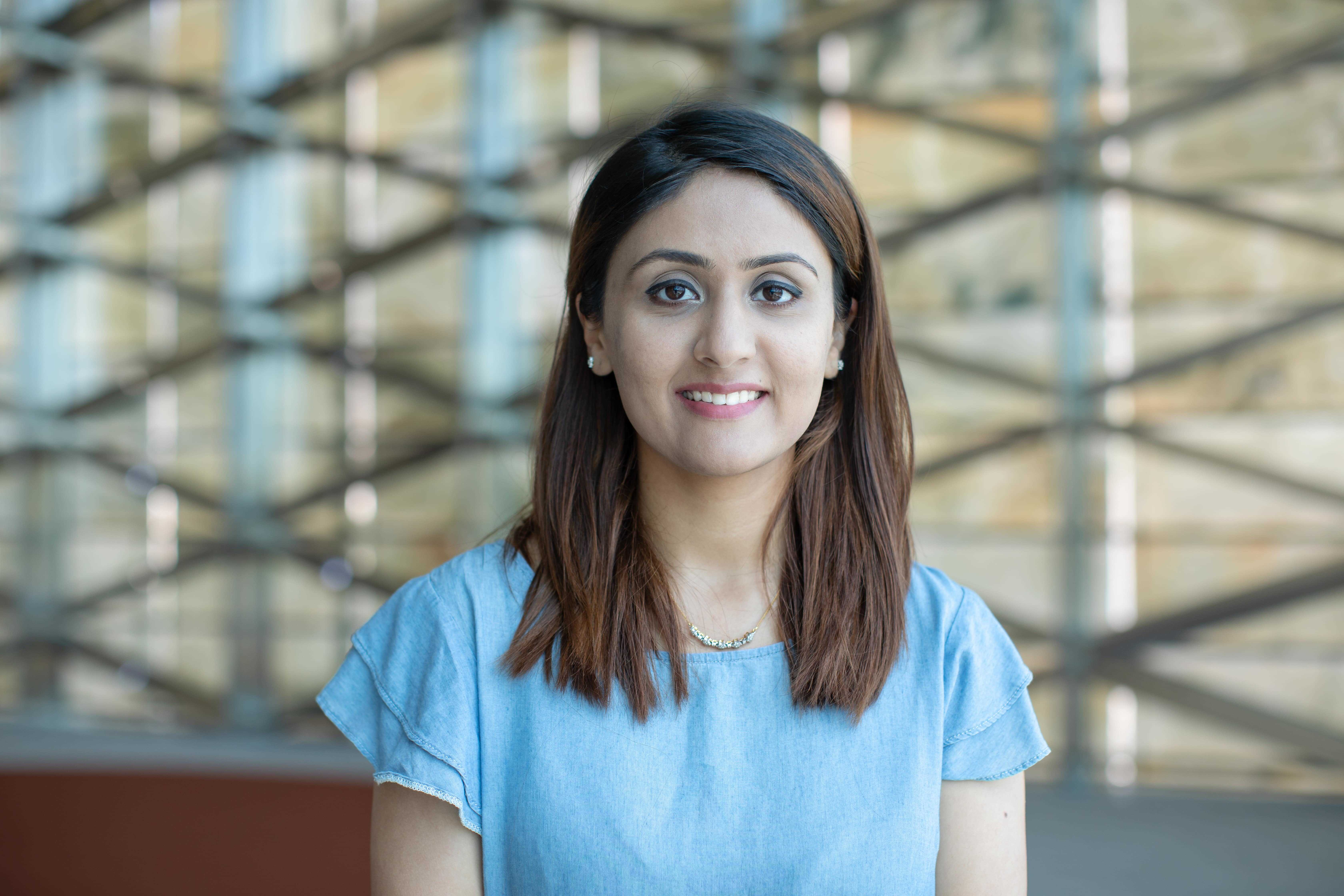}}]{Qurrat-Ul-Ain Nadeem}
(S'15, M'19) was born in Lahore, Pakistan. She received the B.S. degree in Electrical Engineering from Lahore University of Management Sciences (LUMS), Pakistan in 2013 and the M.S. and Ph.D. degrees in Electrical Engineering from King Abdullah University of Science and Technology (KAUST), Thuwal, Makkah Province, Saudi Arabia in 2015 and 2018 respectively. She is currently a Postdoctoral Research Fellow in the Electrical Engineering department at the University of British Columbia. She was selected as the Paul Baran Young Scholar by The Marconi Society in 2018 for her work on full-dimension MIMO. Her research interests include random matrix theory, channel modeling and performance analysis of wireless communication systems.
\end{IEEEbiography}

\begin{IEEEbiography}[{\includegraphics[width=1 in, height=1.35 in,clip,keepaspectratio ]{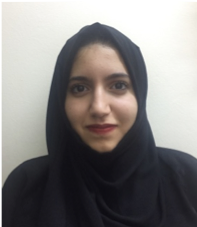}}]{Hibatallah Alwazani} (S'18) received her Bachelors of Science degree in Electrical and Computer Engineering from Effat University, Saudi Arabia, in 2019. She worked as an electrical engineer with Nanomedicine Lab in King Abdullah University of Science and Technology, Saudi Arabia, in 2019. She joined the University of British Columbia as a Masters of Applied Science in Electrical Engineering student in 2020. Her research interests are MIMO communications, machine learning, and optimization.

\end{IEEEbiography}

\begin{IEEEbiographynophoto}{Abla Kammoun} (M'10) was born in Sfax, Tunisia. She received the Engineering degree in signal and systems from the Tunisia Polytechnic School, La Marsa, and the Master's and Ph.D. degrees in digital communications from Telecom Paris Tech [then {\'E}cole Nationale Sup{\'e}rieure des T{\'e}l{\'e}communications (ENST)]. From 2010 to 2012, she was a Post-Doctoral Researcher at the TSI Department, Telecom Paris Tech. Then, she was at Supelec, Alcatel-Lucent Chair on Flexible Radio, until 2013. She is currently a Research Scientist at KAUST. Her research interests include performance analysis of wireless communication systems, random matrix theory, and statistical signal processing. 
\end{IEEEbiographynophoto}

\begin{IEEEbiography}[{\includegraphics[width=1 in, height=1.35 in,clip,keepaspectratio ]{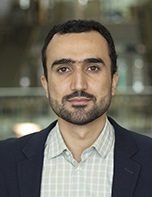}}]{Anas Chaaban} (S'09 - M'14 - SM'17) received the Ma{\^i}trise {\`e}s Sciences degree in electronics from Lebanese University, Lebanon, in 2006, the M.Sc. degree in communications technology and the Dr. Ing. (Ph.D.) degree in electrical engineering and information technology from the University of Ulm and the Ruhr-University of Bochum, Germany, in 2009 and 2013, respectively. From 2008 to 2009, he was with the Daimler AG Research Group On Machine Vision, Ulm, Germany. He was a Research Assistant with the Emmy-Noether Research Group on Wireless Networks, University of Ulm, Germany, from 2009 to 2011, which relocated to the Ruhr-University of Bochum in 2011. He was a Postdoctoral Researcher with the Ruhr-University of Bochum from 2013 to 2014, and with King Abdullah University of Science and Technology from 2015 to 2017. He joined the School of Engineering at the University of British Columbia as an Assistant Professor in 2018. His research interests are in the areas of information theory and wireless communications.
\end{IEEEbiography}

\begin{IEEEbiography}[{\includegraphics[width=1 in, height=1.5 in,clip,keepaspectratio ]{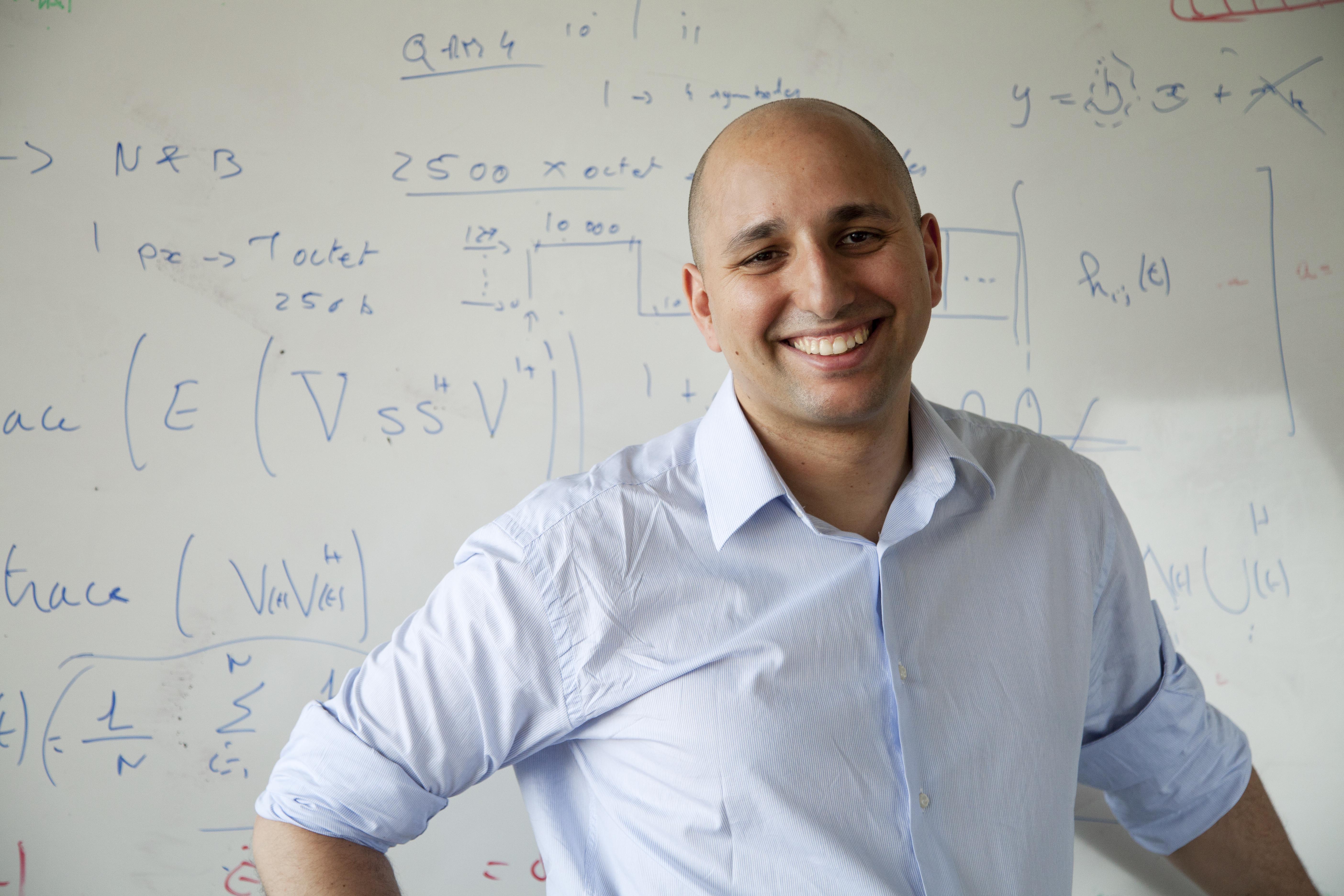}}]{M{\'e}rouane Debbah} (S’01–M’04–SM’08–F’15) received the M.Sc. and Ph.D. degrees from the Ecole Normale Sup{\'e}rieure Paris-Saclay, France. He was with Motorola Labs, Saclay, France, from 1999 to 2002, and also with the Vienna Research Center for Telecommunications, Vienna, Austria, until 2003. From 2003 to 2007, he was an Assistant Professor with the Mobile Communications Department, Institut Eurecom, Sophia Antipolis, France. In 2007, he was appointed Full Professor at CentraleSupelec, Gif-sur-Yvette, France. From 2007 to 2014, he was the Director of the Alcatel-Lucent Chair on Flexible Radio. Since 2014, he has been Vice-President of the Huawei France Research Center. He is jointly the director of the Mathematical and Algorithmic Sciences Lab as well as the director of the Lagrange Mathematical and Computing Research Center. He has managed 8 EU projects and more than 24 national and international projects. His research interests lie in fundamental mathematics, algorithms, statistics, information, and communication sciences research. He is an IEEE Fellow, a WWRF Fellow, and a Membre {\'e}m{\'e}rite SEE. He was a recipient of the ERC Grant MORE (Advanced Mathematical Tools for Complex Network Engineering) from 2012 to 2017. He was a recipient of the Mario Boella Award in 2005, the IEEE Glavieux Prize Award in 2011, the Qualcomm Innovation Prize Award in 2012 and the 2019 IEEE Radio Communications Committee Technical Recognition Award. He received 20 best paper awards, among which the 2007 IEEE GLOBECOM Best Paper Award, the Wi-Opt 2009 Best Paper Award, the 2010 Newcom++ Best Paper Award, the WUN CogCom Best Paper 2012 and 2013 Award, the 2014 WCNC Best Paper Award, the 2015 ICC Best Paper Award, the 2015 IEEE Communications Society Leonard G. Abraham Prize, the 2015 IEEE Communications Society Fred W. Ellersick Prize, the 2016 IEEE Communications Society Best Tutorial Paper Award, the 2016 European Wireless Best Paper Award, the 2017 Eurasip Best Paper Award, the 2018 IEEE Marconi Prize Paper Award, the 2019 IEEE Communications Society Young Author Best Paper Award and the Valuetools 2007, Valuetools 2008, CrownCom 2009, Valuetools 2012, SAM 2014, and 2017 IEEE Sweden VT-COM-IT Joint Chapter best student paper awards. He is an Associate Editor-in-Chief of the journal Random Matrix: Theory and Applications. He was an Associate Area Editor and Senior Area Editor of the IEEE TRANSACTIONS ON SIGNAL PROCESSING from 2011 to 2013 and from 2013 to 2014, respectively.
\end{IEEEbiography}

\begin{IEEEbiography}[{\includegraphics[width=1 in, height=1.35 in,clip,keepaspectratio ]{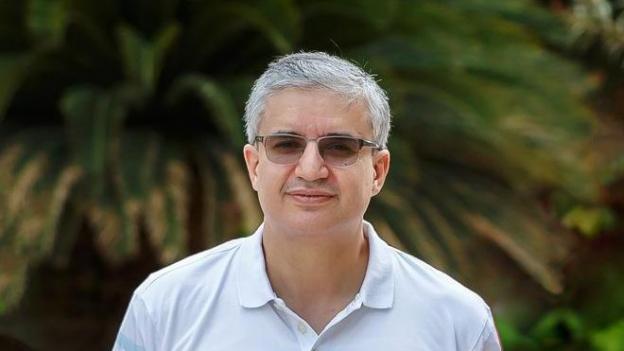}}]{Mohamed-Slim Alouini} 
(S'94, M'98, SM'03, F'09) was born in Tunis, Tunisia. He received the Ph.D. degree in Electrical Engineering from the California Institute of Technology (Caltech), Pasadena, CA, USA, in 1998. He served as a faculty member in the University of Minnesota, Minneapolis, MN, USA, then in the Texas A\&M University at Qatar, Education City, Doha, Qatar before joining King Abdullah University of Science and Technology (KAUST), Thuwal, Makkah Province, Saudi Arabia as a Professor of Electrical Engineering in 2009. His current research interests include the modeling, design, and
performance analysis of wireless communication systems.
\end{IEEEbiography}

\end{document}